\documentclass[preprints,review,accept,pdftex,moreauthors]{Definitions/mdpi} 

\firstpage{1} 
\pubvolume{1}
\issuenum{1}
\articlenumber{0}
\pubyear{2025}
\copyrightyear{2025}
\externaleditor{Firstname Lastname}
\datereceived{28 July 2025} 
\daterevised{26 August 2025} 
\dateaccepted{8 September 2025} 
\datepublished{} 
\hreflink{https://\linebreak doi.org/} 

\usepackage{luatex85}

\Title{Experimental Review of the Quarkonium Physics at the LHC}

\TitleCitation{Experimental Review of the Quarkonium Physics at the LHC}

\Author{Yiyang Zhao $^{\dagger}$\orcidB{}, Jinfeng Liu $^{\dagger}$\orcidC{}, Xing Cheng\orcidD{}, Chi Wang\orcidE{} and Zhen Hu $^{*}$\orcidA{}}

\AuthorNames{Firstname Lastname, Firstname Lastname and Firstname Lastname}

\AuthorCitation{Zhao, Y.; Liu, J.; Cheng, X.; Wang, C.; Hu, Z.}

\address[1]{%
Department of Physics \& Center for High Energy Physics, Tsinghua University, Beijing 100084, China; 

yiyang.zhao@cern.ch (Y.Z.); jinfeng.liu@cern.ch (J.L.); xing.cheng@cern.ch (X.C.); chi.w@cern.ch (C.W.)
}

\corres{\hangafter=1 \hangindent=1.05em \hspace{-0.82em} Correspondence: zhenhu@tsinghua.edu.cn}

\firstnote{\hangafter=1 \hangindent=1.05em \hspace{-0.82em} These authors contributed equally to this work. }  
\secondnote{}

\abstract{\textls[-15]{We review recent heavy quarkonium measurements in $pp$, $p\mathrm{Pb}$, and $\mathrm{PbPb}$ collisions at the LHC by the ALICE, ATLAS, CMS, and LHCb collaborations using Run‑2 and early Run‑3 data. Production studies include present differential cross-sections and polarization measurements of charmonium and bottomonium, providing precise tests of QCD theoretical calculations and unveiling symmetry relations among spin and orbital configurations. Notably, a $t\bar{t}$ quasi-bound-state has been observed at the LHC recently. Suppression analyses quantify the sequential melting of bottomonium states in $\mathrm{PbPb}$ collisions, serving as a probe of the deconfined quark--gluon plasma. Cold nuclear matter effects are constrained through comparisons of quarkonium yields in $p\mathrm{Pb}$ and $pp$ collisions. Furthermore, multi‑quarkonium investigations observe di‑ and tri‑quarkonium production processes and resonances, exploring multi‑parton interactions and the symmetry structure underlying exotic hadron states.}}


\begin{document}



\section{Introduction}

\subsection{Theoretical Motivation}

Heavy quarkonium is the bound state of a heavy quark (\(c,b,t\)) and its antiquark. While light‐quark (\(u,d,s\)) mesons are inherently relativistic systems, heavy‐quark pairs move at approximately non‐relativistic velocities (typical squared velocities are \(v^2 \approx 0.3\) for charm quark and \(v^2 \approx 0.1\) for bottom quark) within the bound state. This separation of scales (specifically the hierarchy \(m_q \gg m_q v \gg m_q v^2\), corresponding to hard, soft, and ultrasoft scales) enables the effective field theories to describe non-perturbative eﬀects in Quantum Chromodynamics (QCD)~\cite{Bodwin:1994jh}. Therefore, heavy quarkonium serves as a unique probe for investigating the QCD‐governed strong interaction within the Standard Model (SM) and uncovering potential new phenomena.

\textls[-15]{Experimental investigations of heavy quarkonium usually focus on charmonium (\(c\bar c\), especially \(J/\psi\), \(\psi(2\mathrm{S})\)) and bottomonium (\(b\bar b\), especially \(\Upsilon(1\mathrm{S})\), \(\Upsilon(2\mathrm{S})\), \(\Upsilon(3\mathrm{S})\)). The top quark, being the heaviest of all quarks, was long believed to decay before it could hadronize into a bound state~\cite{Bernreuther:2008ju}. Nevertheless, recent analyses of the \(t\bar t\) invariant‑mass differential cross-section have uncovered an excess indicative of toponium (\(t\bar t\)) resonances~\cite{CMS:2025kzt,ATLAS:2025mvr}, thereby opening promising avenues for experimental studies of the toponium family (details in Section~\ref{toponium}).}

Historically, quarkonium production has been challenging to describe from first principles. A variety of effective QCD frameworks have been developed to model the non-perturbative evolution of a heavy $q\bar q$ pair into a color‐neutral bound state. Among these, the color‐singlet model (CSM)~\cite{Cho:1995ce} treats the $q\bar q$ as produced directly in the same quantum state as the physical quarkonium; the color‐octet mechanism (COM) embedded in NRQCD~\cite{Bodwin:1994jh,Petrelli:1997ge} allows intermediate color‐octet $q\bar q$ configurations whose transition to the bound state is encoded in universal long‐distance matrix elements; and the color‐evaporation model (CEM)~\cite{Fritzsch:1977ay,Amundson:1996em} integrates the $q\bar q$ cross-section over the invariant‐mass range below open‐heavy‐flavor threshold. In NRQCD, perturbative short‐distance coefficients are factorized from non-perturbative long‐distance matrix elements, which are assumed to be process independent. Various effective QCD models of quarkonium production yield distinct predictions for production cross-sections and spin alignments. Consequently, precise experimental determinations of these observables are essential for refining our theoretical comprehension of quarkonium production mechanisms.

Under extreme temperature and density, QCD predicts a phase transition from hadronic matter to a deconfined quark–gluon plasma (QGP)~\cite{Matsui:1986dk}. In such a medium, color screening weakens the binding potential between the heavy quark pair. Heavy quarkonium thus serves as a powerful probe of the deconfined QGP.

Heavy quarkonium also serves as ``standard candles’’ for detector calibration and sensitive probes of exotics or physics beyond the SM. 
Precision measurements of quarkonium decay rates into leptonic or exclusive hadronic final states can reveal small departures from Standard Model couplings. For example, \( R_{\tau\mu}^{\Upsilon} = \Gamma(\Upsilon(1S)\to \tau^+\tau^-)\big/\Gamma(\Upsilon(1S)\to \mu^+\mu^-) \) tests lepton-flavor universality, while searches for \(\mathcal{Q}\to \gamma + A'\) or \(\mathcal{Q}\to \text{invisible}\) probe weakly coupled dark states~\cite{Brambilla:2010cs,BaBar:2010pvu,Essig:2013lka,BaBar:2009cka}.
Exotic configurations (tetraquarks~\cite{LHCb:2020bwg,CMS:2025xwt,CMS:2025vnq,ATLAS:2023bft,CMS:2020qwa}, glueballs) are also accessible because many quarkonium decays are gluon rich (\(Q\bar Q\to gg,\, ggg\)) or proceed via radiative transitions \(\mathcal{Q}\to \gamma + X\).

\begin{figure}[H]
 \includegraphics[width=0.75\linewidth]{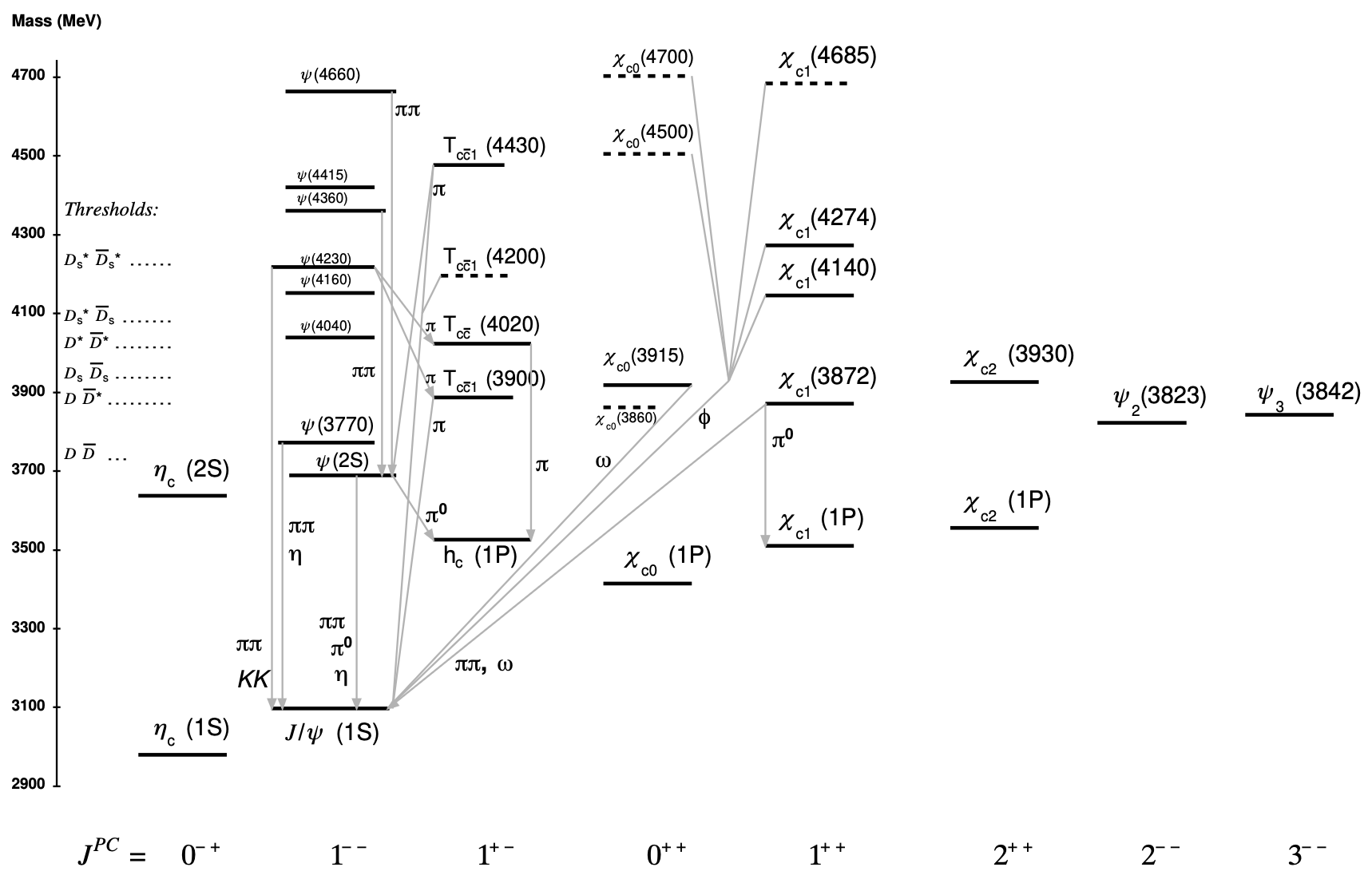}
 \caption{Representation 
 of 
 the charmonium spectrum, adopted from the Particle Data Group 2024 Review~\cite{PDG2024}. Dashed lines denote \(c\bar c\) states not yet experimentally established. Arrows denote the dominant hadronic transitions. For clarity, single-photon transitions such as
\( \psi(\mathrm{n}\mathrm{S})\to\gamma\,\eta_c(\mathrm{mS}), 
 \psi(\mathrm{n}\mathrm{S})\to\gamma\,\chi_{cJ}(\mathrm{1P}), 
 \chi_{cJ}(\mathrm{1P})\to\gamma\,J/\psi \)
have been omitted. The thresholds corresponding to a pair of ground‐state open‐charm mesons are marked in the figure. The $J^P$ quantum numbers of $T_{c\bar c}(4020)$ are still undetermined. Decays into open‐flavor final states are not shown.}
 \label{fig:charmonium}
\end{figure}

The first heavy quarkonium, $J/\psi$, was discovered by the SLAC E598 and Brookhaven experiments in November 1974 almost simultaneously~\cite{PhysRevLett.33.1404,PhysRevLett.33.1406}, which marked the “November Revolution”. Since then, charmonium spectroscopy has expanded rapidly: the SPEAR storage ring resolved the first radial excitation $\psi(2\mathrm{S})$~\cite{Baltrusaitis:1975fn} and, through radiative transitions, the $P$-wave triplet $\chi_{c0,1,2}(\mathrm{1P})$~\cite{Biddick:1977sv,Tanenbaum:1975ef,Whitaker:1976hb, Eichten:1978tg}; subsequent experiments at DORIS and ADONE uncovered the vector state $\psi(3770)$~\cite{Rapidis:1977cv} and higher excitations $\psi(4040)$~\cite{DASP:1978dns}, $\psi(4160)$~\cite{DASP:1978dns} and $\psi(4415)$~\cite{Siegrist:1976br,Eichten:1978tg}. The known charmonium spectrum is summarized in Figure~\ref{fig:charmonium}.

\begin{figure}[H]
 \includegraphics[width=0.9\linewidth]{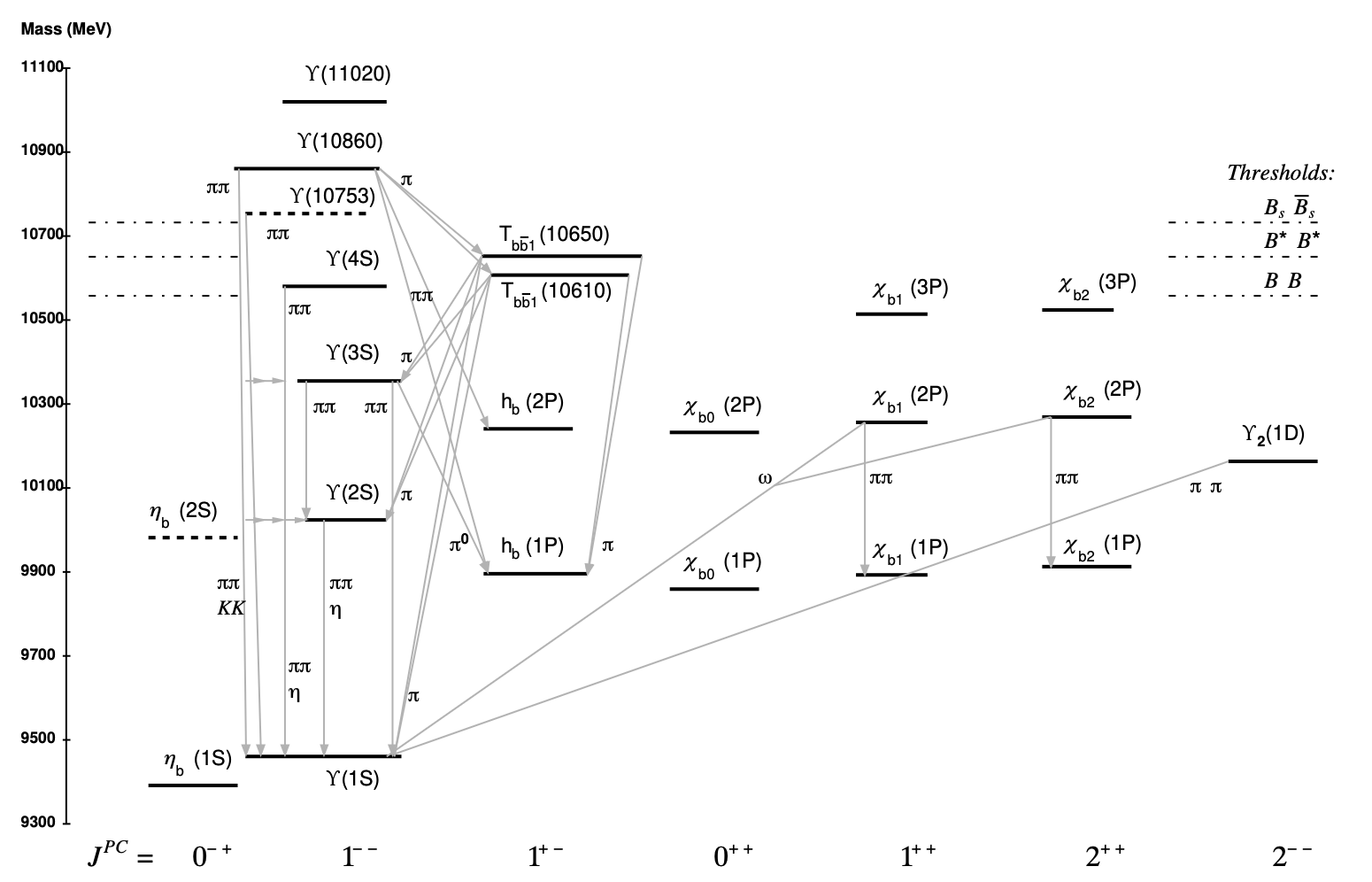}
 \caption{Representation
 of 
 the bottomonium spectrum, adopted from the Particle Data Group 2024 Review~\cite{PDG2024}. Dashed lines denote \(b\bar b\) states not yet experimentally established. Arrows denote the dominant hadronic transitions. For clarity, single-photon transitions such as
\(
 \Upsilon(\mathrm{n}\mathrm{S})\to\gamma\,\eta_b(\mathrm{mS}), 
 \Upsilon(\mathrm{n}\mathrm{S})\to\gamma\,\chi_{bJ}(\mathrm{mP}), 
 \chi_{bJ}(\mathrm{nP})\to\gamma\,\Upsilon(\mathrm{mS})
\)
have been omitted. The thresholds corresponding to a pair of ground‐state open‐bottom mesons are indicated in the figure. Decays into open‐flavor final states are not shown. Moreover, the transitions
\(
 \Upsilon(10753)\to\omega\,\chi_{b1}(\mathrm{1P}), 
 \omega\,\chi_{b2}(\mathrm{1P})
\)
have also been reported.}
 \label{fig:bottomonium}
\end{figure}

Shortly thereafter, in 1977 the Fermilab E288 experiment reported the bottomonium ground state $\Upsilon(1\mathrm{S})$ in proton--nucleus collisions~\cite{Herb:1977prl}, followed by CLEO’s resolution of $\Upsilon(2\mathrm{S})$ and $\Upsilon(3\mathrm{S})$ at CESR~\cite{Andrews:1980prl}, the identification of $\chi_{b1,2}(\mathrm{1P,2P})$~\cite{Haas:1984prl}, and later the discovery of $\Upsilon(\mathrm{4S})$~\cite{CUSB:1980prl} and the vector states $\Upsilon(\mathrm{5S})$, $\Upsilon(\mathrm{6S})$ at CESR and SLAC PEP~\cite{Besson:1985prl}. The asymmetric‐energy $B$ factories (PEP-II, KEKB) and dedicated charm- and bottom-factory programs (CLEO-c, BES) in the 1990s and 2000s improved mass resolution and mapped numerous hadronic and radiative decay channels~\cite{Bevan:2014epjc} as summarized in Figure~\ref{fig:bottomonium}.

With the advent of high‐luminosity runs at the LHC, comprehensive measurements of quarkonium production cross-sections, polarizations, nuclear modification factors and rare transitions have been realized, completing the conventional charmonium and bottomonium spectra and opening a new era for the study of exotic quarkonium‐like states.

\subsection{Datasets and the LHC}

\begin{figure}[H]
 \includegraphics[width=0.55\linewidth]{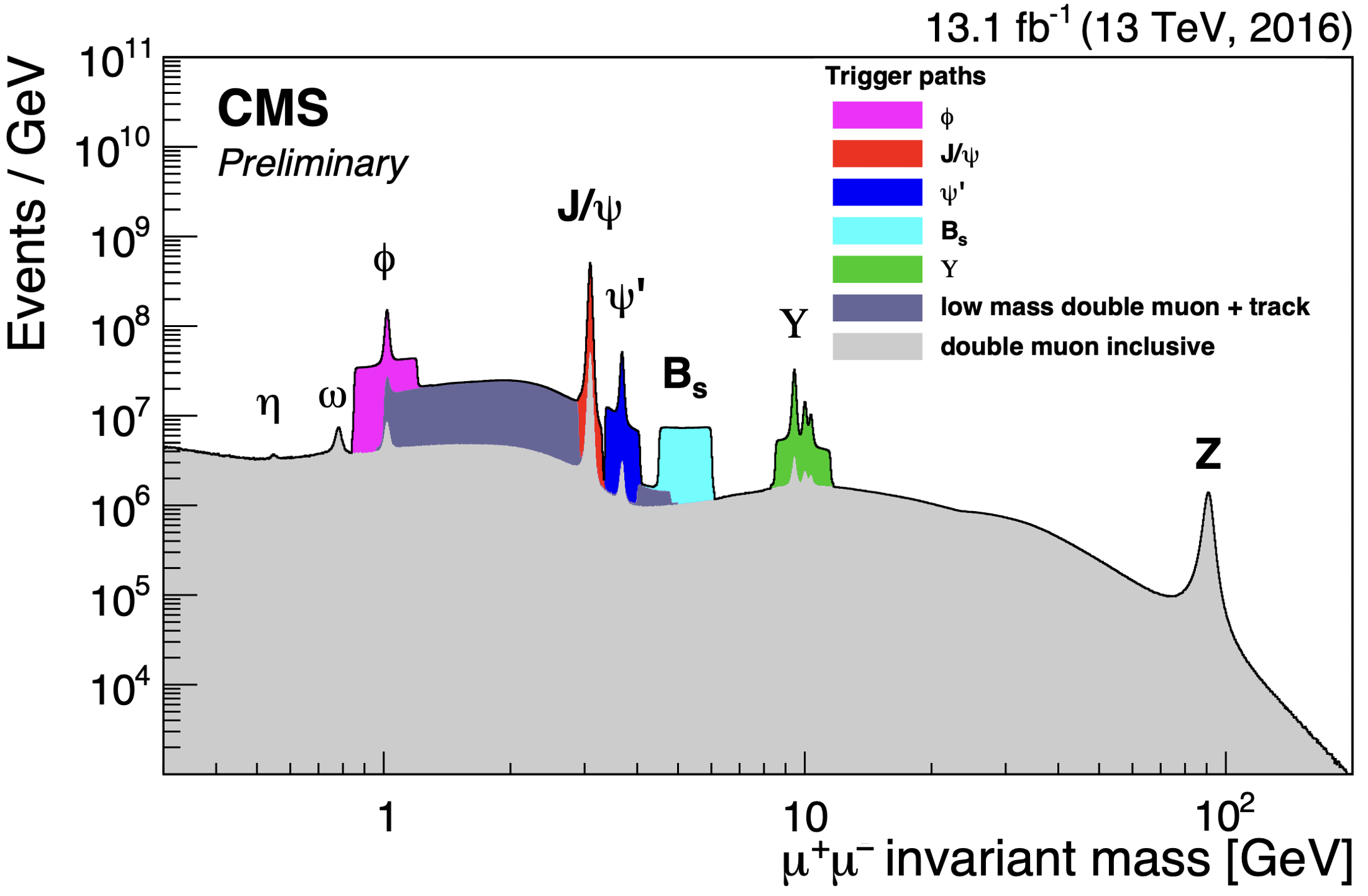}
 \caption{Dimuon
 mass spectrum of CMS in 2016 $pp$ collisions at {13} {TeV}, corresponding to an integrated luminosity of {13.1} {fb$^{-1}$}. Colored shaded bands indicate the mass windows of dedicated trigger paths: $\phi$ (magenta), $J/\psi$ (red), $\psi'$ (blue), $B_s$ (cyan), and $\Upsilon$ (green). The dark and light gray bands show the acceptances of the low‑mass dimuon+track triggers and inclusive dimuon triggers, respectively~\cite{CMS:dimuon}.}
 \label{fig:dimuon_cms}
\end{figure}

\begin{figure}[H]
 \includegraphics[width=0.55\linewidth]{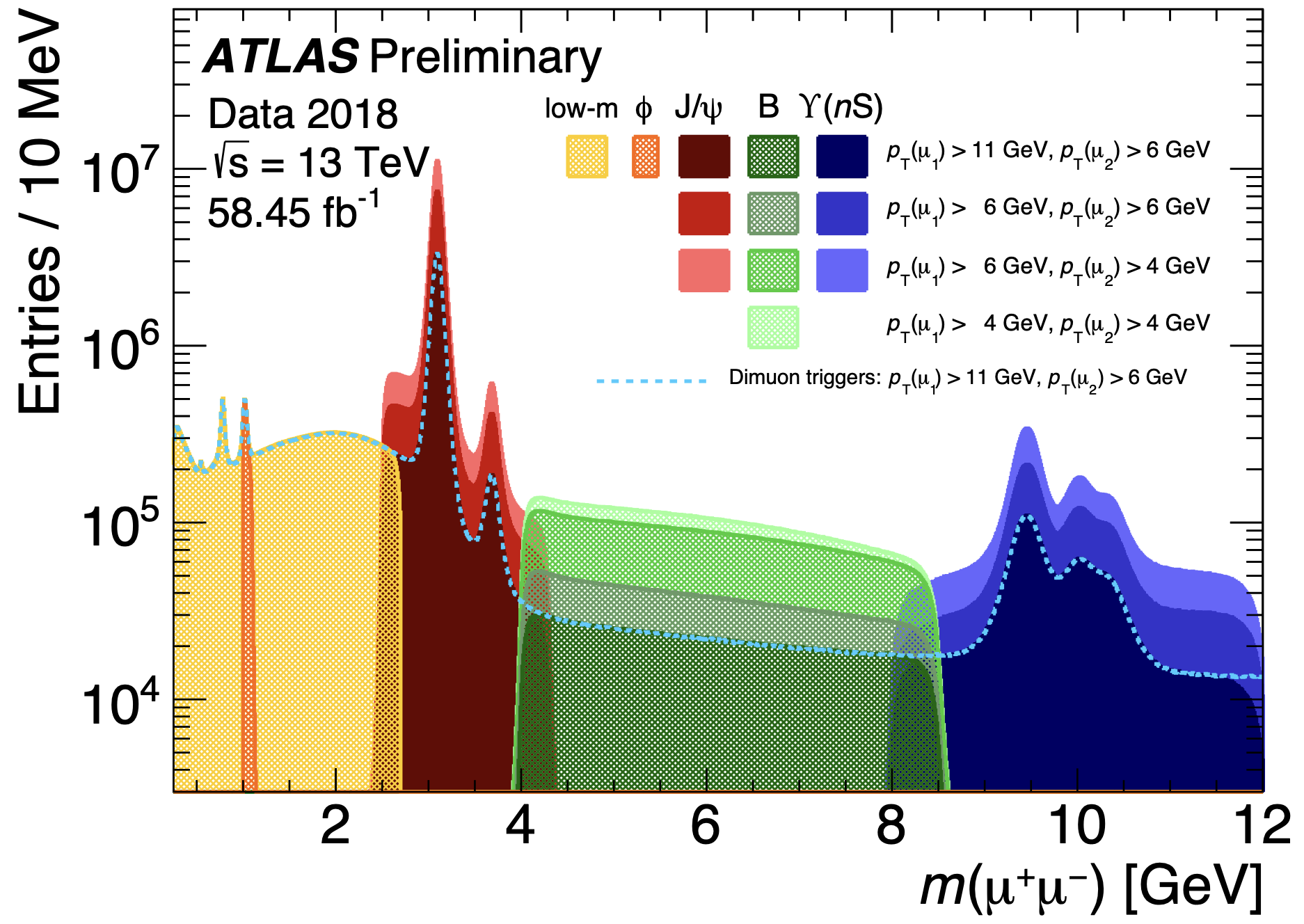}
 \caption{Dimuon 
 mass spectrum of ATLAS in 2018 $pp$ collisions at {13} {TeV}, corresponding to an integrated luminosity of {58.45} {fb$^{-1}$}. Shaded histograms show contributions from low‑mass $\phi$ (yellow), $J/\psi$ (red) and $\Upsilon(\mathrm{nS})$ (green and blue) resonances under different muon transverse‑momentum selections. The cyan dashed curve indicates the dimuon trigger thresholds~\cite{ATLAS:dimuon}.}
 \label{fig:dimuon_atlas}
\end{figure}

\begin{figure}[H]
 \includegraphics[width=0.75\linewidth]{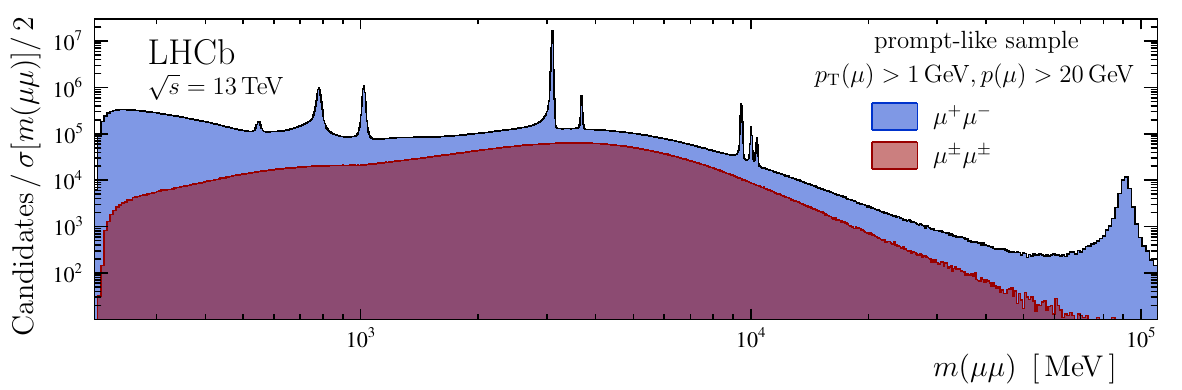}
 \caption{Dimuon 
 mass spectrum of LHCb at 13 TeV divided by opposite or same sign muon pair~\cite{LHCb:dimuon_mass}.}
 \label{fig:dimuon_lhcb}
\end{figure}

\begin{figure}[H]
 \includegraphics[width=0.5\linewidth]{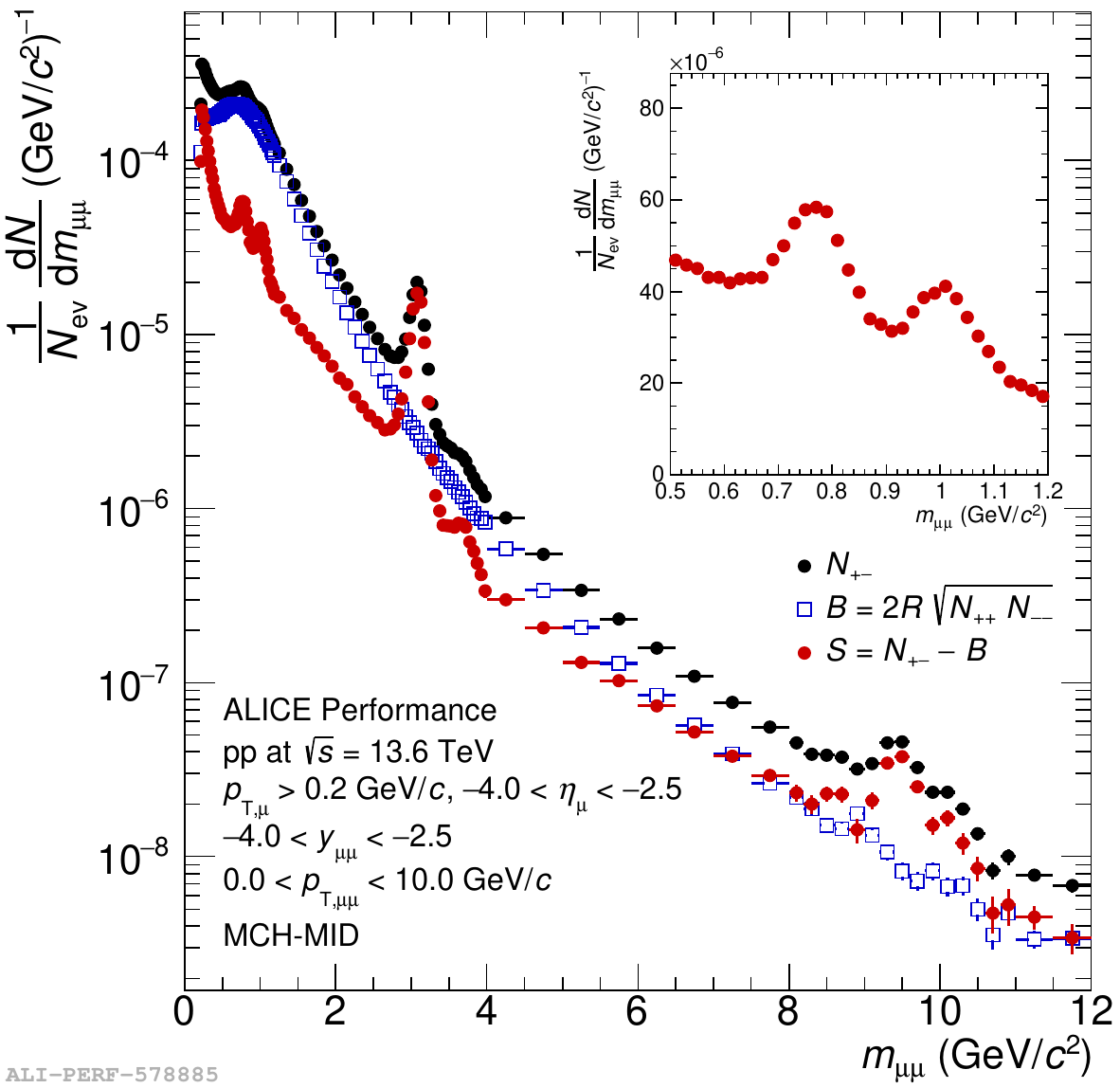}
 \caption{Dimuon mass spectrum in $pp$ collisions at {13.6} {TeV} recorded by ALICE. Black points show opposite-sign pairs $N_{+-}$. Blue open squares denote the like-sign combinatorial background $B=2R\sqrt{N_{++}N_{--}}$. Red circles are the signal $S=N_{+-}-B$. The inset zooms the low-mass region ($0.5<m_{\mu\mu}<{1.2}{GeV}/c^2$)~\cite{Jung:2025fbx}.}
 \label{fig:dimuon_alice}
\end{figure}

The Large Hadron Collider (LHC) at CERN delivers unprecedented collision energies and luminosities for the study of heavy‐quarkonium production and suppression. Four major experiments (ALICE, ATLAS, CMS, and LHCb) provide complementary coverage and measurement capabilities. CMS and ATLAS feature powerful central detectors with high‐resolution silicon trackers, large‐acceptance muon systems and strong solenoidal and toroid magnets, enabling precision dimuon spectroscopy over \(|\eta|<2.4\) and \(|\eta|<2.5\), respectively~\cite{CMS:2008jinst,ATLAS:2008jinst} (Figures~\ref{fig:dimuon_cms} and \ref{fig:dimuon_atlas}). LHCb focuses on the forward region (\(2<\eta<5\)) as a spectrometer optimized for particle identification and vertex reconstruction~\cite{LHCb:2008jinst} (Figure~\ref{fig:dimuon_lhcb}). ALICE combines a low‐magnetic‐field (around \({0.5} {T}\)) in the central barrel and specialized forward muon arm for heavy‐ion physics, allowing measurements down to very low transverse momentum down to \({0.1}{GeV}\) in proton--proton ($pp$), proton--lead (\(p\mathrm{Pb}\)) and lead--lead (\(\mathrm{Pb}\mathrm{Pb}\)) collisions (Figure \ref{fig:dimuon_alice})~\cite{ALICE:2008jinst}.

\vspace{-6pt}

\vspace{-6pt}

\vspace{-6pt}

In this review, we concentrate on results from the most recent data-taking periods of Run-2 and Run-3. 
Section \ref{sec2} is devoted to the production characteristics, including cross-sections and polarization observables of both individual charmonium and bottomonium states. Section \ref{toponium} provides a brief overview of the latest theoretical and experimental studies on toponium production and its potential observability, highlighting prospects for future exploration. 
Sections \ref{sec31} to \ref{sec32} address the suppression of quarkonium in heavy‑ion collisions, examining how the quark--gluon plasma (QGP) modifies quarkonium yields and kinematic distributions. Sections \ref{sec33} and \ref{sec34} focus on multiplicity dependent production and production in ultraperipheral collisions respectively.
Section \ref{sec4} addresses the associated production of multiple quarkonium, exploiting double and triple quarkonium final states as probes of parton correlations and double parton scattering in the \mbox{high‐energy regime}.

\begin{figure}[H]
 \includegraphics[width=\linewidth]{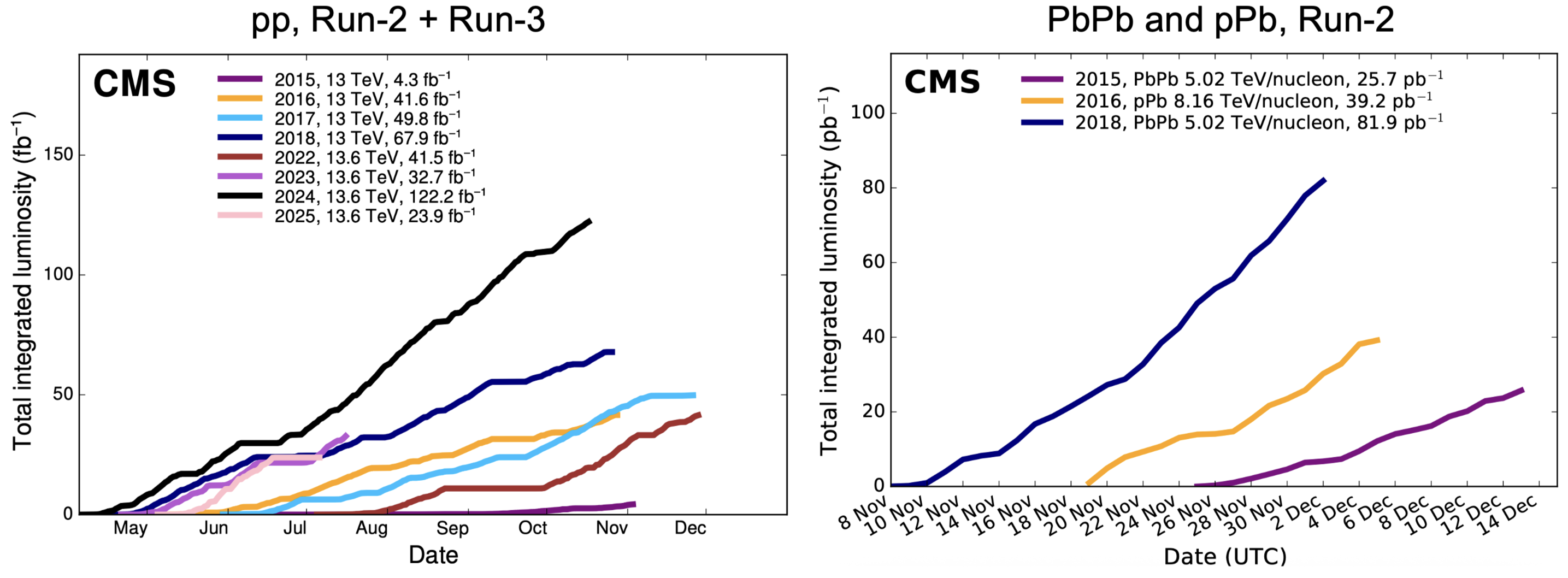}
 \caption{Integrated luminosity recorded by the CMS experiment in \(pp\), \(p\mathrm{Pb}\) and \(\mathrm{Pb}\mathrm{Pb}\) collisions during LHC Run-2 and Run-3. Left: the integrated luminosity delivered by the LHC and recorded by CMS during $pp$ collisions in Run-2 and Run-3. Right: the integrated luminosity for $\mathrm{Pb}\mathrm{Pb}$ and $p\mathrm{Pb}$ collisions recorded by CMS in Run-2~\cite{CMSLUM2015-16,CMS-PAS-LUM-22-001}.}
 \label{fig:cms-lumi}
\end{figure}

The datasets surveyed in this review encompass \(pp\), \(p\mathrm{Pb}\) and \(\mathrm{Pb}\mathrm{Pb}\) collisions at several center‐of‐mass energies. During Run-2 (2015–2018), the LHC delivered \(pp\) collisions at \(\sqrt{s}={13}~\text{TeV}\), \(p\mathrm{Pb}\) collisions at \(\sqrt{s_{\mathrm{NN}}}={5.02}~\text{TeV}\) and \(\sqrt{s_{\mathrm{NN}}}={8.16}~\text{TeV}\), and \(\mathrm{Pb}\mathrm{Pb}\) collisions at \(\sqrt{s_{\mathrm{NN}}}={2.76}~\text{TeV}\) and \(\sqrt{s_{\mathrm{NN}}}={5.02}~\text{TeV}\). In Run-3 (2022–2026), the \(pp\) collision energy has been  increased to \(\sqrt{s}={13.6}~\text{TeV}\), while \(\mathrm{Pb}\mathrm{Pb}\) operations have continued at \(\sqrt{s_{\mathrm{NN}}}={5.36}~\text{TeV}\). The integrated luminosity recorded by CMS is shown in Figure~\ref{fig:cms-lumi}.

\section{Production Properties Measurements}\label{sec2}

Theoretical frameworks, such as NRQCD, yield distinct predictions for observables like differential cross-sections and spin-alignment~\cite{Bodwin:1994jh,Petrelli:1997ge,Fritzsch:1977ay,Amundson:1996em}. Discriminating among these theoretical models requires precise experimental measurements of quarkonium \mbox{production properties}.

Recent experimental campaigns at the Tevatron and the LHC have provided a wealth of quarkonium data over a broad range of center-of-mass energies. CDF and D0 first reported measurements of $J/\psi$, $\psi(2\mathrm{S})$ and $\Upsilon(\mathrm{n}\mathrm{S})$ cross-sections and polarizations in $p\bar p$ collisions at $\sqrt{s}=1.8$–1.96 TeV~\cite{Acosta:2001gv,Abulencia:2007us,Affolder:2000nn}. At the LHC, ALICE, ATLAS, CMS, and LHCb have extended these studies in $pp$ collisions at $\sqrt{s}=7$, 8, 13, and 13.6 TeV, reconstructing quarkonium via dilepton decays ($q\bar q\to\mu^+\mu^-, e^+e^-$) with branching fractions of a few percent. Despite these high-precision data, no single theoretical model simultaneously describes both the production cross-section and polarization patterns observed, underscoring the need for comprehensive reviews~\cite{Chatrchyan:2011kc,Aad:2011aea,Aaij:2011jh,Chatrchyan:2017xx,Aaboud:2016kef,Aaij:2015rla,Abelev:2017tra}. In this section, we summarize the current status of quarkonium production in $pp$ collisions, organizing the discussion into measurements of differential cross-sections (Section \ref{sec21}) and polarization observables (Section \ref{sec22}), with emphasis on the most recent LHC Run-2 and Run-3 results.

\subsection{Production Cross-Section}\label{sec21}
The production cross-section of a given quarkonium state \({q \bar q}\) in \(pp\) collisions is defined as the probability per unit luminosity for producing \({q \bar q}\) and is most often expressed differentially in transverse momentum \(p_T\) and rapidity \(y\). Experimentally, the double-differential cross-section times the di-lepton branching fraction is measured as 
\begin{equation}
    \frac{d^2\sigma(pp\to {q \bar q})}{dp_T dy}\;\mathrm{Br}({q \bar q}\to l^+l^-)
\;=\;\frac{N_{q \bar q}}{\mathcal{L} \Delta p_T \Delta y}\;\frac{1}{A(p_T,y) \epsilon(p_T,y)} ,
\end{equation}
where \(N_{q \bar q}\) is the signal yield in the fiducial region extracted from fits to the di-lepton invariant mass spectrum, \(\mathcal{L}\) the integrated luminosity, \(\Delta p_T\) and \(\Delta y\) the bin widths, \(A\) the geometrical acceptance, and \(\epsilon\) the combined trigger and reconstruction efficiency. Corrections for feed-down from higher mass states are included when quoting prompt production. Precise measurements of these cross-sections over a wide kinematic range at \(\sqrt{s}=7\)–13.6 TeV have provided stringent tests of NRQCD factorization and alternative models (COM, CEM), which predict distinct \(p_T\) and \(y\) dependencies~\cite{Bodwin:1994jh,Petrelli:1997ge,Fritzsch:1977ay,Amundson:1996em}.

\subsubsection{Charmonium Production Cross-Section}
The inclusive production cross-sections of the quarkonium states $J/\psi$ and $\psi(2\mathrm{S})$ in $pp$ collisions were determined by several LHC experiments over a wide kinematic range. 

All analyses reconstructed charmonium through its dilepton (${q \bar q} \to\mu^+\mu^-/e^+e^-$) or hadronic ($\psi(2\mathrm{S}) \to J/\psi\,\pi^+\pi^-$) decay channels, extracted raw yields through unbinned maximum-likelihood fits to invariant mass (and pseudo-proper time) distributions, and corrected for detector acceptance $A(p_T,y)$ and efficiency $\varepsilon(p_T,y)$ using detailed Monte Carlo (MC) simulations or data-driven tag-and-probe methods. Backgrounds from incorrect combination of final state particles and feed-down decays were modeled and subtracted, and systematic uncertainties were evaluated by varying fit models, efficiency corrections, luminosity determinations, and branching fraction inputs~\cite{Aaboud:2016kef,Chatrchyan:2017xx}. The exclusive measurement further requires the final-state particles to be sufficiently isolated to suppress inelastic contamination, whereas the cross-section ratio analysis benefits from cancellation of correlated uncertainties.

In $pp$ collisions at $\sqrt{s}={13} {TeV}$, ATLAS, CMS, and LHCb have measured differential cross-sections of $J/\psi$ and $\psi(2\mathrm{S})$, which were reconstructed via their dimuon \mbox{decays~\cite{Aaboud:2016kef,Chatrchyan:2017xx,LHCbCEPCharmonium13TeV2018}}. In ATLAS and CMS measurements, the differential production cross-sections are measured as functions of rapidity $y$ and transverse momentum $p_T$ and compared to theoretical predictions. In contrast, the LHCb analysis provides the cross-section as a function of $y$ only.

\begin{figure}[H]
 \includegraphics[width=0.75\linewidth]{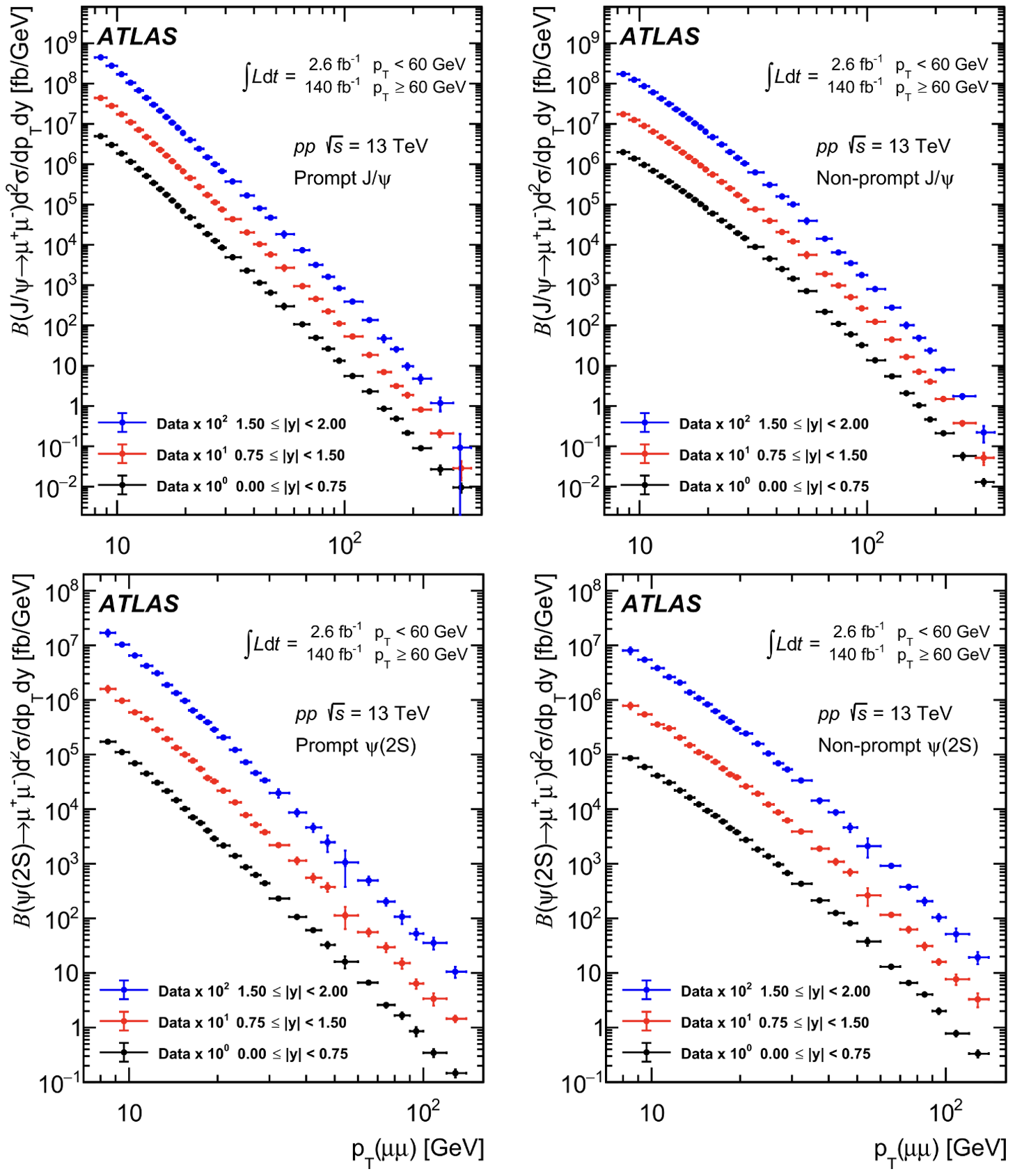}
 \caption{Prompt 
 and non-prompt $J/\psi$ (top) and $\psi(2\mathrm{S})$ (bottom) double-differential cross-sections measured by ATLAS in $pp$ collisions at $\sqrt{s}={13}~\text{TeV}$~\cite{Aaboud:2016kef}.}
 \label{fig:atlas-ccbar}
\end{figure}

\begin{figure}[H]
 \includegraphics[width=0.5\linewidth]{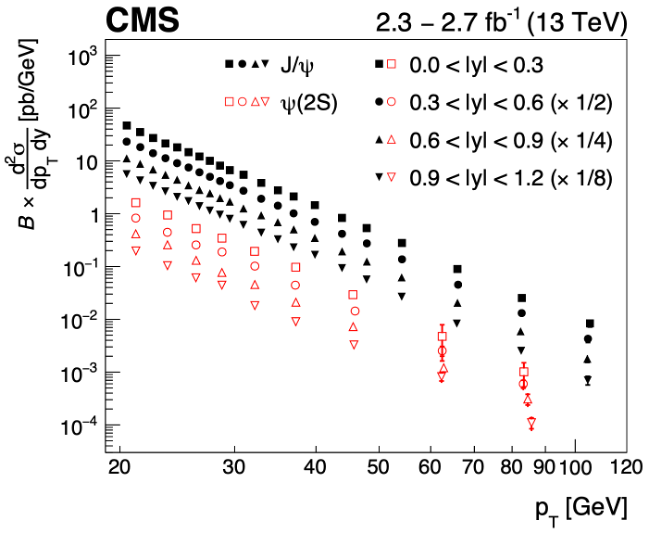}
 \caption{Prompt 
 $J/\psi$ and $\psi(2\mathrm{S})$ double-differential cross-sections measured by CMS in $pp$ collisions at $\sqrt{s}={13}~\text{TeV}$~\cite{Chatrchyan:2017xx}.}
 \label{fig:cms-ccbar}
\end{figure}

\begin{figure}[H]
 \includegraphics[width=\linewidth]{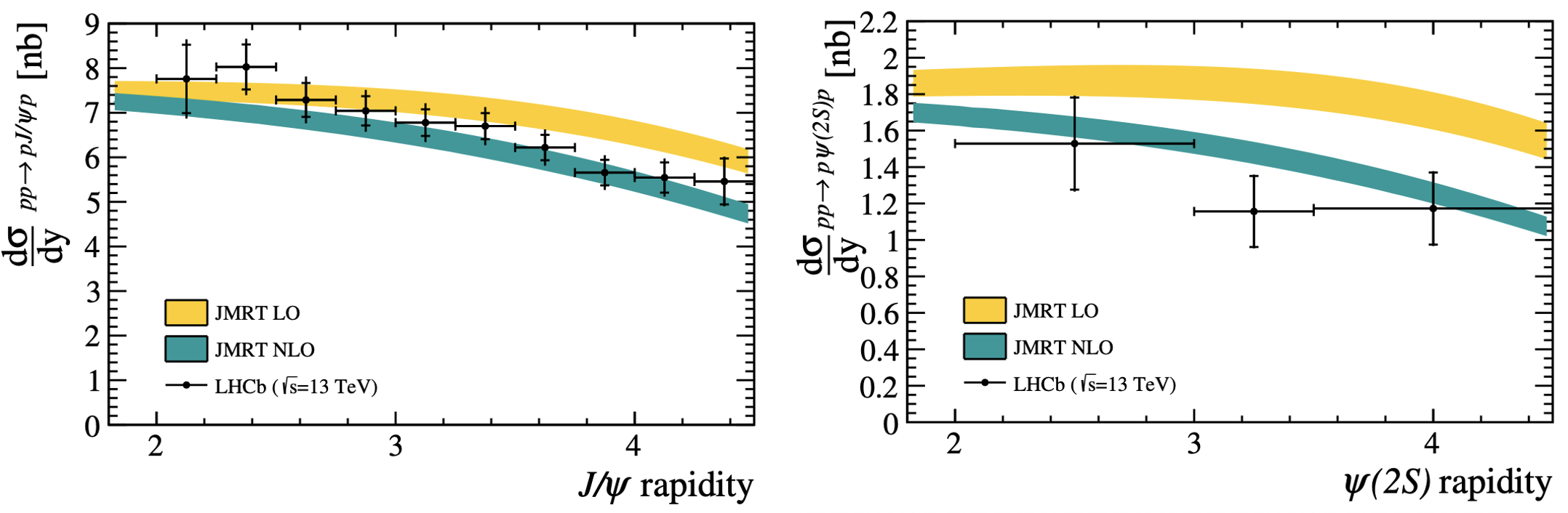}
 \caption{Inclusive $J/\psi$ (\textbf{left}) and $\psi(2\mathrm{S})$ (\textbf{right}) differential cross-sections measured by LHCb in $pp$ collisions at $\sqrt{s}={13}~\text{TeV}$~\cite{LHCbCEPCharmonium13TeV2018}.}
 \label{fig:lhcb-ccbar}
\end{figure}

Among those analyses, ATLAS measures prompt and non-prompt contributions separately (Figure~\ref{fig:atlas-ccbar}) by fitting the pseudo-proper decay length $\ell = L_{xy}\,\frac{m}{p_T}$, where $L_{xy}$ is the transverse displacement between the primary vertex and the dimuon vertex, to distinguish directly produced quarkonium from those originating in $b$-hadron decays. CMS (Figure~\ref{fig:cms-ccbar}) only measures the prompt production cross-section, while LHCb (Figure~\ref{fig:lhcb-ccbar}) focuses on the inclusive production. This difference largely reflects detector vertexing and trigger capabilities: CMS and ATLAS can separate prompt and non-prompt using lifetime information. In terms of observables, ATLAS ($8<p_T<{150}~\text{GeV}$) and CMS ($20<p_T<{100}~\text{GeV}$) report double-differential cross-sections $\mathrm{d}^2\sigma/(\mathrm{d}p_T\,\mathrm{d}y)$, LHCb reports $\mathrm{d}\sigma/\mathrm{d}y$, and ALICE reports the $p_T$-dependent cross-section ratio.

\vspace{-6pt}

\vspace{-6pt}

\begin{figure}[H]
 \includegraphics[width=\linewidth]{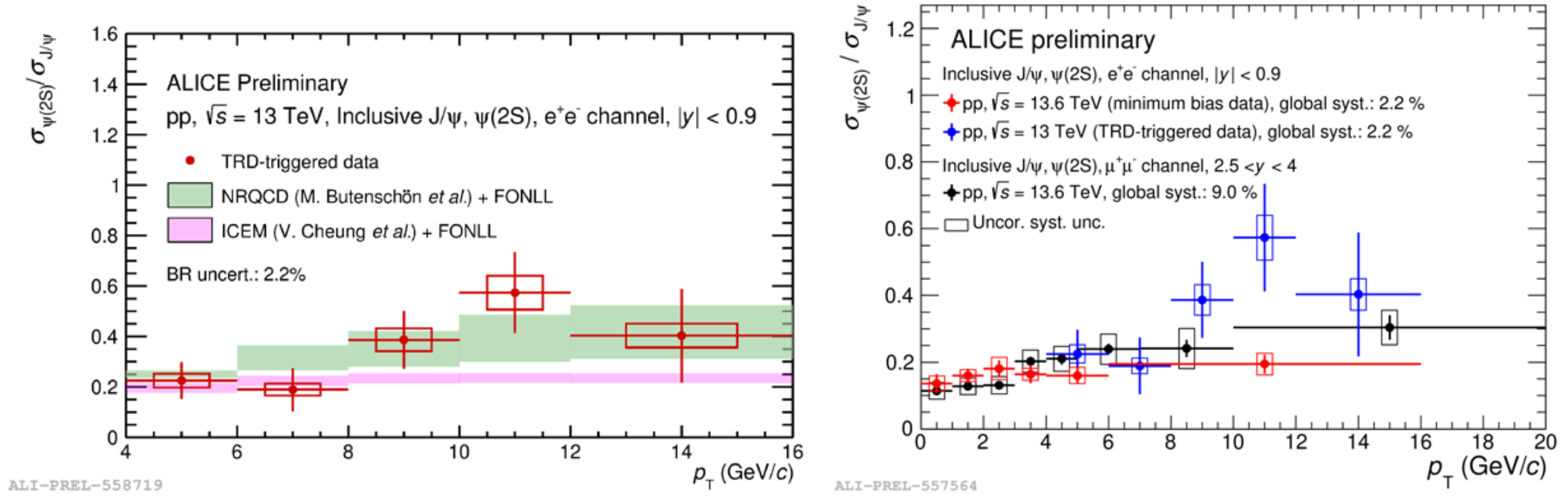}
 \caption{\textbf{Left}: Inclusive 
 $\psi(2\mathrm{S})/J/\psi$ cross-section ratio as a function of transverse momentum $p_{T}$ in $pp$ collisions at $\sqrt{s}={13}~\text{TeV}$, compared with NRQCD~\cite{Butenschoen:2011yh} and ICEM~\cite{Cheung:2018ev} predictions.\linebreak \textbf{Right}: $p_{T}$-dependent measurements of $\sigma_{\psi(2\mathrm{S})}/\sigma_{J/\psi}$ at midrapidity ($|y|<0.9$) and forward rapidity ($2.5<|y|<4.0$) in $pp$ collisions at $\sqrt{s}={13}~\text{TeV}$ and $\sqrt{s}={13.6}~\text{TeV}$, shown alongside the Run 2 midrapidity results at $\sqrt{s}={13}~\text{TeV}$.~\cite{ALICEQuarkoniaSmallSystems2023}}
 \label{fig:alice-ccbar}
\end{figure}

In $pp$ collisions at $\sqrt{s}={13}~\text{TeV}$ and ${13.6}~\text{TeV}$, ALICE measured the inclusive cross-section ratio 
\(\frac{\sigma_{\psi(2\mathrm{S})}}{\sigma_{J/\psi}}\)
at midrapidity ($|y|<0.9$) in the $e^+e^-$ channel and forward rapidity ($2.5<|y|<4.0$) in the $\mu^+\mu^-$ channel~\cite{ALICEQuarkoniaSmallSystems2023}. The results are shown in Figure~\ref{fig:alice-ccbar}.

In $pp$ collisions at $\sqrt{s}={13}~\text{TeV}$ and ${13.6}~\text{TeV}$, ALICE measured the inclusive cross-section ratio 
\(\frac{\sigma_{\psi(2\mathrm{S})}}{\sigma_{J/\psi}}\)
as a function of $p_T$ at midrapidity ($|y|<0.9$) in the $e^+e^-$ channel and forward rapidity ($2.5<|y|<4.0$) in the $\mu^+\mu^-$ channel~\cite{ALICEQuarkoniaSmallSystems2023}. The results are shown in Figure~\ref{fig:alice-ccbar}.

Overall, the measurements are mutually consistent across their respective phase-space regions and broadly compatible with theoretical calculations within the quoted uncertainties. Further improvements in precision and extended kinematic coverage will sharpen global data constraints, e.g., those entering NRQCD extractions.

\subsubsection{Bottomonium Production Cross-Section}

A series of precise measurements of bottomonium production cross-sections in $pp$ collisions has been carried out by CMS, ALICE and LHCb. These studies span center-of-mass energies from \(\sqrt{s}={5.02}~\text{TeV}\) to \({13.6}~\text{TeV}\), and include ground and excited states \(\Upsilon(1\mathrm{S})\), \(\Upsilon(2\mathrm{S})\) and \(\Upsilon(3\mathrm{S})\)~\cite{Hu:2017pat,ALICEInclusiveQuarkonium5p02TeV2023,LHCbUpsilonMultiplicity2025}. Collectively, these measurements provide stringent tests of non-relativistic QCD factorization and alternative quarkonium production models across a wide kinematic range.

All analyses reconstruct \(\Upsilon(\mathrm{n}\mathrm{S})\to\mu^+\mu^-\) decays and extract signal yields by performing maximum-likelihood fits to the dimuon invariant-mass distributions. The experiments complement each other in their kinematic coverages. CMS extends up to \(p_T<100\,\text{GeV/c}\) in the central ($|y|<1.2$) region, while ALICE and LHCb focus on low \(p_T\) at forward region ($2.5<y<4$ for ALICE and $2<y<4.5$ for LHCb). All the measurements only consider the prompt contribution.

\begin{figure}[H]
 \includegraphics[width=0.5\linewidth]{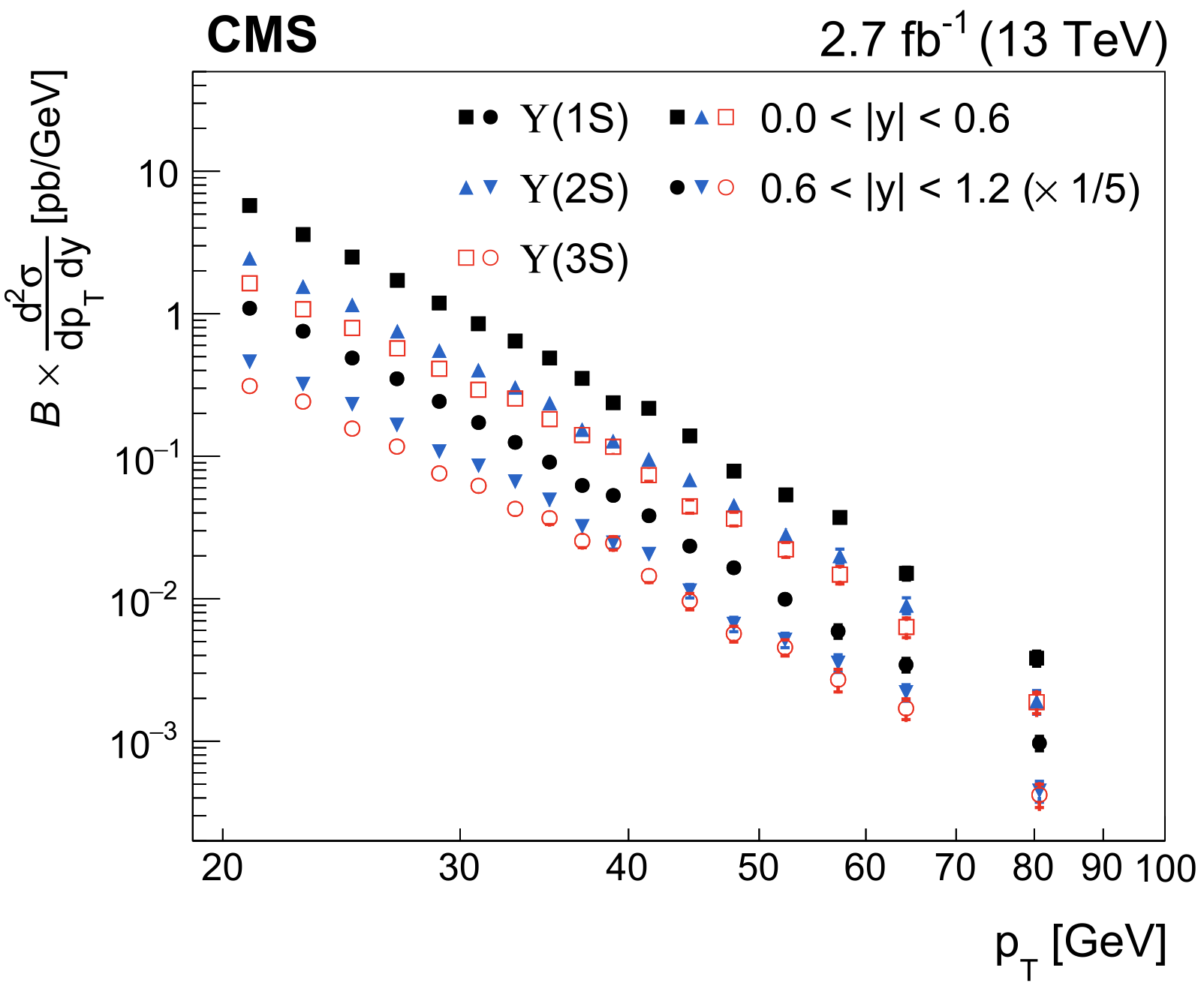}
 \caption{\(\Upsilon(\mathrm{n}\mathrm{S}) (\mathrm{n}=1,2,3)\) double-differential 
 cross-sections measured by CMS at \(\sqrt{s}={13}~\text{TeV}\)~\cite{Hu:2017pat}.}
 \label{fig:cms-ups}
\end{figure}

\begin{figure}[H]
 \includegraphics[width=\linewidth]{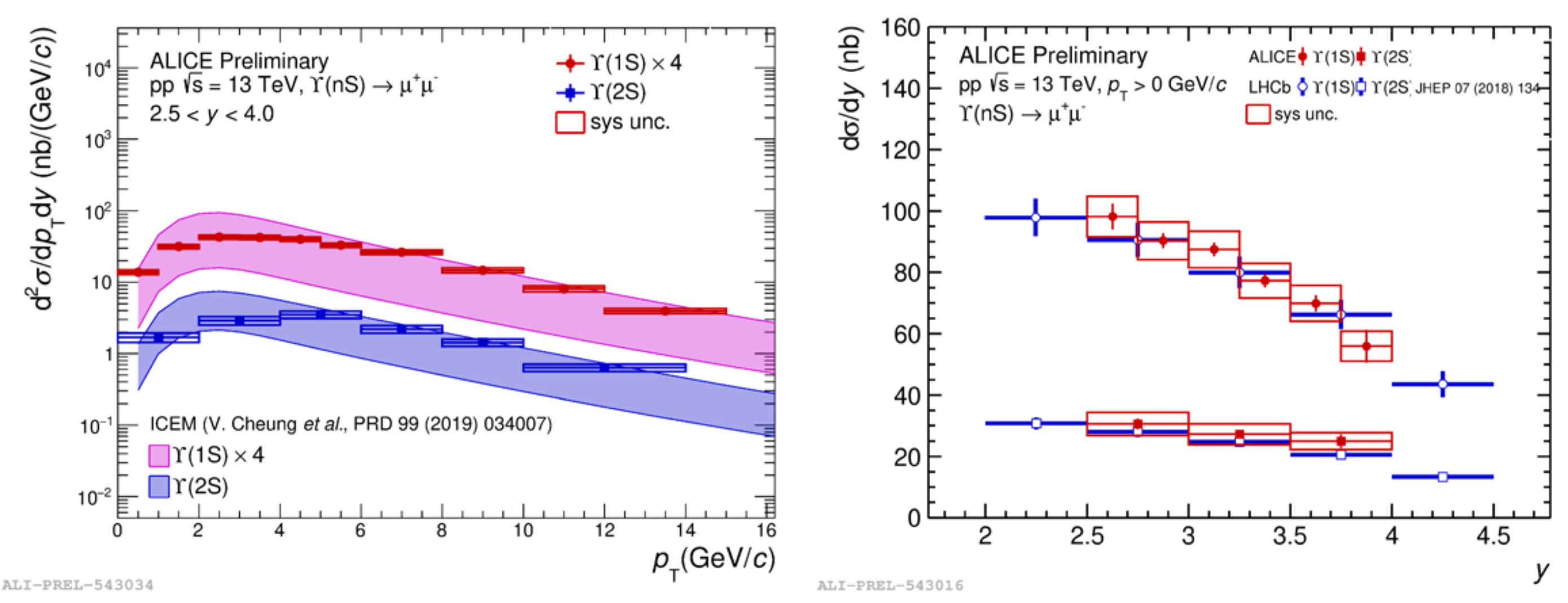}
 \caption{\textbf{Left}: 
 inclusive $\Upsilon(1\mathrm{S})$ and $\Upsilon(2\mathrm{S})$ production cross-sections at forward rapidity as a function of $p_{T}$ in $pp$ collisions at $\sqrt{s}={13}~{\text{TeV}}$, compared to ICEM+FONLL calculations~\cite{Cheung:2019a,Cacciari:1998it}.\linebreak \textbf{Right}: rapidity dependence of the inclusive $\Upsilon(1\mathrm{S})$ and $\Upsilon(2\mathrm{S})$ cross-sections at forward $y$ in $pp$ collisions at $\sqrt{s}={13}~{\text{TeV}}$ shown alongside LHCb measurements at the same energy~\cite{ALICEInclusiveQuarkonium5p02TeV2023}.}
 \label{fig:alice-ups}
\end{figure}

\begin{figure}[H]
 \includegraphics[width=\linewidth]{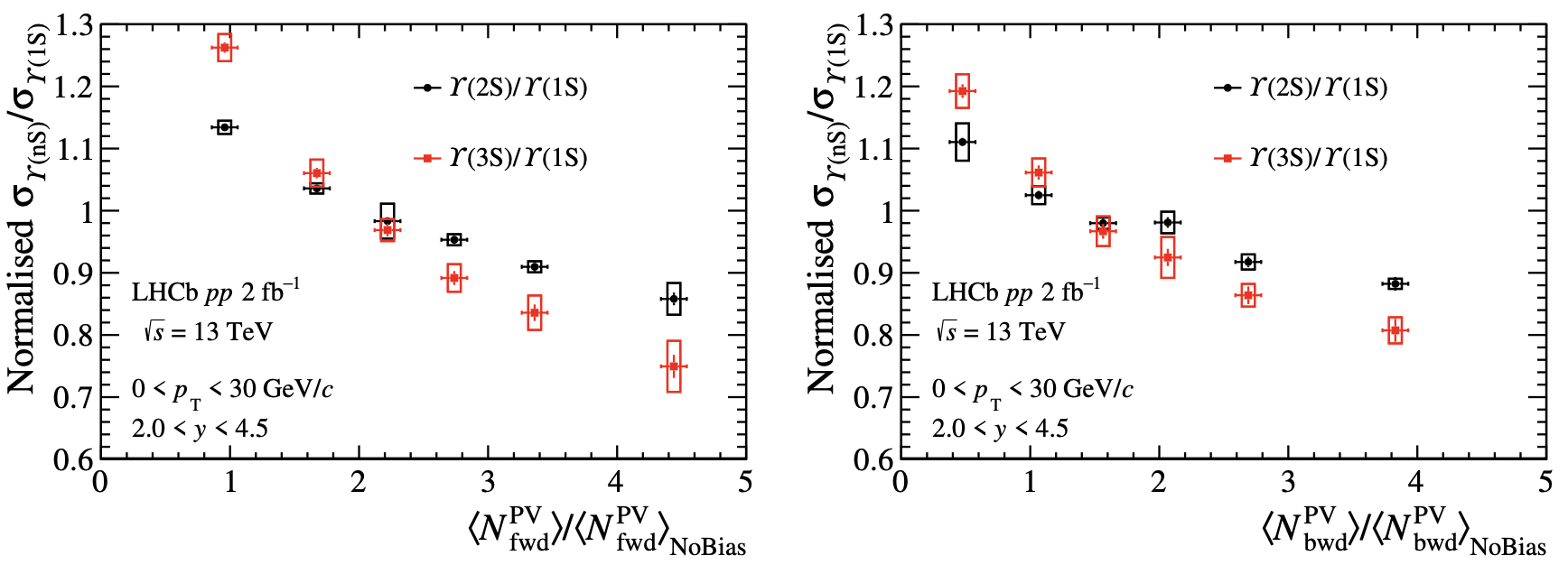}
 \caption{Normalized ratios $\frac{\sigma_{\Upsilon(2\mathrm{S})}}{\sigma_{\Upsilon(1\mathrm{S})}}$ and $\frac{\sigma_{\Upsilon(3S)}}{\sigma_{\Upsilon(1\mathrm{S})}}$ as functions of the self-normalised multiplicity $N^{\mathrm{PV}}_{\mathrm{fwd}}$ (\textbf{left}) and $N^{\mathrm{PV}}_{\mathrm{bwd}}$ (\textbf{right}) for $2.0<y<4.5$ and $0<p_{T}<{30}{~{\text{GeV/c}}}$~\cite{LHCbUpsilonMultiplicity2025}.}
 \label{fig:lhcb-ups}
\end{figure}

CMS measurements (Figure~\ref{fig:cms-ups}) focus on the double-differential quarkonium production cross-sections $\frac{d^{2}\sigma}{dy\,dp_{T}}$, whereas ALICE (Figure~\ref{fig:alice-ups}) reports single-differential cross-sections $\frac{d\sigma}{dp_{T}}$ and $\frac{d\sigma}{dy}$ separately. LHCb (Figure~\ref{fig:lhcb-ups}) investigates the excited-to-ground yield ratios $\frac{\sigma(\Upsilon(\mathrm{n}\mathrm{S}))}{\sigma(\Upsilon(1\mathrm{S}))},n>1$ as a function of event multiplicity. Here the multiplicity is defined by the self-normalized primary-vertex counts $N^{\mathrm{PV}}_{\mathrm{fwd}}$ and $N^{\mathrm{PV}}_{\mathrm{bwd}}$ to characterize the forward and backward charged-particle activity. 

Those measurements exhibit coherent dependencies on $p_T$, $y$, and event variables, with cross-experiment agreement within uncertainties. Increased statistics and reduced systematics, especially at a very high $p_T$ region, will strengthen the data inputs used by theory frameworks such as NRQCD while remaining agnostic to any particular model in this summary.
\vspace{-6pt}

\subsubsection{Search for $t\bar{t}$ Quasi-Bound-State} \label{toponium}

While charmonium and bottomonium have been thoroughly studied during the LHC Run-2 period, searches for toponium states have continued in parallel. According to traditional views, top quarks would decay through weak interactions before forming bound states due to their excessively large mass~\cite{Bernreuther:2008ju}. However, owing to the unprecedented luminosity, studies of top quark pairs ($t\bar{t}$) have received widespread interest, especially regarding the possible quasi-bound-states ($\eta_{t\bar{t}}$). Exciting evidence has been reported by the CMS~\cite{CMS:2025kzt} and ATLAS~\cite{ATLAS:2025mvr} collaborations in analyses of the $t\bar{t}$ invariant mass spectrum ($m(t\bar{t})$), as a significant event excess has been observed near the production threshold.

\begin{figure}[H]

\centering 
	\includegraphics[width=\textwidth]{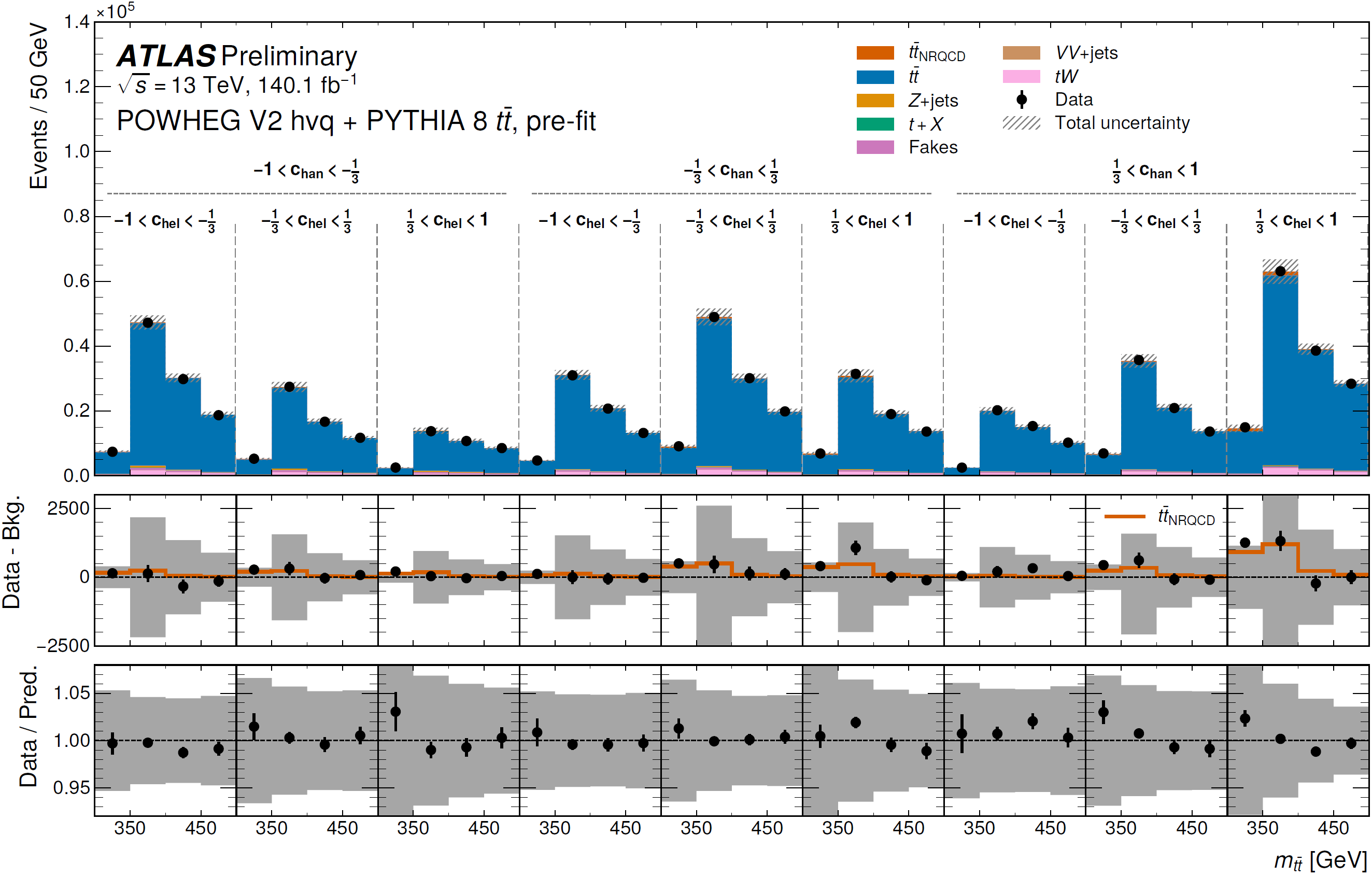}
	\caption{Fits 
    on the $m(t\bar{t})$ dimension with the ATLAS $\eta_{t\bar{t}}$ search. \textbf{Top panel}: event distributions (black dots) and fits (filled area) in nine categories. \textbf{Middle panel}: event distributions with the background contributions subtracted (distributions of the excess) and the simulated distributions of possible $\eta_{t\bar{t}}$ state (orange solid line) in nine categories. \textbf{Bottom panel}: the ratio of the data and the extended $t\bar{t}$ model including $t\bar{t}_{NRQCD}$ contributions~\cite{ATLAS:2025mvr}.}
	\label{F ATLAS toponium}
\end{figure}

The CMS and ATLAS collaborations followed a similar analysis strategy. In their analyses, $t\bar{t}$ candidates were reconstructed by the channel $t\bar{t}\to e^{+}e^{-}/\mu^{+}\mu^{-}/e^{\pm}\mu^{\mp}+jets$ using the $pp$ collision dataset collected at 13 TeV, and the $m(t\bar{t})$ distribution was examined. Several background sources were considered, including non-resonant $t\bar{t}$ production, misidentified $tW$ candidates, and other processes. All background component distributions were modeled by MC simulations. The possible $\eta_{t\bar{t}}$ state was also simulated to estimate its production cross-section. To measure the spin of possible quasi-bound-states, candidates were separated into nine categories according to two angular observables $c_{\mathrm{hel}}$ and $c_{\mathrm{han}}$, which characterize the angle between two charged leptons in the final state~\cite{ATLAS:2025mvr}.  Fits were applied across all nine categories incorporating contributions from both backgrounds and the hypothetical $\eta_{t\bar{t}}$ state. The fits conducted by the ATLAS collaboration are illustrated in Figure~\ref{F ATLAS toponium}.

A similar phenomenon has been observed by the CMS and ATLAS collaborations: a significant event excess above the background expectation has been identified near the pair production threshold. Both collaborations announced a significance exceeding 5 standard deviations and the production cross-sections of the $\eta_{t\bar{t}}$ state as

\begin{equation}
    \begin{aligned}
        \sigma^{\mathrm{CMS}}(pp\to\eta_{t\bar{t}})&=8.8\pm0.5(\mathrm{stat.})^{+1.1}_{-1.3}(\mathrm{syst.})\rm\ pb\\
        \sigma^{\mathrm{ATLAS}}(pp\to\eta_{t\bar{t}})&=9.0\pm1.2(\mathrm{stat.})\pm0.6(\mathrm{syst.})\rm\ pb
    \end{aligned}
\end{equation}

The production cross-section results from both collaborations were consistent and comparable to the theory calculation of $6.43\rm\ pb$~\cite{Fuks:2021xje}. Regarding the spin of the state, a more significant excess has been observed by both collaborations in larger $c_{\mathrm{hel}}$ and $c_{\mathrm{han}}$ categories, supporting the $^{1}\mathrm{S}^{[1]}_{0}$ pseudo-scalar configuration hypothesis. 

The CMS and ATLAS collaborations have obtained remarkably consistent results, announcing the first observation of the $t\bar{t}$ quasi-bound-state. It provides an explanation for the previously observed differences between MC and data at the $t\bar{t}$ mass threshold, as observed by the CMS~\cite{CMS:2024ybg} and ATLAS~\cite{ATLAS:2019hau,ATLAS:2023gsl} collaborations. As a groundbreaking discovery, it provides a new insight into the study of NRQCD and the top quark. This is the first time that an unstable bound state formed by the QCD Coulomb potential, predicted by NRQCD, has been observed experimentally~\cite{Fadin:1990wx}. Both its existence and its property indicate that the QCD Coulomb potential is strong enough to form a broad enhancement structure. This state also provides a great verification of the Sommerfeld effect~\cite{Kiyo:2008bv}.

Furthermore, it adds valuable information for the top Yukawa coupling~\cite{CMS:2020djy}. In addition, it is worth noting that this state is also of great significance in the research of quantum entanglement. A $t\bar{t}$ system exhibits the maximum entanglement near the threshold~\cite{Afik:2020onf, ATLAS:2023fsd,CMS:2024pts}. And the $^{1}\mathrm{S}^{[1]}_{0}$ pseudo-scalar nature of the state is a classic example of the maximum entanglement between two particles in theory~\cite{D0:1995jca}. It provides us with an extremely high energy entanglement source generated by strong interaction. It also indicates that quantum entanglement can be generated and maintained until it is detected, even in extremely high-energy environments.

However, it cannot be ignored that the simulation near the top pair production threshold remains challenging, necessitating additional theoretical calculations and experimental verification. The precise measurement of the properties of the $\eta_{t\bar{t}}$ state would also be a priority during the LHC Run-3 period.

\subsection{Polarization}\label{sec22}

Quarkonium polarization probes the spin alignment of the produced \(q\bar q\) bound state. The polarization of quarkonium states decaying to $\mu^+\mu^-$ can be extracted from the \mbox{angular distribution} 
\begin{equation}
    W(\theta,\phi)\;\propto\;\frac{1}{3+\lambda_{\theta}}\Bigl(1+\lambda_{\theta}\cos^{2}\theta+\lambda_{\phi}\sin^{2}\theta\cos2\phi+\lambda_{\theta\phi}\sin2\theta\cos\phi\Bigr)\,,
\end{equation}
where $\theta$ and $\phi$ are the polar and azimuthal angles of the decay lepton in a chosen polarization frame, and the parameters $\lambda_{\theta}$, $\lambda_{\phi}$, and $\lambda_{\theta\phi}$ characterize the degree of longitudinal versus transverse alignment~\cite{CMS:2013CharmoniumPol7TeV,CMS:2013UpsilonPol7TeV,LHCb:2013JpsiPol7TeV,LHCb:2014Psi2SPol7TeV,ALICE:2012JpsiPol7TeV,Bodwin:1994jh,Brambilla:2004jw}.

The most common reference frames are the helicity (HX) frame, in which the $z$-axis is aligned with the quarkonium momentum in the collision center-of-mass frame; the Collins–Soper (CS) frame, where the $z$-axis bisects the angle between the two beam directions in the quarkonium rest frame; and the perpendicular helicity (PX) frame, defined orthogonal to the CS frame. In each case, the $y$-axis is taken along the direction of the vector product of the two-beam momenta in the quarkonium rest frame, ensuring a right-handed coordinate system~\cite{Faccioli:2010kd}.

Different quarkonium production mechanisms predict markedly different values of \(\lambda_\theta\), especially at high \(p_T\), making polarization a sensitive discriminator among models. Measurements by CMS, ATLAS and LHCb at \(\sqrt{s}={7}~\text{TeV}\) have so far revealed small or transverse-longitudinal mix polarizations in apparent tension with leading-order NRQCD expectations~\cite{Braaten:1996nh,Butenschoen:2012px} . However, due to the lack of statistics, further studies are needed to confirm such discrepancy and provide guidance for theoretical improvements.

\textls[-25]{For quarkonium states reconstructed through $\mu^+\mu^-$ decay channel, the angular distribution}
\begin{equation}
    W(\cos\theta)\;\propto\;1 + \lambda_\theta\cos^2\theta
\end{equation}
in the quarkonium rest frame can be integrated over the azimuthal angle in the helicity frame where acceptance effects factorize. Fits to the $|\cos\theta|$ distributions yield $\lambda_\theta$ in each $p_T$ bin after subtraction of continuum background and, for prompt samples, feed-down from higher states or $B$-hadron decays.

\begin{figure}[H]
 \includegraphics[width=0.6\linewidth]{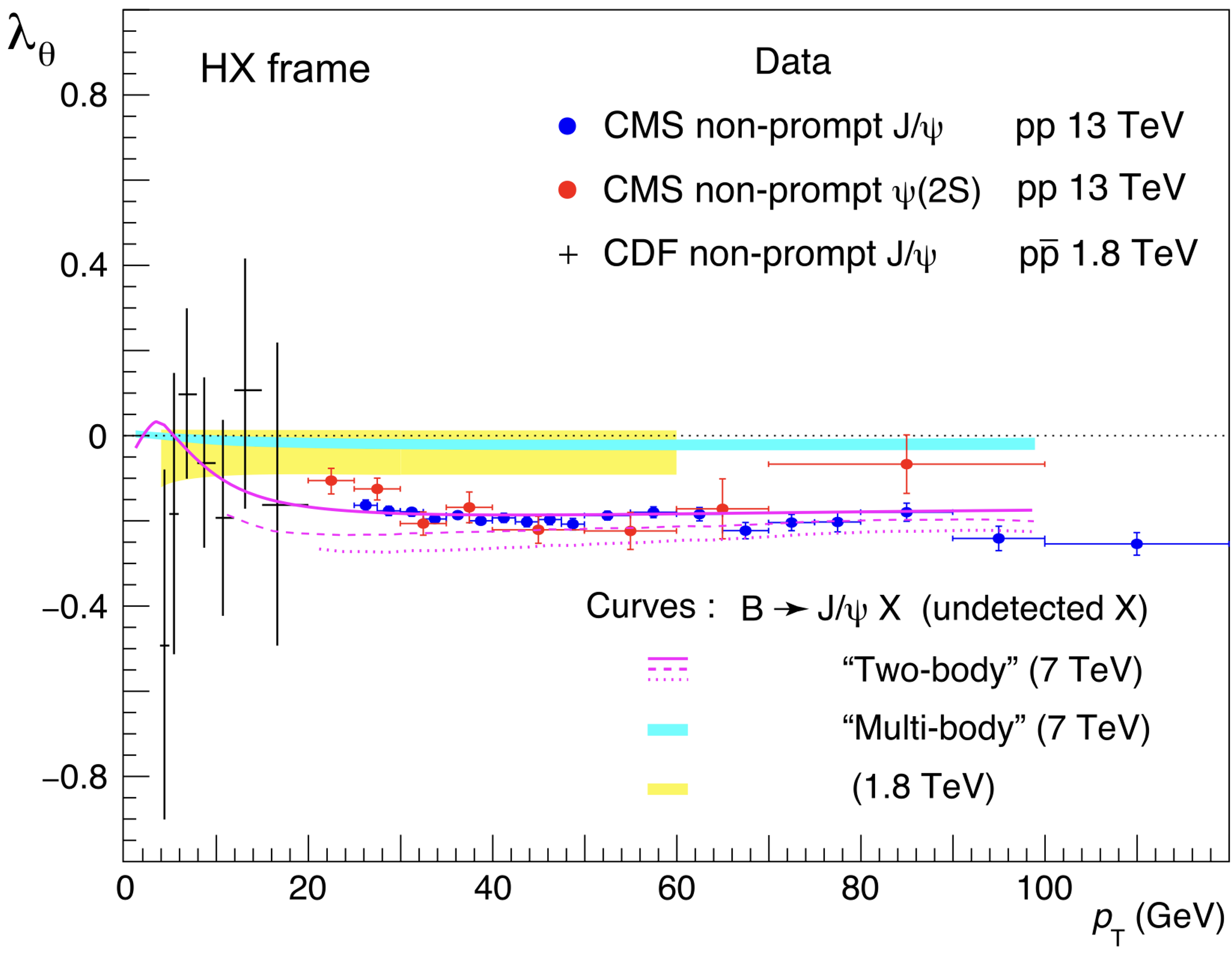}
 \caption{Non-prompt 
 $J/\psi$ polarization parameter $\lambda_{\theta}$ as a function of $p_{T}$ in $pp$ collisions at $\sqrt{s}={13}~{\text{TeV}}$ measured by CMS~\cite{CMSJpsiPolarization13TeV2024}. Vertical bars indicate total uncertainties. Predictions for $J/\psi$ polarization from $B\to J/\psi X$ decays are shown for three theoretical calculations~\cite{Faccioli:2022nh,Krey:2003ej}, and the low-$p_{T}$ CDF measurement~\cite{Abulencia:2007us} is also included.}
 \label{fig:cms-pol-np-13}
\end{figure}

\begin{figure}[H]
\centering
 \includegraphics[width=0.9\linewidth]{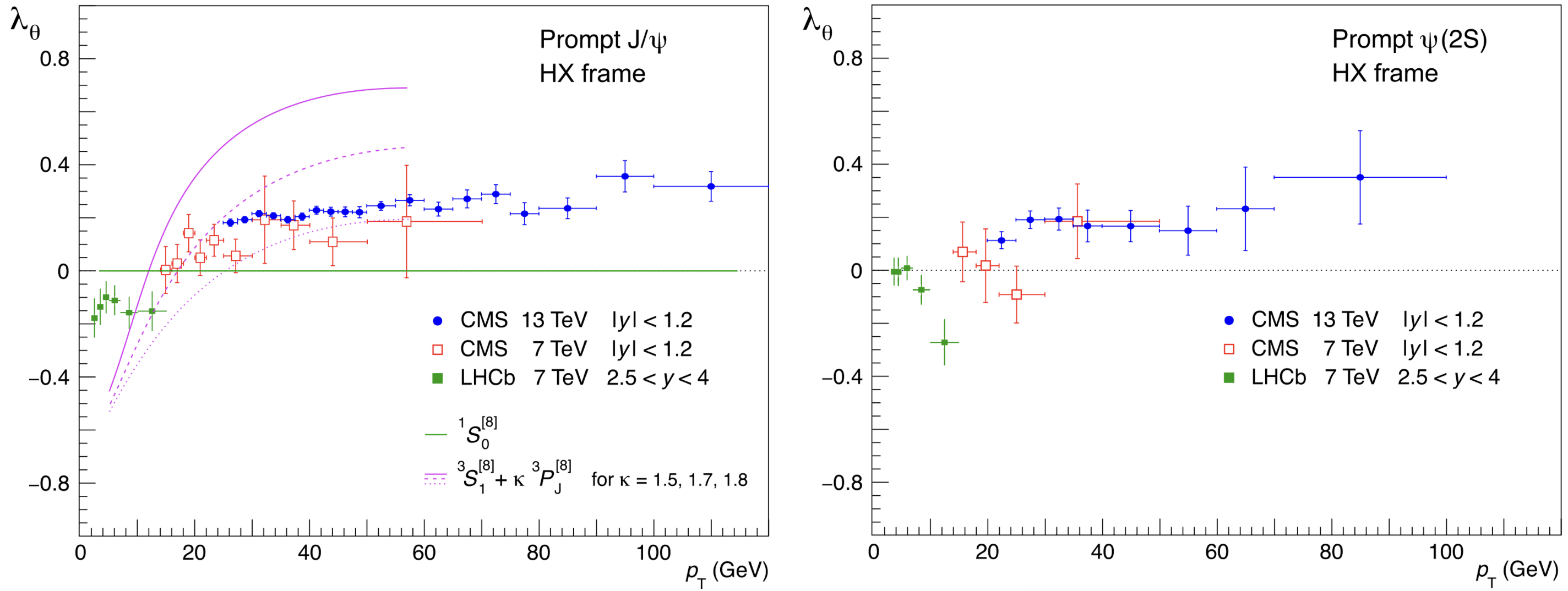}
 \caption{Prompt $J/\psi$ (\textbf{left}) and $\psi(2\mathrm{S})$ (\textbf{right}) polarization parameter $\lambda_{\theta}$ as a function of $p_{T}$ in $pp$ collisions at $\sqrt{s}={13}~{\text{TeV}}$ measured by CMS~\cite{CMSJpsiPolarization13TeV2024}. Compared with CMS~\cite{CMS:2013CharmoniumPol7TeV} and LHCb~\cite{LHCb:2013JpsiPol7TeV} results at $\sqrt{s}={7}~{\text{TeV}}$. Vertical bars indicate total uncertainties.}
 \label{fig:cms-pol-p-13}
\end{figure}

In $pp$ collisions at $\sqrt{s}={13}~{\text{TeV}}$, CMS analyzed 2017–2018 data corresponding to an integrated luminosity of {103.3} {fb$^{-1}$} to extract the polar anisotropy parameter $\lambda_{\theta}$ for prompt and non-prompt $J/\psi$ and $\psi(2\mathrm{S})$ mesons in the helicity frame from dimuon decay angular distributions (Figure~\ref{fig:cms-pol-np-13} and \ref{fig:cms-pol-p-13})~\cite{CMSJpsiPolarization13TeV2024}. The prompt and non-prompt components were separated by fitting the pseudo-proper decay length.

The $p_{T}$ dependence of $\lambda_{\theta}$ was measured for $J/\psi$ in the range 25--120 GeV/c and for $\psi(2\mathrm{S})$ in 20--100 GeV/c. Non-prompt polarizations agree with the two-body $B$-hadron decay hypothesis for $p_{T}>{25}{~{\text{GeV/c}}}$, where $\lambda_{\theta}$ approaches an asymptotic value of $\approx$0.3. When combined with previous CMS and LHCb measurements at $\sqrt{s}={7}~{\text{TeV}}$, the prompt $\lambda_{\theta}$ displays a significant $p_{T}$ dependence at low transverse momentum.

\vspace{-6pt}

\begin{figure}[H]
 \includegraphics[width=0.7\linewidth]{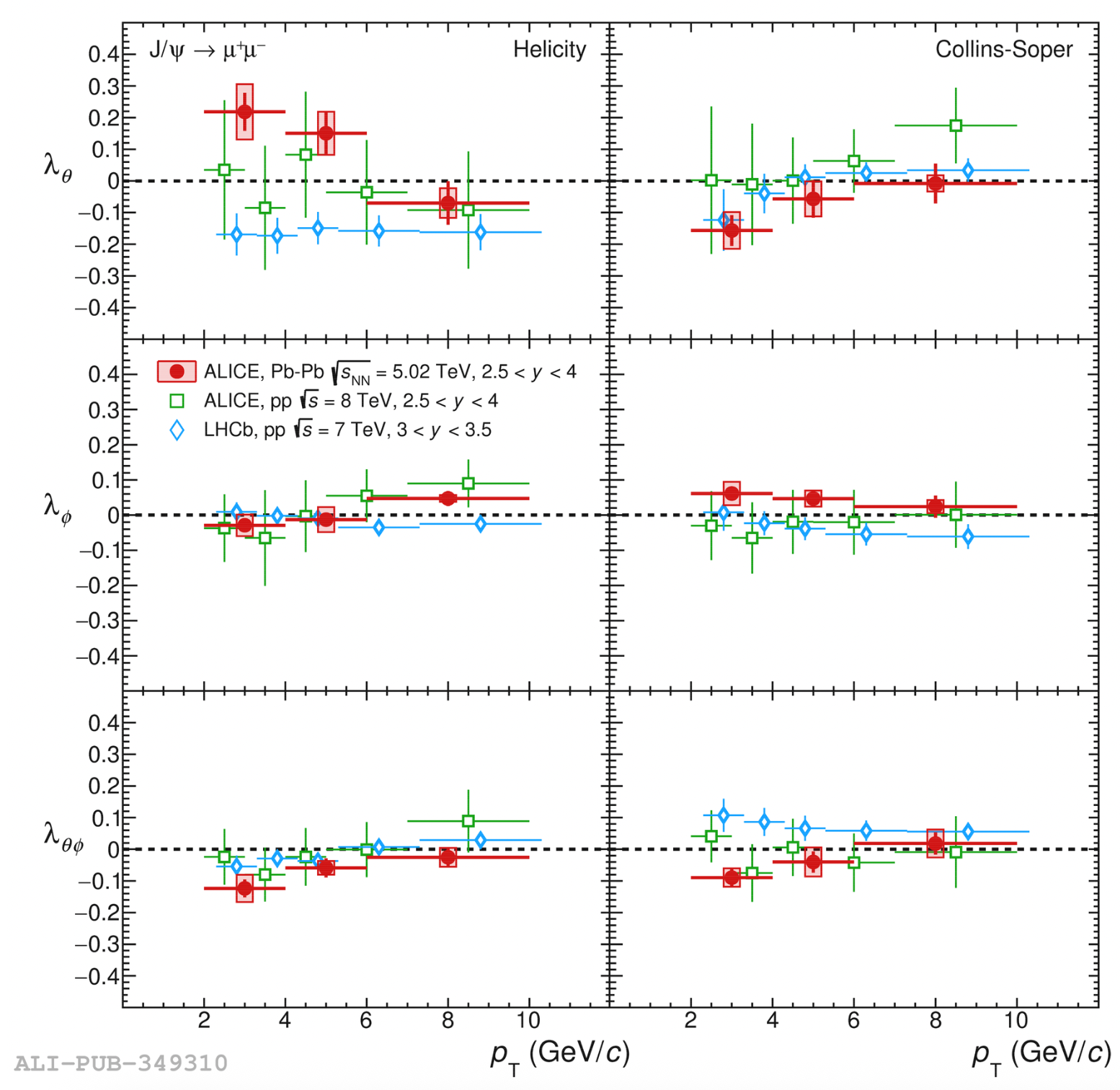}
 \caption{$J/\psi$ polarization parameters as functions $p_{T}$ in $\mathrm{Pb}\mathrm{Pb}$ collisions at $\sqrt{s_{\mathrm{NN}}}={5.02}~{\text{TeV}}$ measured by ALICE~\cite{ALICEQuarkoniumPolarization2020}, compared with the ALICE inclusive $pp$ measurement at $\sqrt{s}={8}~{\text{TeV}}$~\cite{ALICE:2012JpsiPol7TeV} and the LHCb prompt $J/\psi$ result in $pp$ at $\sqrt{s}={7}~{\text{TeV}}$~\cite{LHCb:2013JpsiPol7TeV}.}
 \label{fig:alice-pol-jpsi-502}
\end{figure}

\begin{figure}[H]
 \includegraphics[width=0.7\linewidth]{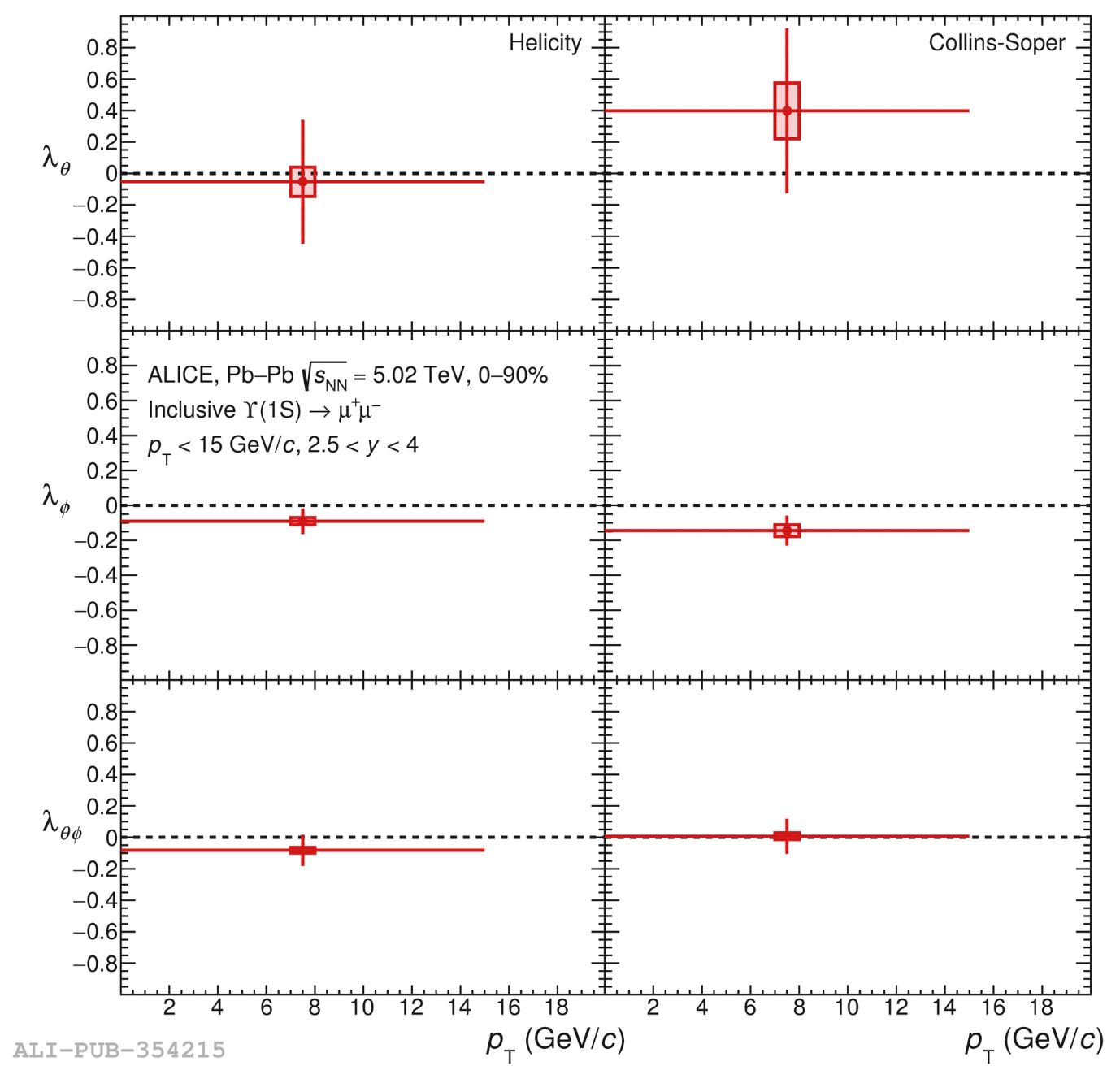}
 \caption{$\Upsilon(1\mathrm{S})$ polarization parameters as functions $p_{T}$ in $\mathrm{Pb}\mathrm{Pb}$ collisions at $\sqrt{s_{\mathrm{NN}}}={5.02}~{\text{TeV}}$ measured by ALICE~\cite{ALICEQuarkoniumPolarization2020}.}
 \label{fig:alice-pol-ups-502}
\end{figure}

In heavy-ion collisions, polarization can also probe the properties of the quark--gluon plasma, and it has been hypothesized that the strong magnetic field in the early stages may align quarkonium spins. In ALICE, the polarization of $J/\psi$ and $\Upsilon(1\mathrm{S})$ mesons is extracted from the anisotropies in the angular distribution of their decay \mbox{muons (Figures~\ref{fig:alice-pol-jpsi-502} and \ref{fig:alice-pol-ups-502})~\cite{ALICEQuarkoniumPolarization2020}.} This contribution reports the $p_T$-differential polarization in $\mathrm{Pb}\mathrm{Pb}$ collisions at $\sqrt{s_{\mathrm{NN}}}={5.02}~{\text{TeV}}$ at forward rapidity ($2.5<y<4$), presented in two reference frames and compared with earlier $pp$ measurements. 

\vspace{-6pt}

\section{Production in Heavy Ion Collisions}

\subsection{Suppression in \(\mathrm{PbPb}\) Collisions}\label{sec31}
 In ultra‐relativistic \(\mathrm{PbPb}\)  
 collisions, the formation of a deconfined QGP modifies the binding potential of heavy \(q\bar q\) pairs via Debye screening, leading to a sequential “melting” of quarkonium states in order of decreasing binding energy~\cite{Matsui:1986dk}. This phenomenon manifests a reduction in observed yields relative to expectations from incoherent superposition of $pp$ interactions, with more loosely bound excitations suppressed at lower medium temperatures than the ground states.

The degree of suppression is commonly quantified by the nuclear modification factor
\begin{equation}
    R_{\mathrm{AA}}(p_T,y) \;=\;\frac{1}{\langle T_{\mathrm{AA}}\rangle}\frac{d^2\sigma_{\mathrm{AA}}/dp_T\,dy}{d^2\sigma_{pp}/dp_T\,dy}\,,
\end{equation}
where \(\langle T_{\mathrm{AA}}\rangle\) is the average nuclear overlap function for a given centrality class, \(d^2\sigma_{\mathrm{AA}}/dp_T\,dy\) the differential yield in \(\mathrm{PbPb}\), and \(d^2\sigma_{pp}/dp_T\,dy\) the corresponding cross-section in \(pp\) collisions~\cite{Adare:2006nq}. The double ratios of excited‐to‐ground yields,
\begin{equation}
    \frac{\bigl[{q \bar q}(\mathrm{nS})/{q \bar q}(1\mathrm{S})\bigr]_{\mathrm{AA}}}{\bigl[{q \bar q}(\mathrm{nS})/{q \bar q}(1\mathrm{S})\bigr]_{pp}}\ (\mathrm{n}>1),
\end{equation}
were also employed, providing the cancellation of common systematic uncertainties and feed‐down fractions~\cite{Chatrchyan:2012lxa}.

Experimentally, suppression measurements proceed by reconstructing quarkonium decays (e.g.,\ \(q\bar q\to\ell^+\ell^-\)) in heavy‐ion and reference \(pp\) datasets under identical kinematic selections and center-mass-energy. Yields are extracted via fits to invariant‐mass spectra in bins of transverse momentum, rapidity, and collision centrality. Detector acceptance and efficiency corrections, derived from detailed MC simulations and validated with data‐driven tag‐and‐probe methods, are applied to obtain corrected yields. Systematic uncertainties from background modeling, yield extraction, luminosity, and nuclear geometry are evaluated by varying fit models and overlap‐function determinations.

\begin{figure}[H]
 \includegraphics[width=0.4\linewidth]{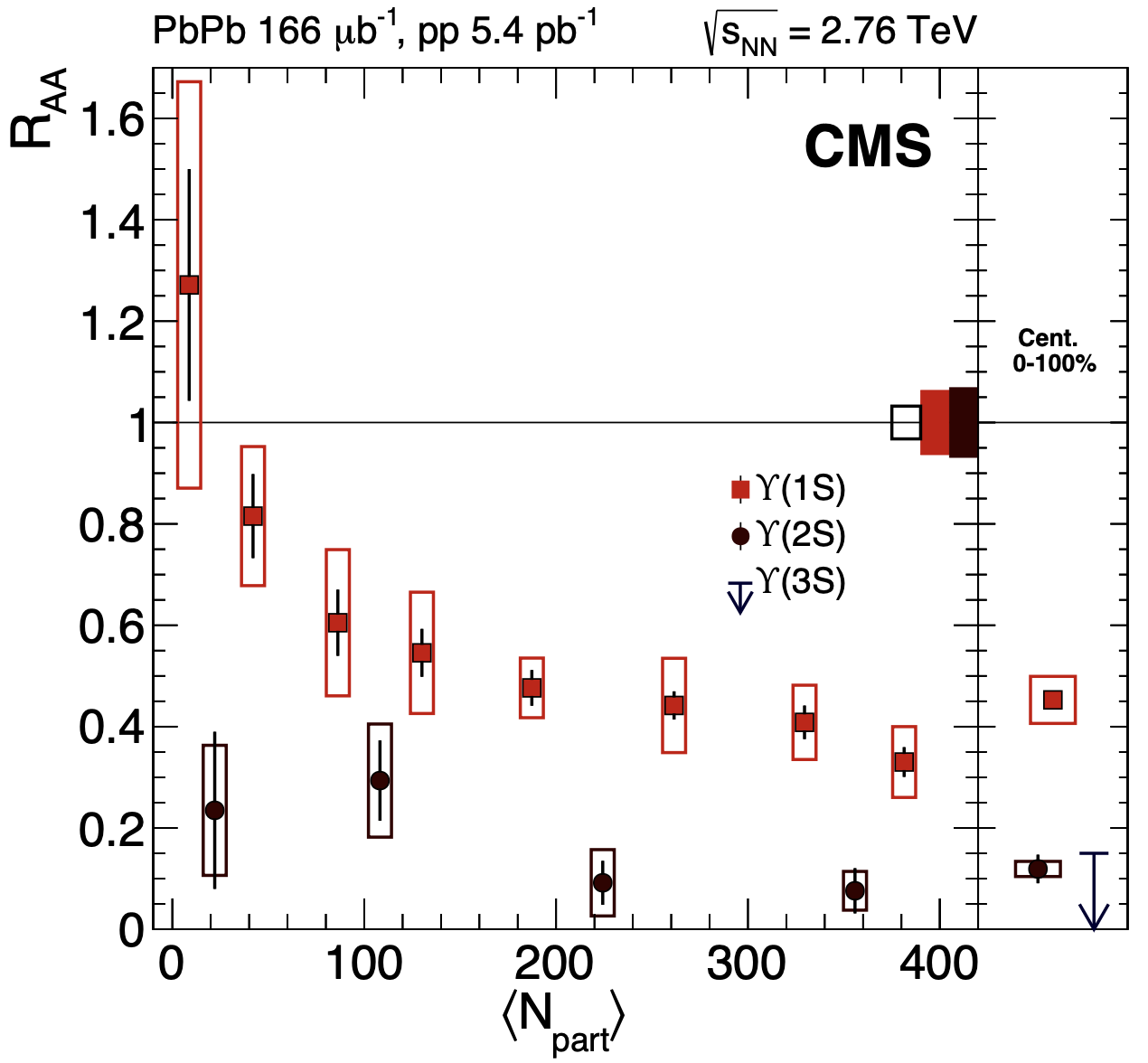}
 \caption{Nuclear 
 modification factors \(R_{\mathrm{AA}}\) of \(\Upsilon(\mathrm{nS})(\mathrm{n}=1,2,3)\) as functions of \(\langle N_{\mathrm{part}}\rangle\) measured by CMS at \({2.76}~\text{TeV}\)~\cite{CMSUpsilonSuppressionPbPb2p76TeV2017}. Statistical (systematic) uncertainties are drawn as vertical bars (boxes), while global normalization uncertainties from the $\mathrm{Pb}\mathrm{Pb}$ measurement ({3.2}{\%}) and the $pp$ reference ({6.3}{\%} for $\Upsilon(1\mathrm{S})$, {6.9}{\%} for $\Upsilon(2\mathrm{S})$) are indicated at unity by open, red‐filled, and black‐filled boxes, respectively.}
 \label{fig:cms-raa-276}
\end{figure}

In $\mathrm{Pb}\mathrm{Pb}$ and $pp$ collisions at $\sqrt{s_{\mathrm{NN}}}={2.76}~\text{TeV}$, CMS measured the yields of $\Upsilon(1\mathrm{S},2\mathrm{S},3\mathrm{S})$ via $\mu^+\mu^-$ decays with integrated luminosities of 166 $\upmu$b$^{-1}$ and 5.4 pb$^{-1}$, respectively (Figure~\ref{fig:cms-raa-276})~\cite{CMSUpsilonSuppressionPbPb2p76TeV2017}. The differential cross-sections $\frac{d^{2}\sigma}{dy\,dp_{T}}$ were obtained for $|y|<2.4$ and $p_{T}<{20}{~{\text{GeV/c}}}$. 

A centrality-dependent suppression was observed in $\mathrm{Pb}\mathrm{Pb}$ collisions relative to $pp$ collisions, which showed no significant dependence on $y$ or $p_{T}$.The $\Upsilon(3\mathrm{S})$ is not observed in $\mathrm{Pb}\mathrm{Pb}$, implying at least a $7$ fold suppression at {95}{\%} confidence. The $R_{\mathrm{AA}}$ results are consistent with sequential melting scenarios in a QGP:
\begin{equation}
    \begin{aligned}
        R_{\mathrm{AA}}(\Upsilon(1\mathrm{S}))& = 0.453 \pm 0.014 \pm 0.046\\
        R_{\mathrm{AA}}(\Upsilon(2\mathrm{S})) &= 0.119 \pm 0.028 \pm 0.015 \\
        R_{\mathrm{AA}}(\Upsilon(3\mathrm{S})) &< 0.145 \quad\mathrm{(95\% C.L.)}.
    \end{aligned}
\end{equation}

\begin{figure}[H]
 \includegraphics[width=\linewidth]{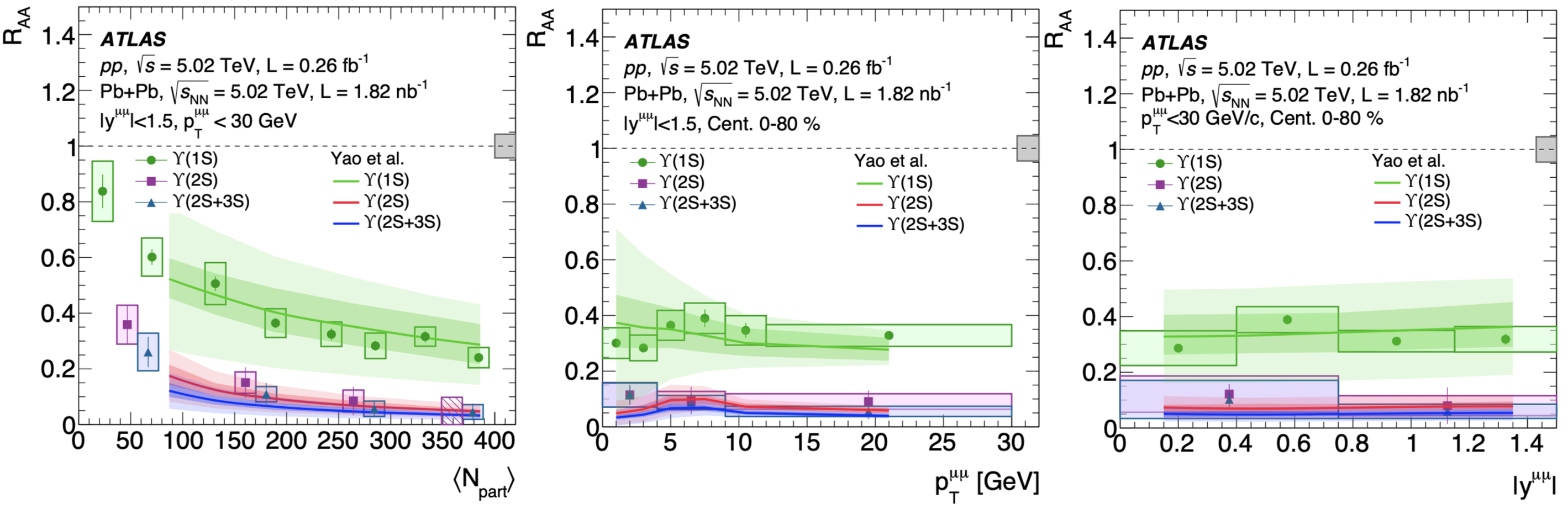}
 \caption{Nuclear 
 modification factors $R_{\mathrm{AA}}$ for $\Upsilon(1\mathrm{S})$, $\Upsilon(2\mathrm{S})$ and $\Upsilon(2\mathrm{S}+3\mathrm{S})$ as a function of \(\langle N_{\mathrm{part}}\rangle\) (\textbf{left}), $p_{T}$ (\textbf{middle}) and $|y|$ (\textbf{right}) in $\mathrm{Pb}\mathrm{Pb}$ at ${5.02}~\text{TeV}$~\cite{ATLASUpsilonPbPb5p02TeV2023} compared to the theoretical calculation~\cite{Du:2017fwk,Brambilla:2021wkt}. Dark bands indicate uncertainties from the nPDF choice, while light bands show model uncertainties from $\pm10\%$ parameter variations.}
 \label{fig:atlas-raa-502}
\end{figure}

In $\mathrm{Pb}\mathrm{Pb}$ and $pp$ collisions at ${5.02}{TeV}$, ATLAS measured $\Upsilon(1\mathrm{S},2\mathrm{S},3\mathrm{S})$ production via $\mu^+\mu^-$ decays using {1.38} {nb$^{-1}$} (2018) and {0.44} {nb$^{-1}$} (2015) of $\mathrm{Pb}\mathrm{Pb}$ data and {0.26} {fb$^{-1}$} of $pp$ data, for $p_{T}<{30}{~{\text{GeV/c}}}$, $|y|<1.5$ and 0–80 \% centrality, extracting $R_{\mathrm{AA}}$ as functions of centrality, $p_{T}$ and $|y|$, and studying excited‐to‐ground suppression ratios~\cite{ATLASUpsilonPbPb5p02TeV2023}. (Figure~\ref{fig:atlas-raa-502})

\begin{figure}[H]
\includegraphics[width=\linewidth]{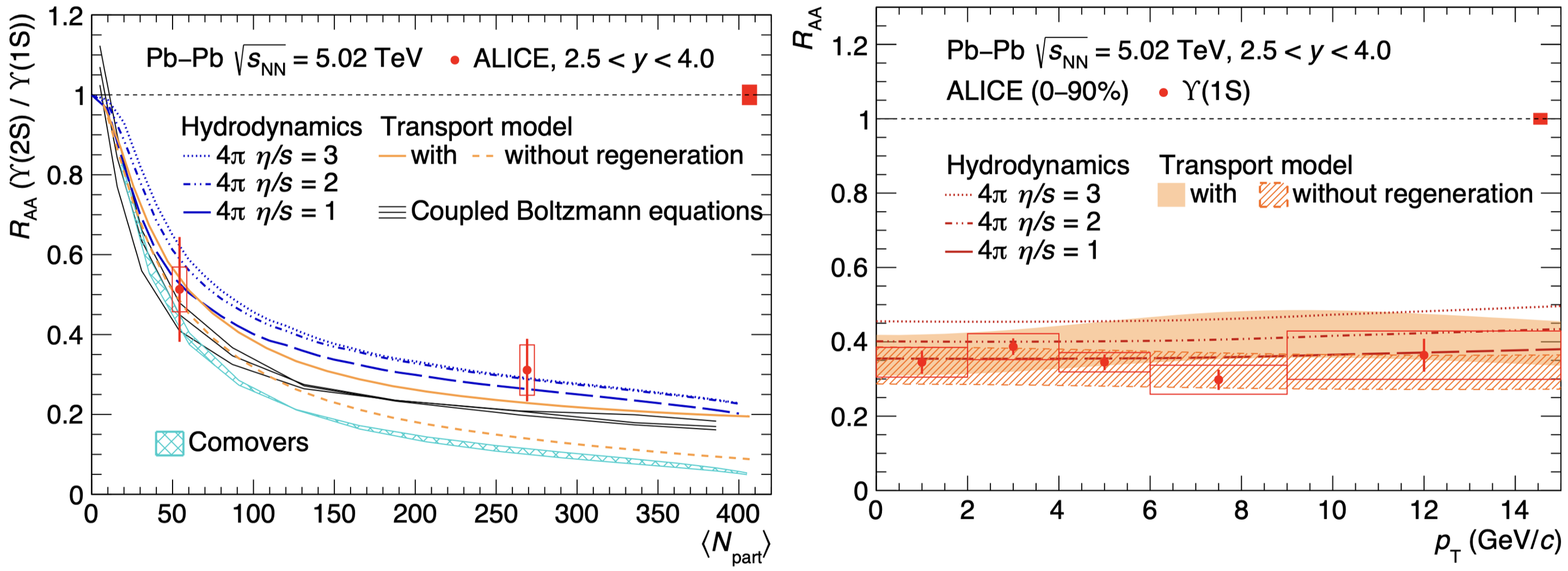}
 \caption{\textbf{Right}: Nuclear modification factor ratio $\frac{R_{\mathrm{AA}}(\Upsilon(2\mathrm{S}))}{R_{\mathrm{AA}}(\Upsilon(1\mathrm{S}))}$ versus the average number of participants $\langle N_{\mathrm{part}}\rangle$ in $\mathrm{Pb}\mathrm{Pb}$ and $pp$ collisions at ${5.02}~\text{TeV}$, shown atop statistical hadronization model values~\cite{Andronic:2018aa} with the red‐filled box at unity indicating the $pp$ reference uncertainty. \textbf{Left}: Measured $R_{\mathrm{AA}}(\Upsilon(1\mathrm{S}))$ as a function of $p_{T}$ in $\mathrm{Pb}\mathrm{Pb}$ collisions, where the red box at unity denotes the global uncertainty correlated with $p_{T}$, compared to hydrodynamic~\cite{Krouppa:2016yjm} and transport~\cite{Du:2017fwk} model predictions~\cite{ALICEUpsilonForwardPbPb5p02TeV2021}.}
 \label{fig:alice-raa-502}
\end{figure}

In the same collisions, ALICE recorded {1.38}{nb$^{-1}$} of $\mathrm{Pb}\mathrm{Pb}$ data with its muon spectrometer at forward rapidity $2.5<y<4.0$, measuring the nuclear modification factor as functions of rapidity, $p_{T}$ and centrality~\cite{ALICEUpsilonForwardPbPb5p02TeV2021}. (Figure~\ref{fig:alice-raa-502})

Both experiments observe sequential melting consistent with theoretical models, with ATLAS covering midrapidity and a broad kinematic range and ALICE extending sensitivity into the forward region.

\vspace{-6pt}

\vspace{-6pt}

\subsection{Suppression in \(pPb\) Collisions}\label{sec32}

In $p\mathrm{Pb}$ 
collisions, quarkonium production is modified predominantly by cold‐nuclear‐matter (CNM) effects, including the shadowing of nuclear parton distribution functions, initial‐state parton energy loss, and nuclear absorption of pre‐resonant \(q\bar q\) pairs~\cite{Vogt:2010aa,Arleo:2012rs}. These effects lead to a modification of yields relative to a simple scaling of \(pp\) cross-sections, and must be disentangled from genuine QGP‐induced suppression.

The nuclear modification factor in \(p\mathrm{A}\) collisions is defined analogously as
\begin{equation}
    R_{p\mathrm{A}}(p_T,y)
=\frac{1}{A}\frac{d^2\sigma_{p\mathrm{A}}/dp_T\,dy}{d^2\sigma_{pp}/dp_T\,dy}\,,
\end{equation}
where \(\mathrm{A}\) is the mass number of the nucleus and \(d^2\sigma_{p\mathrm{A}}/dp_T\,dy\) the differential cross-section in \(p\mathrm{A}\) collisions~\cite{Ferreiro:2018}. Since $R_{p\mathrm{A}}$ is essentially governed by CNM effects, the double ratio $R_{\mathrm{AA}}/R_{p\mathrm{A}}$ can be used to cancel out the CNM contributions in $R_{\mathrm{AA}}$.

\textls[-35]{In addition, another key observable is the forward‐to‐backward ratios, which is \mbox{defined as}}
\begin{equation}
    R_{FB}(p_T,y>0)
=\frac{d^2\sigma_{p\mathrm{A}}(p_T,y)/dp_T\,dy}{d^2\sigma_{p\mathrm{A}}(p_T,-y)/dp_T\,dy}\,,
\end{equation}

The ratio isolates rapidity‐dependent CNM effects without reliance on an external \mbox{\(pp\) reference.}

Methodologically, CNM studies compare yields in \(pPb\) and interpolated or directly measured \(pp\) data at matched \(\sqrt{s_{\mathrm{NN}}}\). Quarkonium states are reconstructed via dilepton channels, with raw yields obtained through invariant‐mass fits in bins of transverse momentum and rapidity. Efficiency and acceptance corrections are applied as for \(PbPb\) measurements. Systematic uncertainties from interpolation of the \(pp\) baseline, luminosity, and model input for nuclear PDFs are assessed by varying proton reference data and shadowing parametrizations. Such studies establish the baseline modifications to be accounted for when interpreting suppression in nucleus–nucleus collisions.

\begin{figure}[H]
 \includegraphics[width=0.9\linewidth]{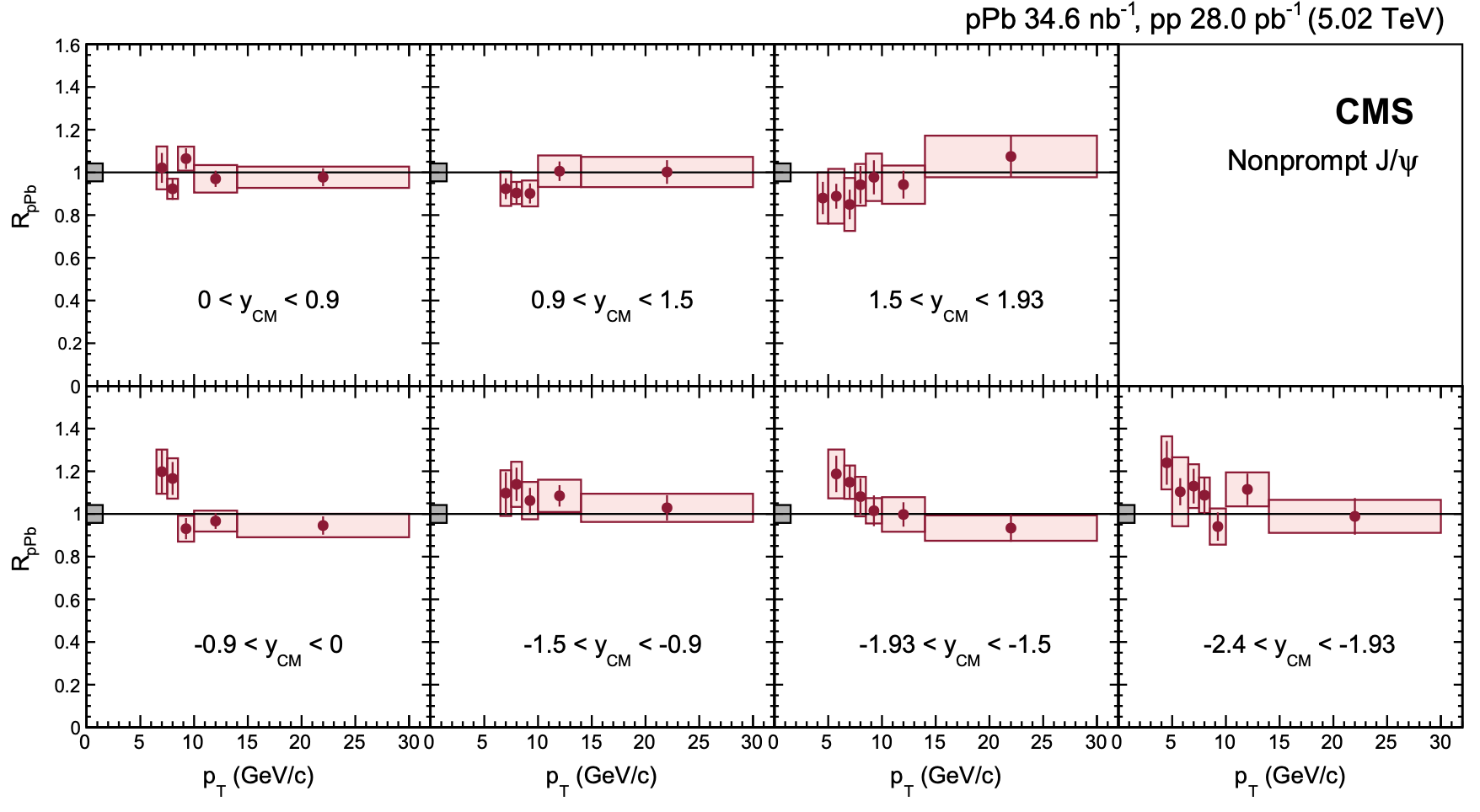}
 \caption{$R_{p\mathrm{Pb}}$ for 
 non‐prompt $J/\psi$ as a function of $p_T$ in seven $y_{\mathrm{CM}}$ intervals, where vertical bars show statistical uncertainties, shaded boxes denote systematic uncertainties, and the fully correlated global uncertainty of {4.2}{\%} is indicated by a grey box at $R_{p\mathrm{Pb}}=1$ next to the left axis~\cite{CMS2017Jpsi_pPb}.}
 \label{fig:cms-RpPb-502}
\end{figure}

\begin{figure}[H]
 \includegraphics[width=0.5\linewidth]{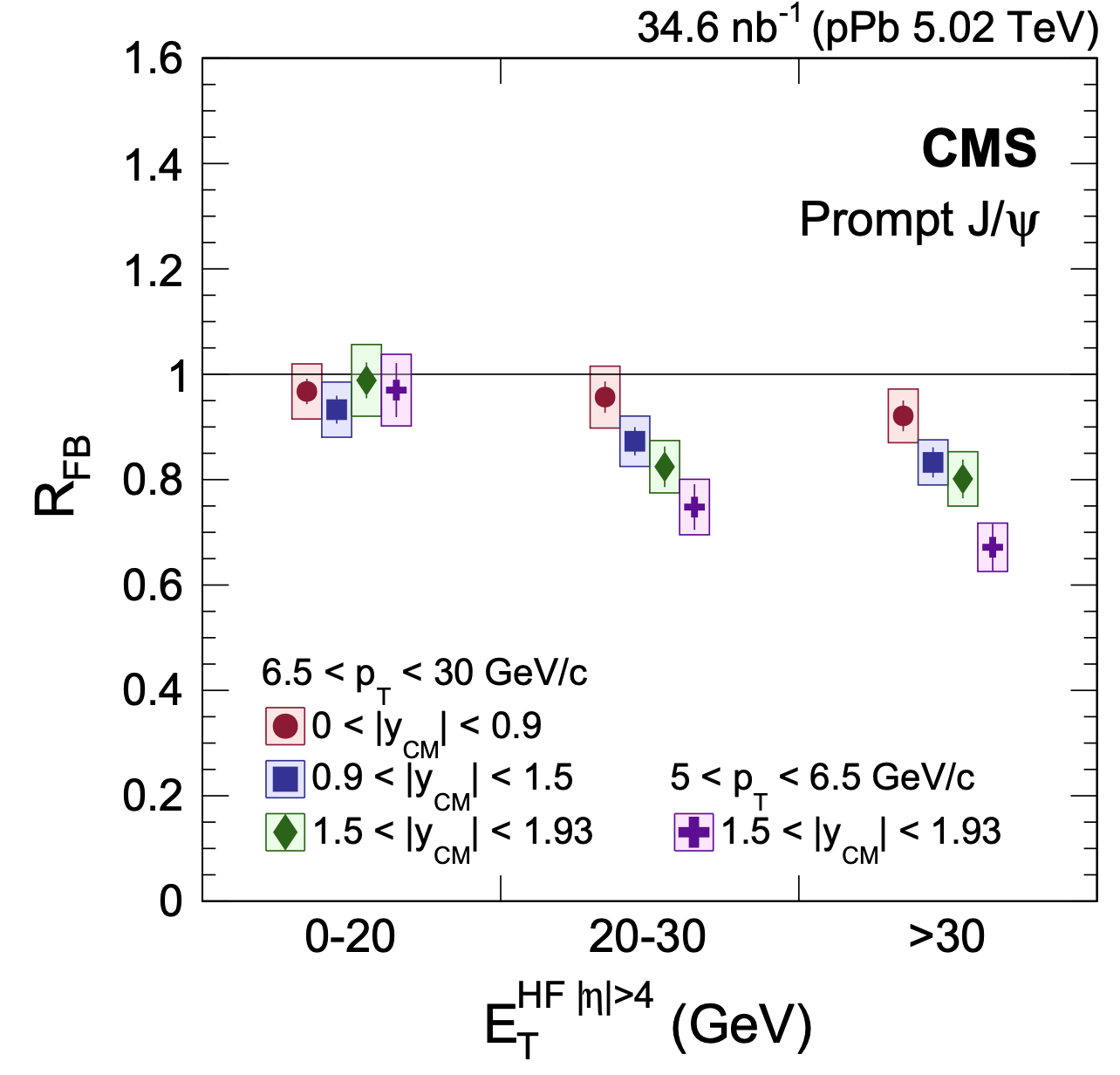}
 \caption{Dependence 
 of the $R_{FB}$ for prompt $J/\psi$ on hadronic activity, quantified by the transverse energy $E_{T}^{\mathrm{HF}}$ deposited at large pseudorapidity $|\eta|>4$. Data points are slightly shifted horizontally to avoid overlap, vertical bars denote statistical uncertainties, and shaded boxes indicate systematic uncertainties~\cite{CMS2017Jpsi_pPb}.}
 \label{fig:cms-RpPb-502-2}
\end{figure}

\begin{figure}[H]
 \includegraphics[width=0.5\linewidth]{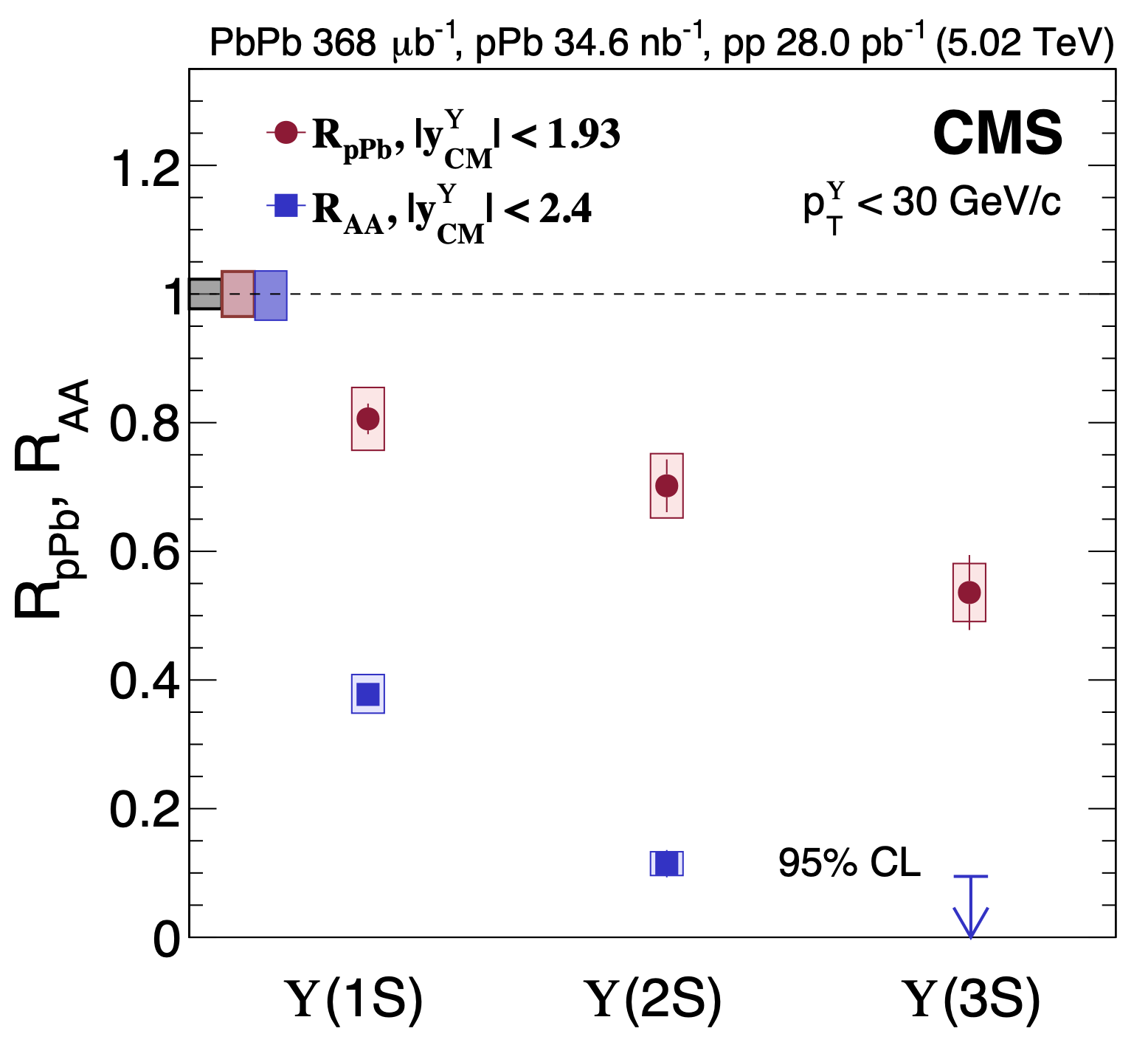}
 \caption{$R_{p\mathrm{Pb}}$ of 
 $\Upsilon(1S)$, $\Upsilon(2S)$, and $\Upsilon(3S)$ (red circles) for the integrated kinematic range $0 < p_T^{\Upsilon} < 30\,\mathrm{GeV}/c$ and $|\eta_{\mathrm{CM}}^{\Upsilon}| < 1.93$. The $R_{p\mathrm{Pb}}$ results are compared to CMS results on $\Upsilon(nS)$ $R_{AA}$ (blue squares for $\Upsilon(1S)$ and $\Upsilon(2S)$) and the blue arrow for the upper limit at 95\% confidence level (CL) on $\Upsilon(3S)$ for $0 < p_T^{\Upsilon} < 30\,\mathrm{GeV}/c$ and $|\eta_{\mathrm{CM}}^{\Upsilon}| < 2.4$ at the same energy~\cite{CMS2022Upsilon_pPb,2019270}. Vertical bars represent statistical and fit uncertainties, and the filled boxes around points depict systematic uncertainties. The gray and red boxes around the unity line represent the uncertainty in the luminosity normalization for pp and $p\mathrm{Pb}$ collisions. The blue box around unity represents the global uncertainty related to $\mathrm{PbPb}$ data.}
 \label{fig:cms-RpPb-502-y}
\end{figure}

In $p\mathrm{Pb}$ and $pp$ collisions at ${5.02}~\text{TeV}$, CMS measured prompt and non-prompt $J/\psi\to\mu^+\mu^-$ for $2<p_{T}<{30}{~{\text{GeV/c}}}$ in $|y|<2.4$ ($pp$) and $-2.87<y<1.93$ ($p\mathrm{Pb}$) with {28} {pb$^{-1}$} and {35} {nb$^{-1}$}, finding $R_{p\mathrm{Pb}}(p_{T},y)$ consistent with 1 (Figure~\ref{fig:cms-RpPb-502}) but a forward-to-backward ratio $R_{FB}$ that decreases with transverse‐energy deposition (Figure~\ref{fig:cms-RpPb-502-2})~\cite{CMS2017Jpsi_pPb}.
Under the same collision conditions, CMS also measured sequential suppression of $\Upsilon$ states. The results show $\Upsilon(1S)$ as being the least suppressed, followed by $\Upsilon(2S)$, and $\Upsilon(3S)$ exhibiting the strongest suppression (Figure~\ref{fig:cms-RpPb-502-y})~\cite{CMS2022Upsilon_pPb}.

\vspace{-6pt}

\vspace{-6pt}

\begin{figure}[H]
 \includegraphics[width=\linewidth]{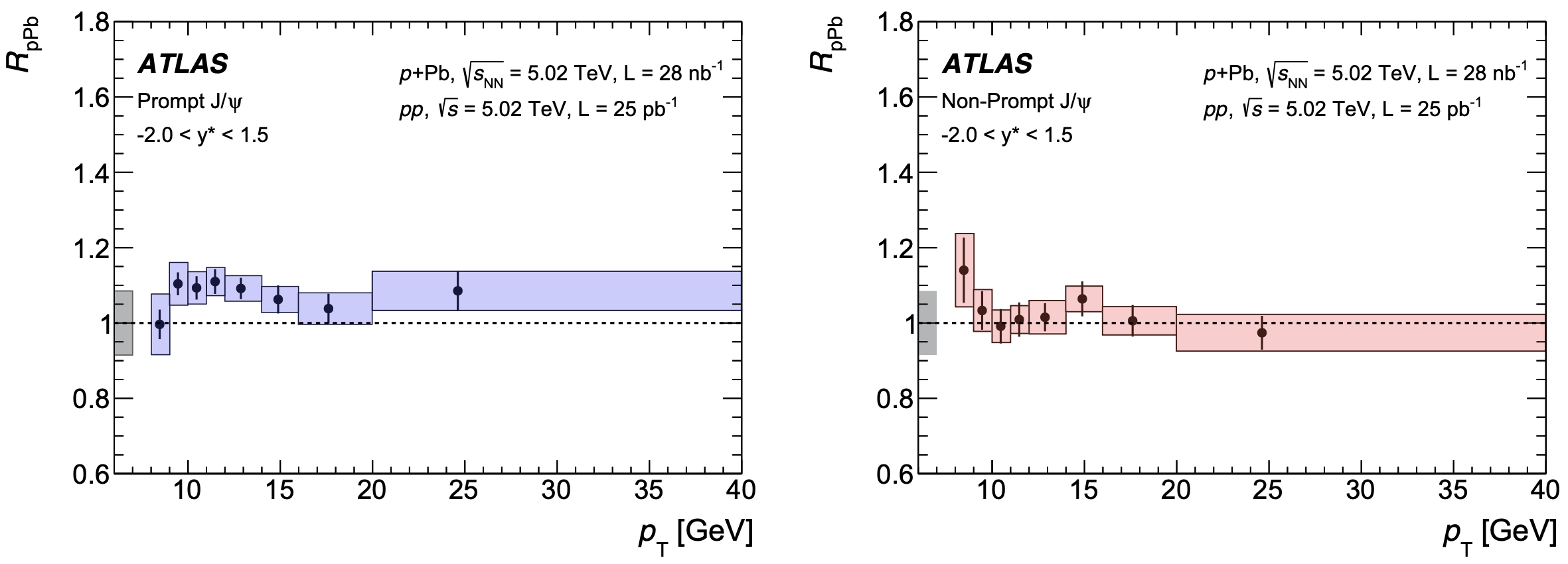}
 \includegraphics[width=\linewidth]{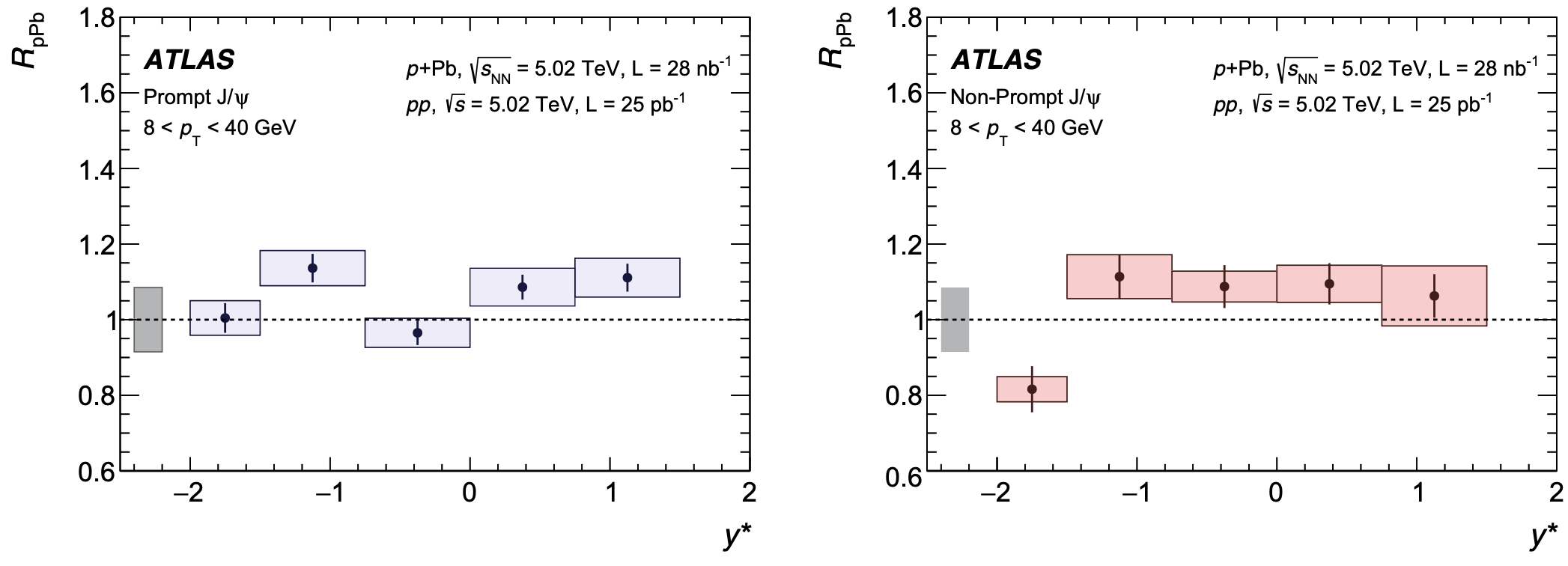}
 \caption{Nuclear 
 modification 
 factor $R_{p\mathrm{Pb}}$ versus $p_{T}$ (\textbf{top}) and $y$ (\textbf{bottom}) for prompt $J/\psi$ (\textbf{left}) and non‐prompt $J/\psi$ (\textbf{right}). The horizontal position of each point is the mean of the weighted $p_{T}$ distribution and the box width is the corresponding $p_{T}$ bin size. Vertical error bars show statistical uncertainties, box heights denote uncorrelated systematic uncertainties, and the leftmost grey boxes at $R_{p\mathrm{Pb}}=1$ indicate the correlated systematic uncertainty~\cite{ATLASQuarkonium_pPb_pp5p02TeV2018}.}
 \label{fig:atlas-RpPb-502}
\end{figure}

\begin{figure}[H]
 \includegraphics[width=0.5\linewidth]{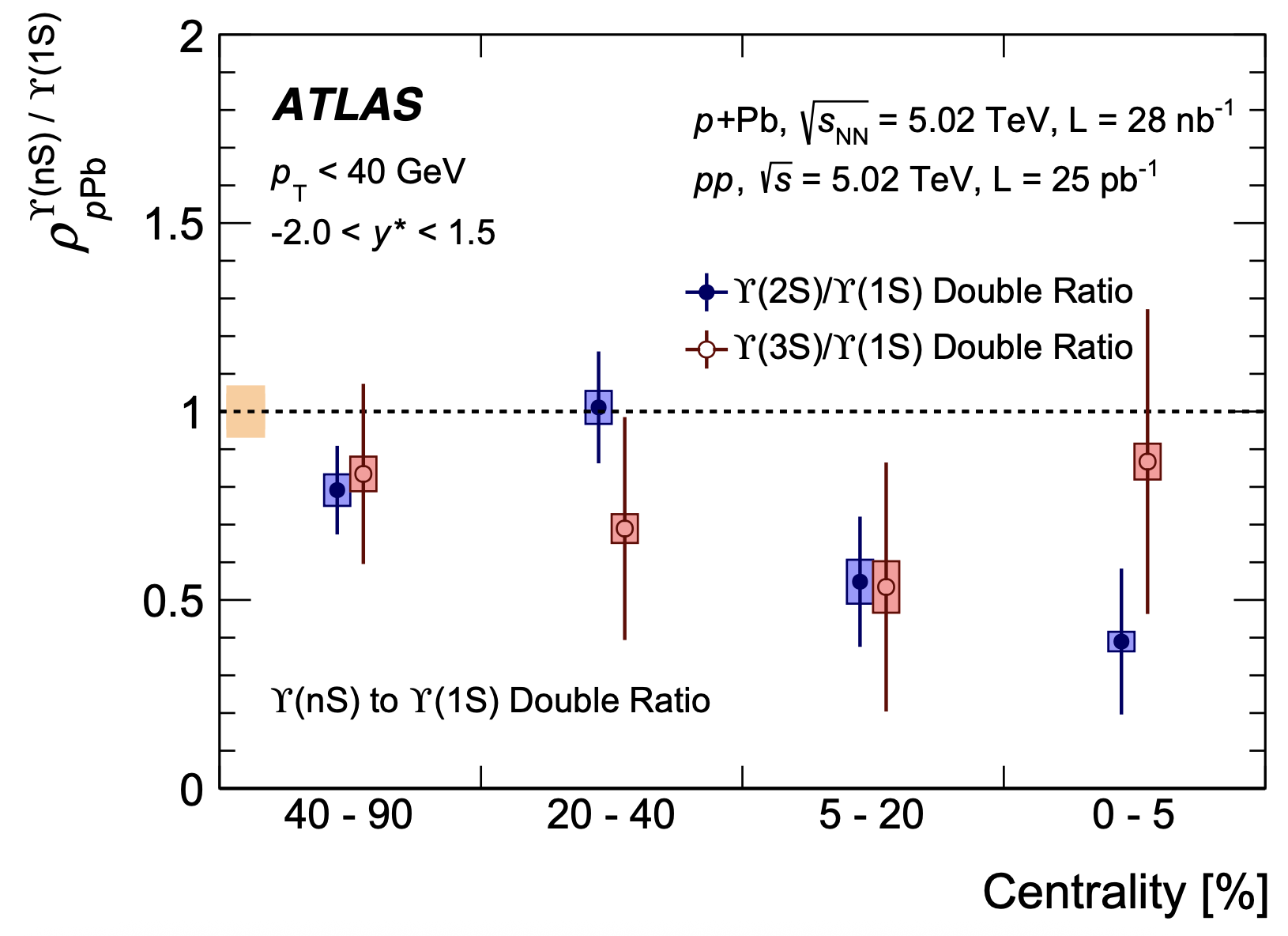}
 \caption{The 
 bottomonium 
 double-ratio  
 $\rho_{p\mathrm{Pb}}^{\Upsilon(\mathrm{n}\mathrm{S})/\Upsilon(1\mathrm{S})}=\frac{R_{p\mathrm{Pb}}(\Upsilon(\mathrm{n}\mathrm{S}))}{R_{p\mathrm{Pb}}(\Upsilon(1\mathrm{S}))}$ as a function of event centrality in $p\mathrm{Pb}$ collisions. Vertical error bars show statistical uncertainties, colored box heights indicate uncorrelated systematic uncertainties, and the leftmost yellow box at $\rho_{p\mathrm{Pb}}^{\Upsilon(\mathrm{n}\mathrm{S})/\Upsilon(1\mathrm{S})}=1$ represents the total uncertainty of the $pp$ reference~\cite{ATLASQuarkonium_pPb_pp5p02TeV2018}.}
 \label{fig:double_ratio_bottomonium}
 \label{fig:atlas-RpPb-502-2}
\end{figure}

ATLAS, using {28}{nb$^{-1}$} of $p\mathrm{Pb}$ and {25}{pb$^{-1}$} of $pp$ data, reconstructed $J/\psi$, $\psi(2\mathrm{S})$ and $\Upsilon(\mathrm{n}\mathrm{S})$ ($n=1,2,3$) for $p_{T}<{30}{~{\text{GeV/c}}}$ and $|y|<1.5$, separated prompt and non-prompt components, and observed no significant suppression of ground‐state $J/\psi$ (Figure~\ref{fig:atlas-RpPb-502})~\cite{ATLASQuarkonium_pPb_pp5p02TeV2018}, which is consistent with the CMS results. However, a stronger low‐$p_{T}$ suppression of excited charmonium and bottomonium in central $p\mathrm{Pb}$ collisions (Figure~\ref{fig:atlas-RpPb-502-2}) was observed.

\vspace{-6pt}

\begin{figure}[H]
 \includegraphics[width=\linewidth]{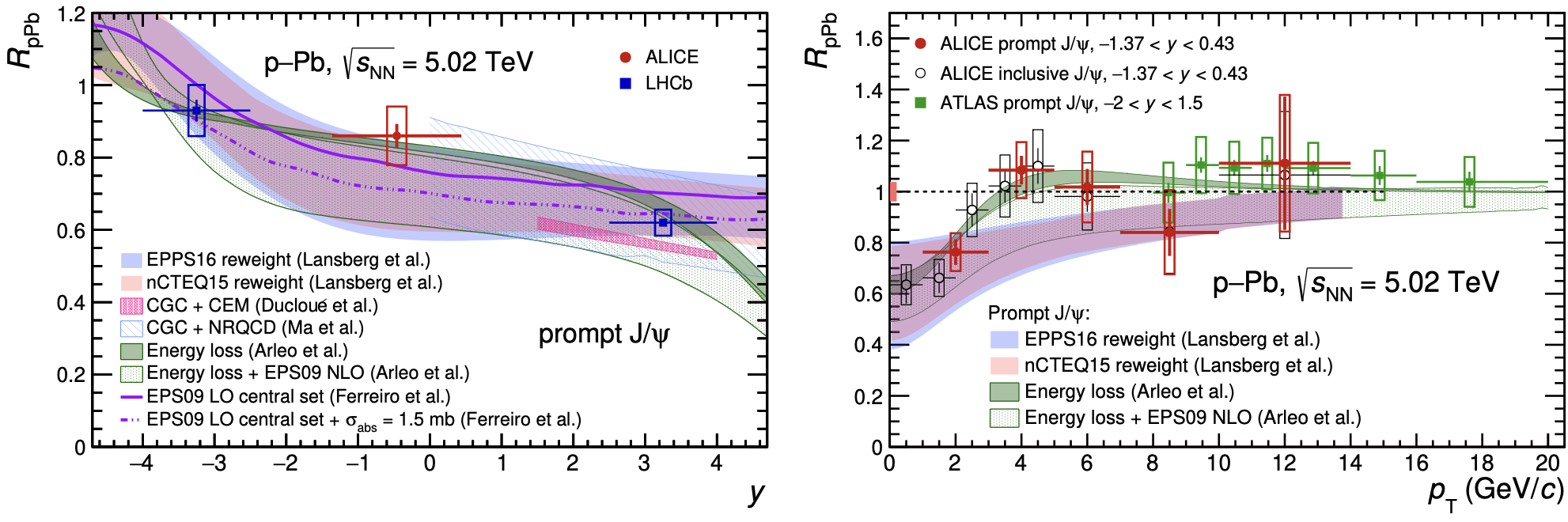}
 \includegraphics[width=\linewidth]{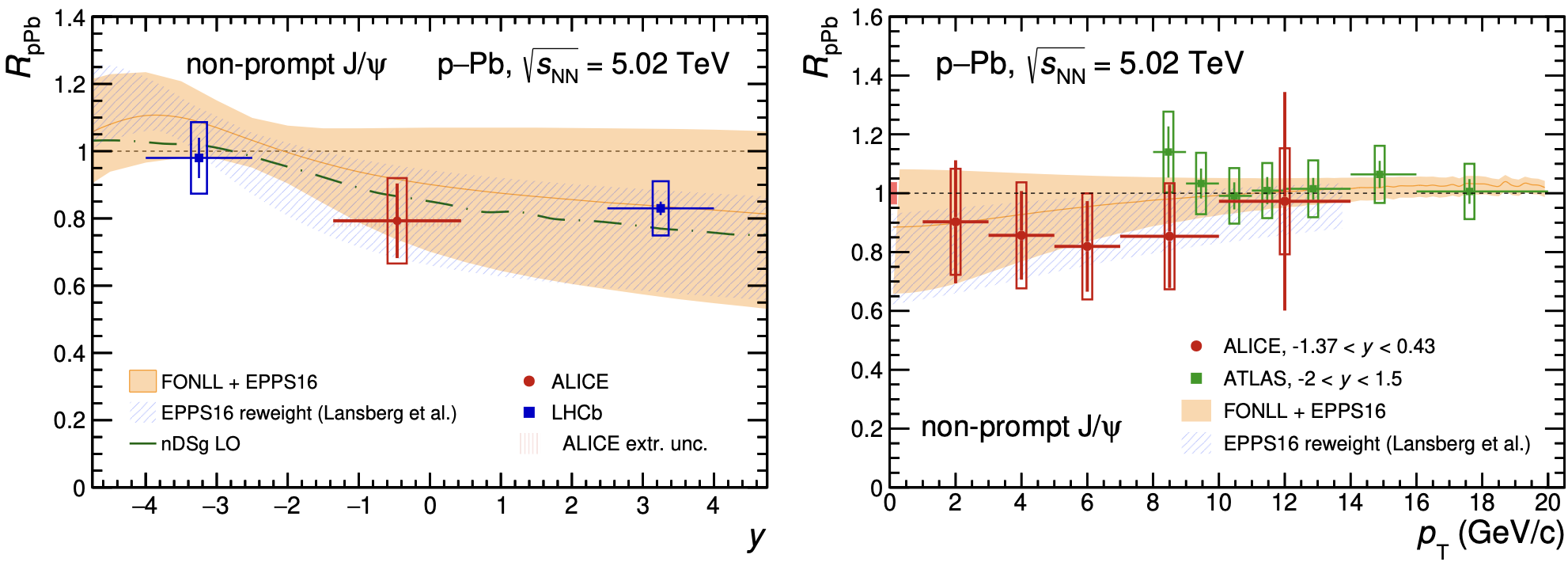}
 \caption{Nuclear modification factor $R_{pPb}$ of prompt $J/\psi$ as a function of rapidity (\textbf{top left}) and versus transverse momentum $p_{T}$ at midrapidity alongside inclusive $J/\psi$ (\textbf{top right}) in $p\mathrm{Pb}$ collisions at ${5.02}~\text{TeV}$, measured by ALICE~\cite{ALICEJpsiMidRapiditypPb5p02TeV2022}, compared with LHCb measurements at backward and forward rapidity in the left panel and ATLAS data up to $p_{T}={20}{~{\text{GeV/c}}}$ in the right. Vertical error bars denote statistical uncertainties and the open boxes systematic uncertainties; the top‐left panel also includes the extrapolation uncertainty to $p_{T}>0$ for ALICE dielectron data, and the filled box at $R_{p\mathrm{Pb}}=1$ in the top‐right panel indicates the global relative uncertainty of the ALICE measurement. The \textbf{bottom‐left} (\textbf{bottom‐right}) 
 panel shows $R_{p\mathrm{Pb}}$ of non-prompt $J/\psi$ as a function of rapidity (transverse momentum) at midrapidity, again compared to LHCb and ATLAS, with vertical bars and open boxes for statistical and systematic uncertainties, the shaded box in the bottom‐left panel for the extrapolation to $p_{T}=0$, and the filled box at $R_{p\mathrm{Pb}}=1$ in the bottom‐right panel marking the global normalization uncertainty. Predictions from the nDSg and EPPS16 nuclear‐PDF parametrizations (including a reweighted calculation) are superimposed on both bottom panels.}
 \label{fig:alice-RpPb-502}
\end{figure}

By using {292} $\upmu$b$^{-1}$ of $p\mathrm{Pb}$ collision data collected in 2016, ALICE measured inclusive, prompt and non-prompt $R_{p\mathrm{Pb}}(p_{T},y)$ of $J/\psi$ at midrapidity in the dielectron channel, and from the non-prompt yield extracted the $b\bar b$ production cross-section at midrapidity~\cite{ALICEJpsiMidRapiditypPb5p02TeV2022}. While CMS delivers a broad kinematic survey of cold‐nuclear‐matter effects, ATLAS extends to excited and bottomonium states with centrality dependence, and ALICE uniquely probes $J/\psi$ down to zero $p_{T}$ with precision dielectron measurements. (Figure~\ref{fig:alice-RpPb-502})

\begin{figure}[H]
 \includegraphics[width=0.8\linewidth]{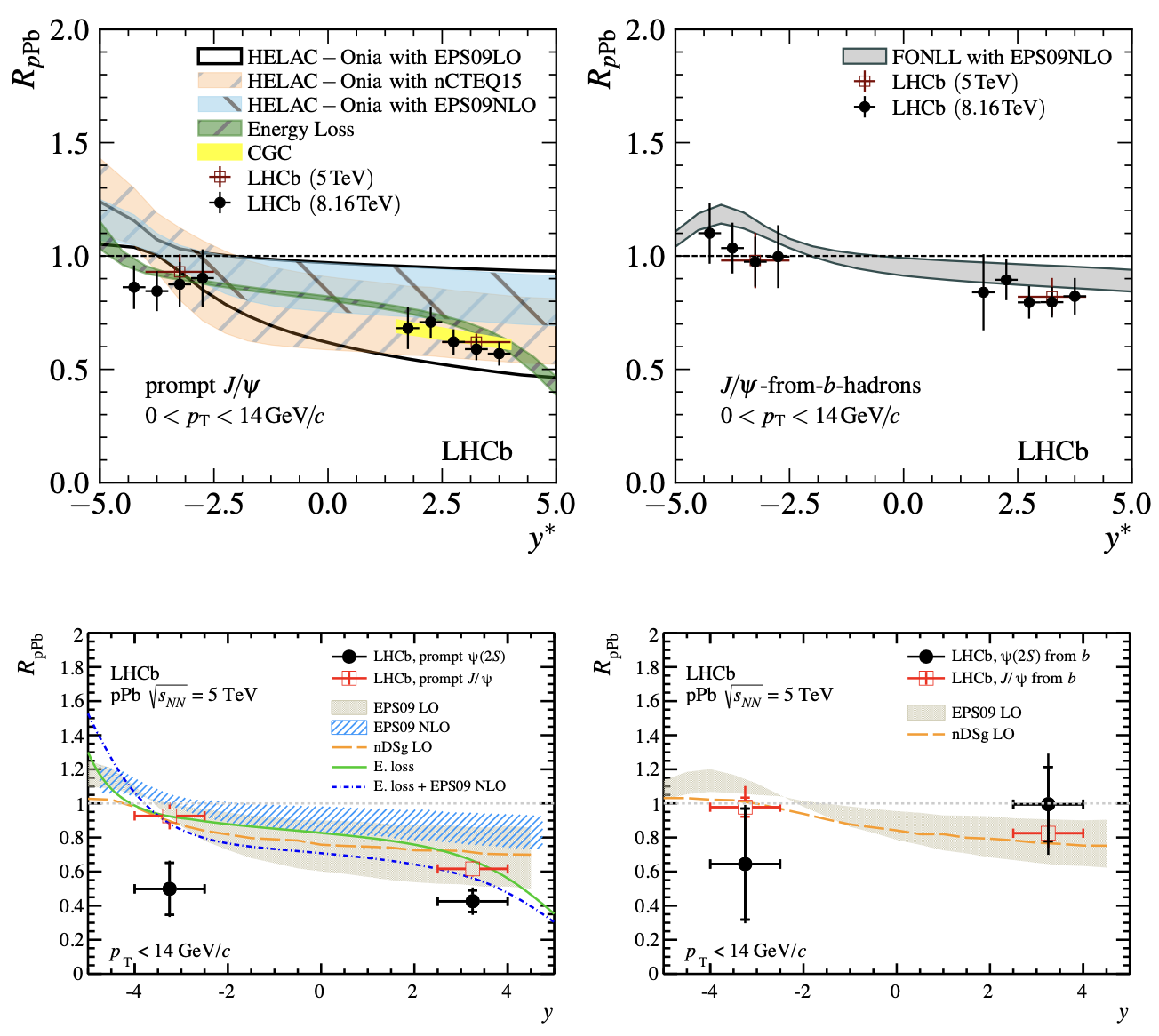}
 \caption{\textbf{Top}: Nuclear 
 modification 
 factor $R_{p\mathrm{Pb}}$ for prompt (left) and non‐prompt (right) $J/\psi$ mesons at ${8.16}~\text{TeV}$ (black points), displayed together with the ${5.02}~\text{TeV}$ measurements and various theoretical predictions. \textbf{Bottom}: $R_{p\mathrm{Pb}}$ for prompt (left) and non‐prompt (right) $\psi(2\mathrm{S})$ at ${5.02}~\text{TeV}$, with the corresponding $J/\psi$ results and model curves overlaid~\cite{LHCbQuarkoniaInpPb2016}.}
 \label{fig:lhcb-RpPb-816}
\end{figure}

\begin{figure}[H]
 \includegraphics[width=0.8\linewidth]{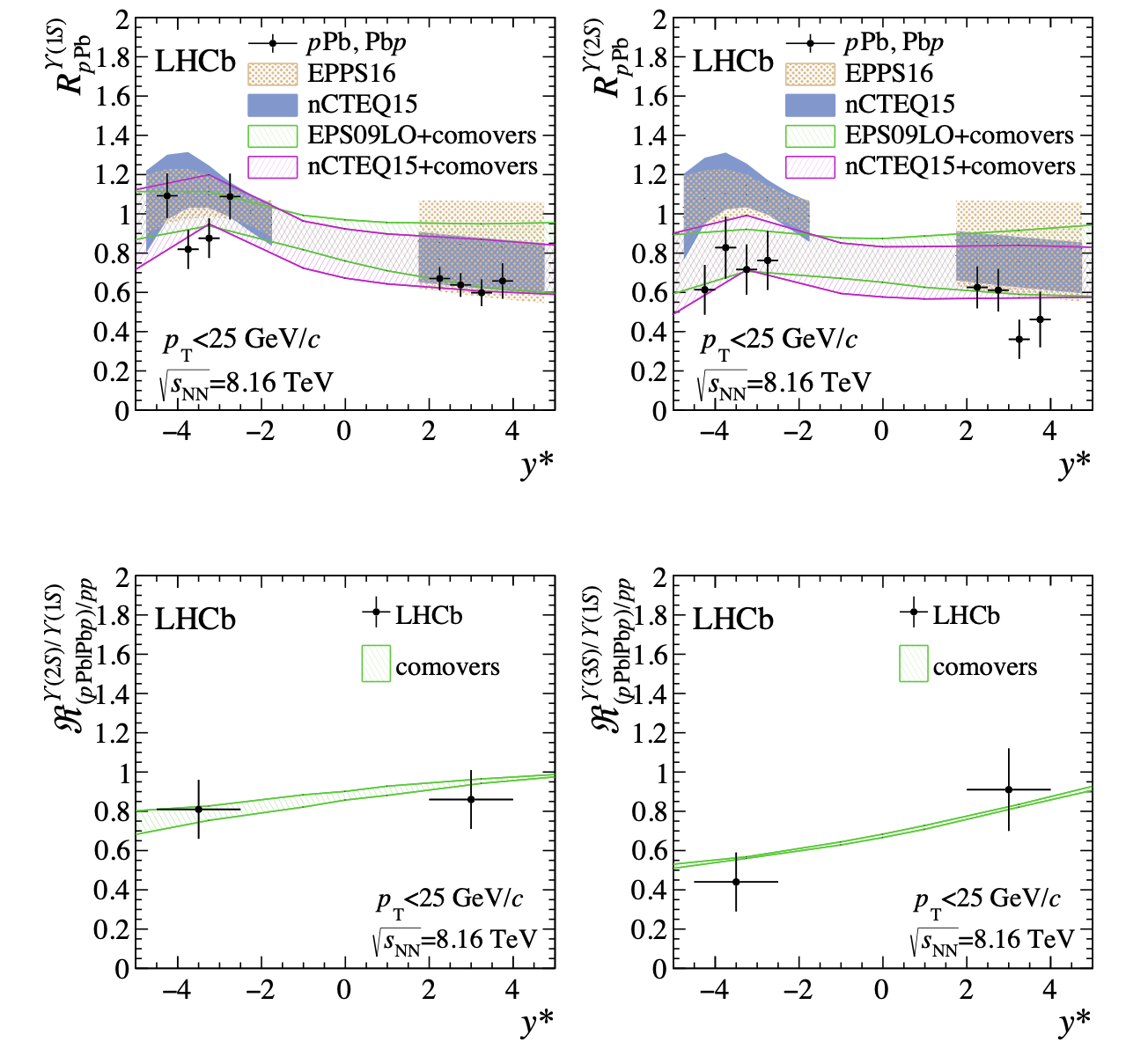}
 \caption{\textbf{Top}: Nuclear 
 modification factor $R_{p\mathrm{Pb}}$ for $\Upsilon(1\mathrm{S})$ (left) and $\Upsilon(2\mathrm{S})$ (right) as a function of rapidity in $p\mathrm{Pb}$ and $\mathrm{Pb}p$ collisions at ${8.16}~\text{TeV}$. \textbf{Bottom}: Nuclear modification factor ratio \(
\frac{R_{p\mathrm{Pb}}(\Upsilon(\mathrm{nS}))}{R_{p\mathrm{Pb}}(\Upsilon(1\mathrm{S}))}
\) for $\Upsilon(2\mathrm{S})$ (left) and $\Upsilon(3\mathrm{S})$ (right) versus rapidity~\cite{LHCbQuarkoniaInpPb2016}.}
 \label{fig:lhcb-RpPb-816-2}
\end{figure}

Complementing these, LHCb has measured quarkonium production in $p\mathrm{Pb}$ collisions at ${8.16}~\text{TeV}$ using the 2016 data sample in the forward ($2<y<5$, $p\mathrm{Pb}$) and backward ($2<y<5$, $\mathrm{Pb}p$) configurations~\cite{LHCbQuarkoniaInpPb2016}. By disentangling prompt and non-prompt $J/\psi$ and $\psi(2\mathrm{S})$ via fits to the dimuon pseudo-proper decay length, LHCb extracted $R_{p\mathrm{Pb}}(p_{T},y)$ for each state separately (Figure~\ref{fig:lhcb-RpPb-816}), achieving marked precision gains over the ${5}{TeV}$ analysis. The collaboration also reports the bottomonium nuclear modification factor ratio 
\(
\frac{R_{p\mathrm{Pb}}(\Upsilon(\mathrm{n}\mathrm{S}))}{R_{p\mathrm{Pb}}(\Upsilon(1\mathrm{S}))},
\) 
as a function of rapidity and event activity (Figure~\ref{fig:lhcb-RpPb-816-2}), thus providing comprehensive constraints on CNM effects across charmonium and bottomonium states.

\subsection{Multiplicity Dependence}\label{sec33}

Multiplicity, defined as the charged-particle yield within a specified rapidity or pseudorapidity window, quantifies event activity and the soft environment accompanying hard production. In \(\mathrm{PbPb}\) and \(p\mathrm{Pb}\) collisions it correlates strongly with geometry and centrality proxies (e.g.,\ \(\langle N_{\mathrm{part}}\rangle,\langle N_{\mathrm{coll}}\rangle\)), whereas in \(pp\) there is no geometric centrality, yet multiplicity remains a powerful classifier of underlying–event intensity. Consequently, multiplicity has become a key for testing quarkonium formation and dissociation.

\begin{figure}[H]
  \includegraphics[width=0.5\linewidth]{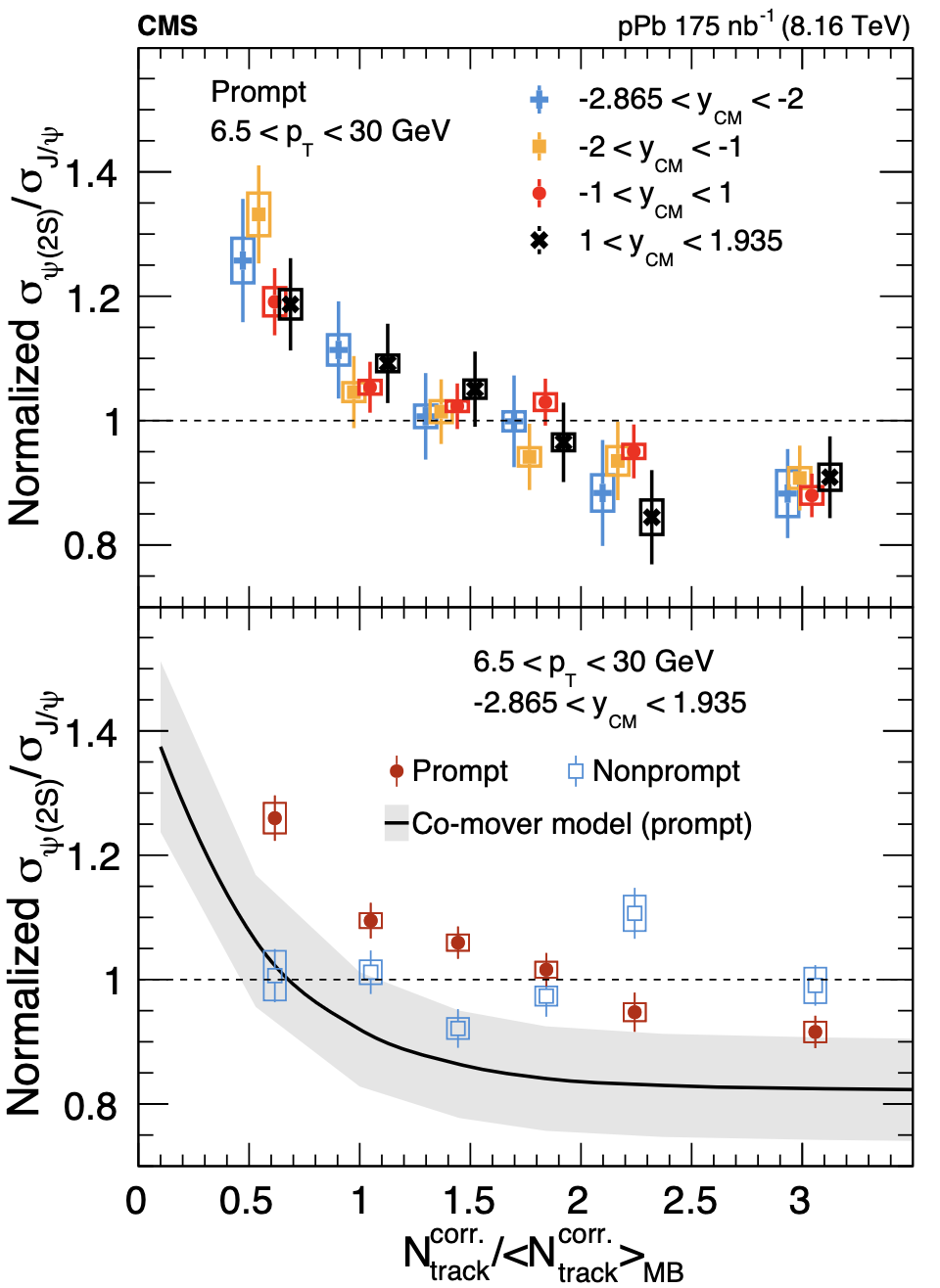}
  \caption{\textbf{Upper}: the 
 prompt normalized cross-section ratio versus normalized event charged-particle multiplicity for four different rapidity selections. The different sets of colored data points have a slight horizontal offset applied for visual clarity. \textbf{Lower}: the same observable for prompt (red circular points) and non-prompt (blue square points) mesons when combining all four rapidity selections into a single measurement. The black line with a gray band shows a model incorporating co-mover effects~\cite{Ferreiro:2014bia} for prompt charmonia. Error bars represent statistical uncertainties, while boxes display systematic uncertainties~\cite{CMS:2025pPbPsi2SJpsiMult816}.}
  \label{fig:cms-pPb}
\end{figure}

\begin{figure}[H]
  \includegraphics[width=\linewidth]{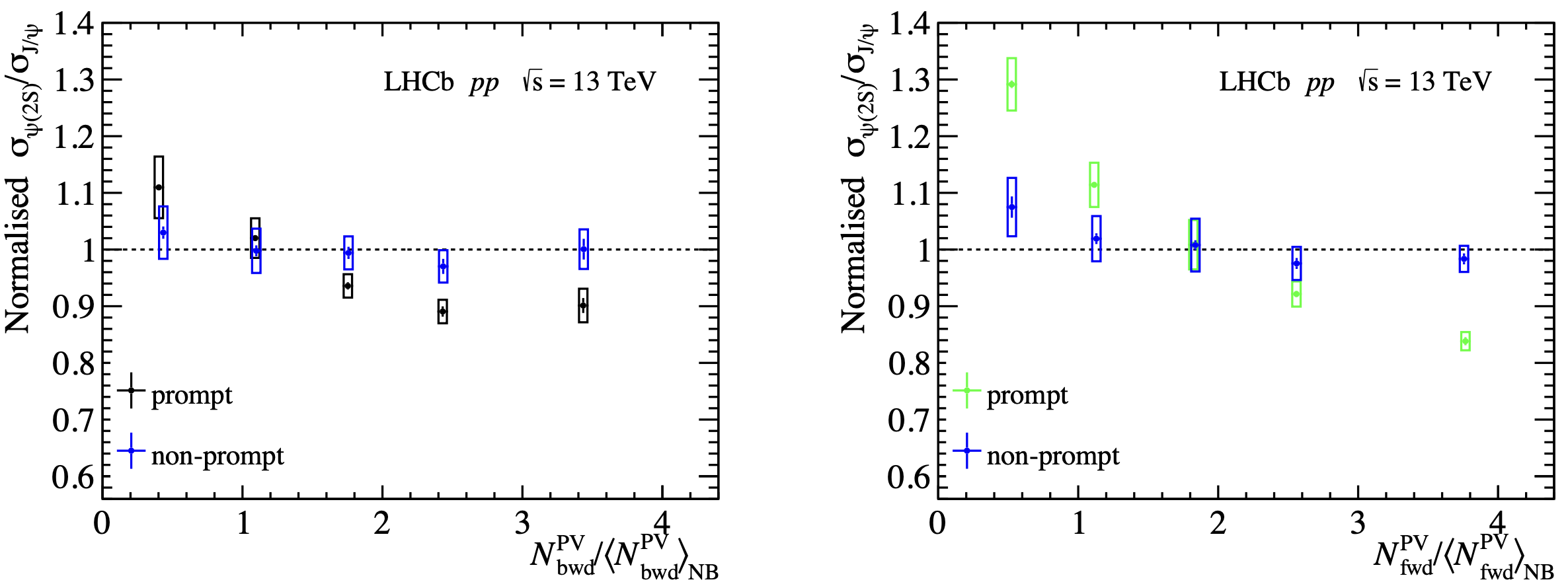}
  \caption{Normalized production ratio as a function of (\textbf{left}) \(N_{\mathrm{bwd}}^{\mathrm{PV}}/\langle N_{\mathrm{bwd}}^{\mathrm{PV}}\rangle_{\mathrm{NB}}\), with \(-30<z_{\mathrm{PV}}<180~\mathrm{mm}\), and (\textbf{right}) \(N_{\mathrm{fwd}}^{\mathrm{PV}}/\langle N_{\mathrm{fwd}}^{\mathrm{PV}}\rangle_{\mathrm{NB}}\), with \(-180<z_{\mathrm{PV}}<180~\mathrm{mm}\), integrated over the full \(p_{T}\)-\(y\) range of \(2.0<y<4.5\) and \(0.3<p_{T}<20~\mathrm{GeV}/c\)~\cite{LHCb:2024ppPsi2SJpsiMult13TeV}.}
  \label{fig:pp-lhcb}
\end{figure}

\begin{figure}[H]
  \includegraphics[width=\linewidth]{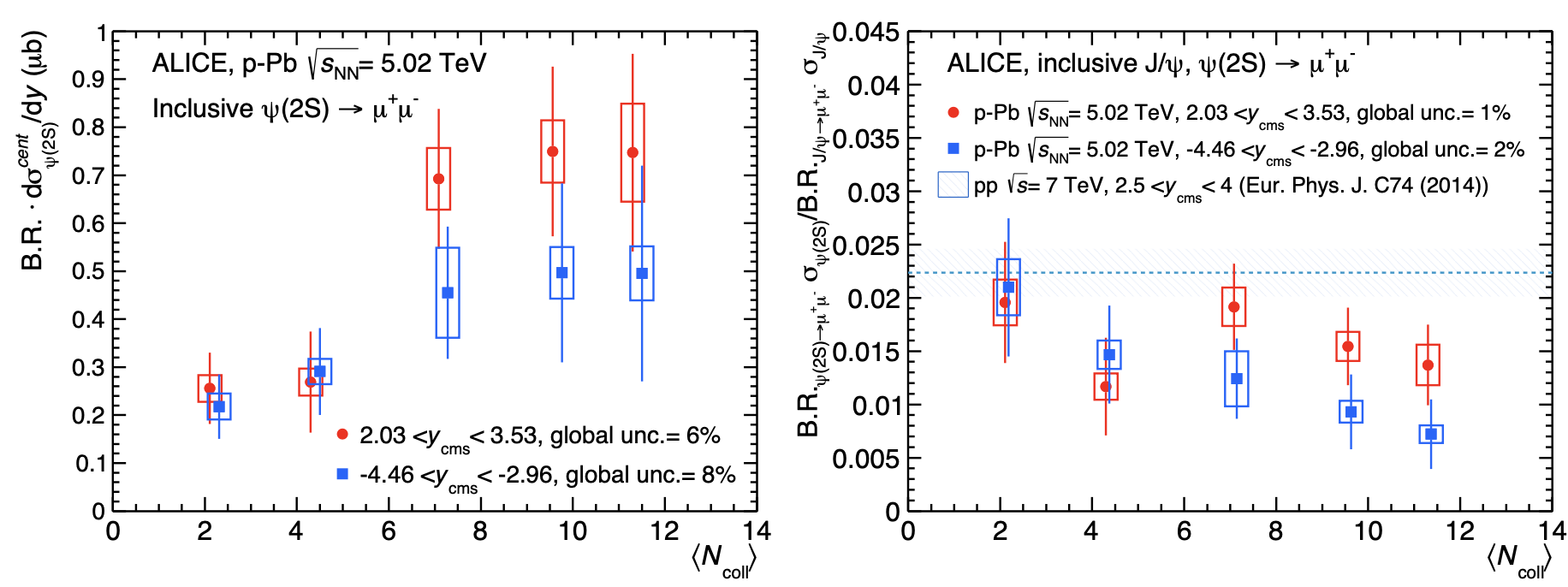}
  \caption{\textbf{Left}: \(\psi(2\mathrm{S})\) production 
 cross-sections shown as a function of centrality of the collision \(\langle N_{\mathrm{coll}}\rangle\) for \(p\mathrm{Pb}\) and \(\mathrm{Pb}p\) collisions. \textbf{Right}: \((\mathrm{BR}_{\psi(2\mathrm{S}) \to \mu^{+}\mu^{-}}\times\sigma_{\psi(2\mathrm{S})})/\bigl(\mathrm{BR}_{J/\psi \to \mu^{+}\mu^{-}}\times\sigma_{J/\psi}\bigr)\) shown as a function of \(\langle N_{\mathrm{coll}}\rangle\) and compared to the \(pp\) value (line), with a band representing its uncertainty. In both figures, vertical error bars correspond to statistical uncertainties, while the open boxes represent the systematic uncertainties. The \(\mathrm{Pb}p\) points are slightly shifted in \(\langle N_{\mathrm{coll}}\rangle\) to improve visibility~\cite{ALICE:2016Psi2SCentralitypPb}.}
  \label{fig:alice-pPb-cent}
\end{figure}

The observable \(\sigma_{\psi(2\mathrm{S})}/\sigma_{J/\psi}\) provides a clean probe of activity-dependent quarkonium dynamics because luminosity, efficiency, and most initial-state effects largely cancel, making the ratio directly sensitive to final–state interactions across \(pp\), \(p\mathrm{Pb}\), and \(\mathrm{PbPb}\). 
In \(p\mathrm{Pb}\) at \({8.16}~{\text{TeV}}\), CMS performs a multiplicity scan with prompt and non-prompt separation and observes a statistically significant multiplicity dependence only for the prompt ratio, with the non-prompt ratio remaining flat (Figure~\ref{fig:cms-pPb})~\cite{CMS:2025pPbPsi2SJpsiMult816}. 
In \(pp\) at \({13}~{\text{TeV}}\), LHCb measures the ratio versus overlapping  and nonoverlapping multiplicity estimators. A clear dependence appears solely for the prompt ratio with the local estimator, while nonoverlapping estimators and the non-prompt ratio are nearly flat (Figure~\ref{fig:pp-lhcb})~\cite{LHCb:2024ppPsi2SJpsiMult13TeV}. 
In a centrality-differential \(p\mathrm{Pb}\) study at \({5.02}~{\text{TeV}}\), ALICE found an enhanced suppression of \(\psi(2\mathrm{S})\) relative to \(J/\psi\) via the double ratio and \(Q_{p\mathrm{Pb}}\), most pronounced at backward (Pb–going) rapidity (Figure~\ref{fig:alice-pPb-cent})~\cite{ALICE:2016Psi2SCentralitypPb}.

\vspace{-6pt}

\vspace{-6pt}

\begin{figure}[H]
  \includegraphics[width=0.5\linewidth]{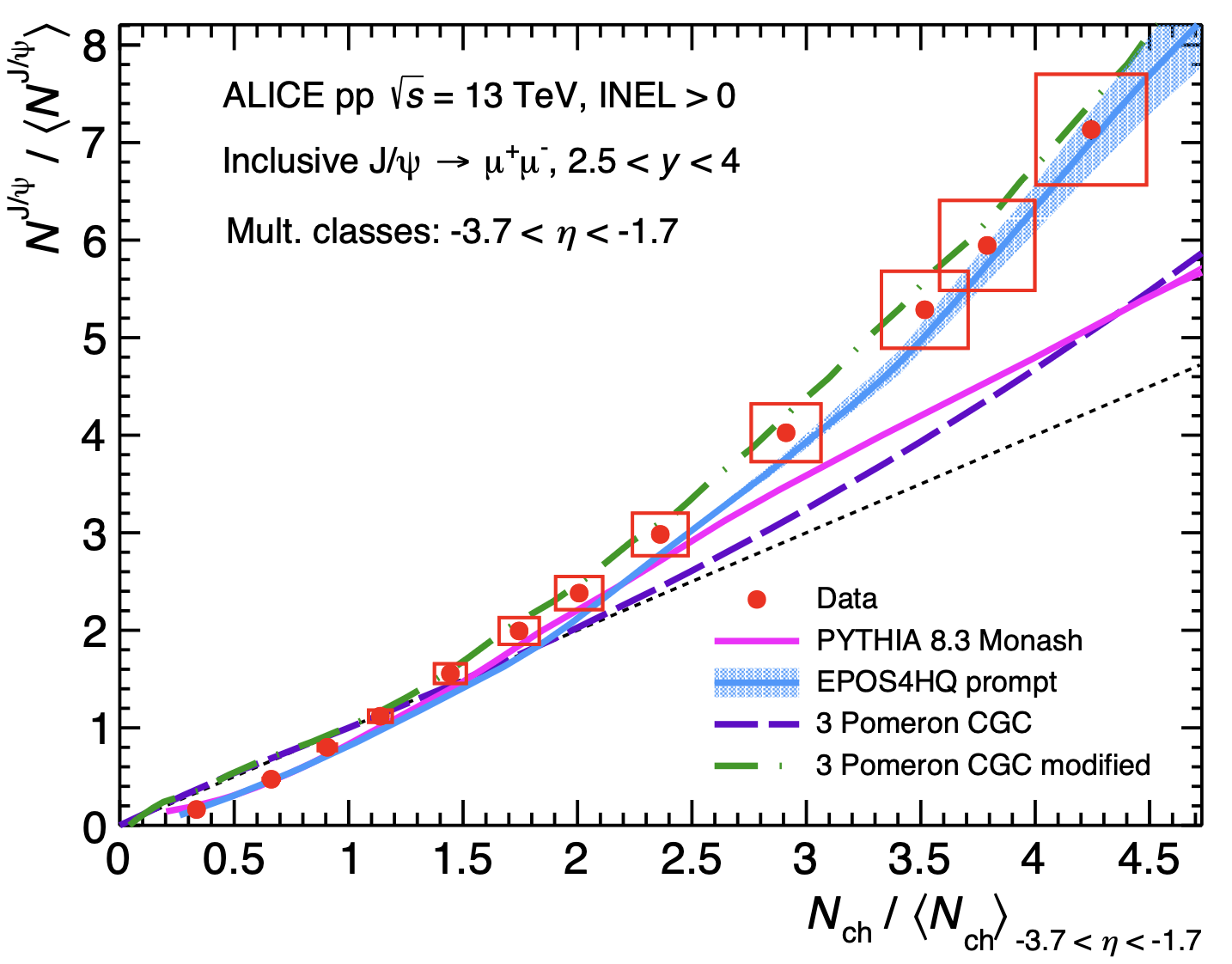}
  \caption{\(J/\psi\) relative 
 yield as a function of the relative multiplicity in the V0C acceptance \(-3.7<\eta<-1.7\), compared with PYTHIA~8.3~\cite{Bierlich:2022pfr}, EPOS4HQ~\cite{Zhao:2023ucp} and two 3-Pomeron CGC model predictions~\cite{Levin:2019fvb,ConesadelValle:2024hse}. The dotted line represents the diagonal~\cite{ALICE:2025ppJpsiForwardMult13TeV}.}
  \label{fig:pp-alice}
\end{figure}

ALICE also measures the multiplicity dependence of inclusive \(J/\psi\) at \(\sqrt{s}={13}~\text{TeV}\) and finds a steeper-than-linear rise in the relative \(J/\psi\) yield as a function of multiplicity~\cite{ALICE:2025ppJpsiForwardMult13TeV} (Figure~\ref{fig:pp-alice}).

\subsection{Production in Ultraperipheral Collisions}\label{sec34}

Ultraperipheral collisions (UPCs) occur at impact parameters exceeding the sum of the nuclear radius, where intense quasireal photon fields from fast ions drive photonuclear and two-photon interactions. In \(\mathrm{PbPb}\) and \(p\mathrm{Pb}\) UPCs, exclusive vector-meson photoproduction is coherent when the photon couples to the full nucleus (small \(p_T\)) and incoherent when it couples to a single nucleon (large \(p_T\)). These measurements probe nuclear gluons at \(x \sim 10^{-5} – 10^{-2}\) under clean, low-multiplicity conditions.

\begin{figure}[H]
  \includegraphics[width=0.5\linewidth]{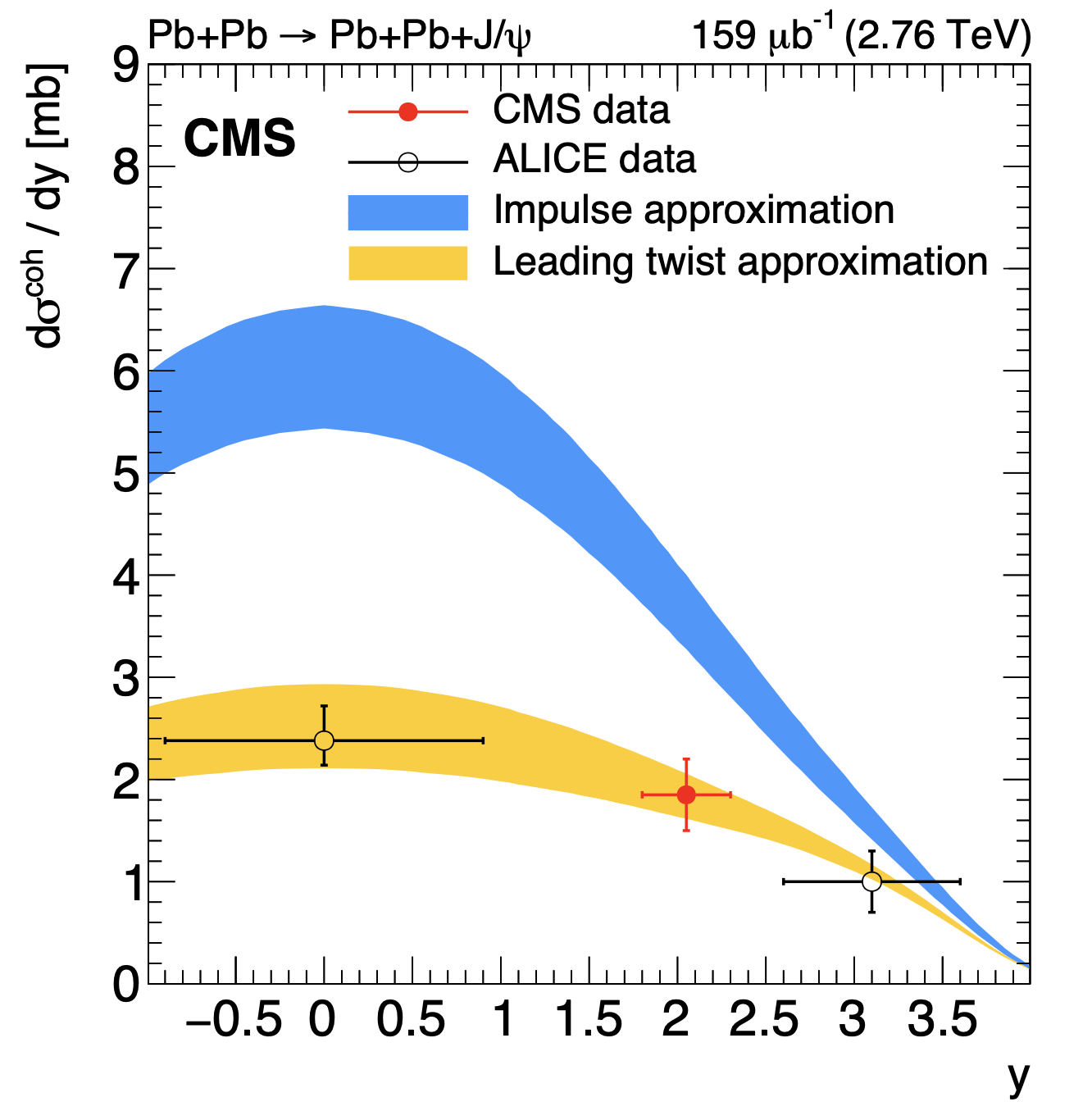}
  \caption{Differential 
 cross-section versus rapidity for coherent \(J/\psi\) production in ultraperipheral \(\mathrm{PbPb}\) collisions at \({2.76}~{\text{TeV}}\), measured by ALICE~\cite{ALICE:2013wjo,ALICE:2012yye} and CMS~\cite{CMS:2017PbPbUPC276}. The vertical error bars include the statistical and systematic uncertainties added in quadrature, and the horizontal bars represent the range of the measurements in \(y\). Also the impulse approximation and the leading twist approximation calculations are shown.}
  \label{fig:upc-cms276}
\end{figure}

\begin{figure}[H]
  \includegraphics[width=0.6\linewidth]{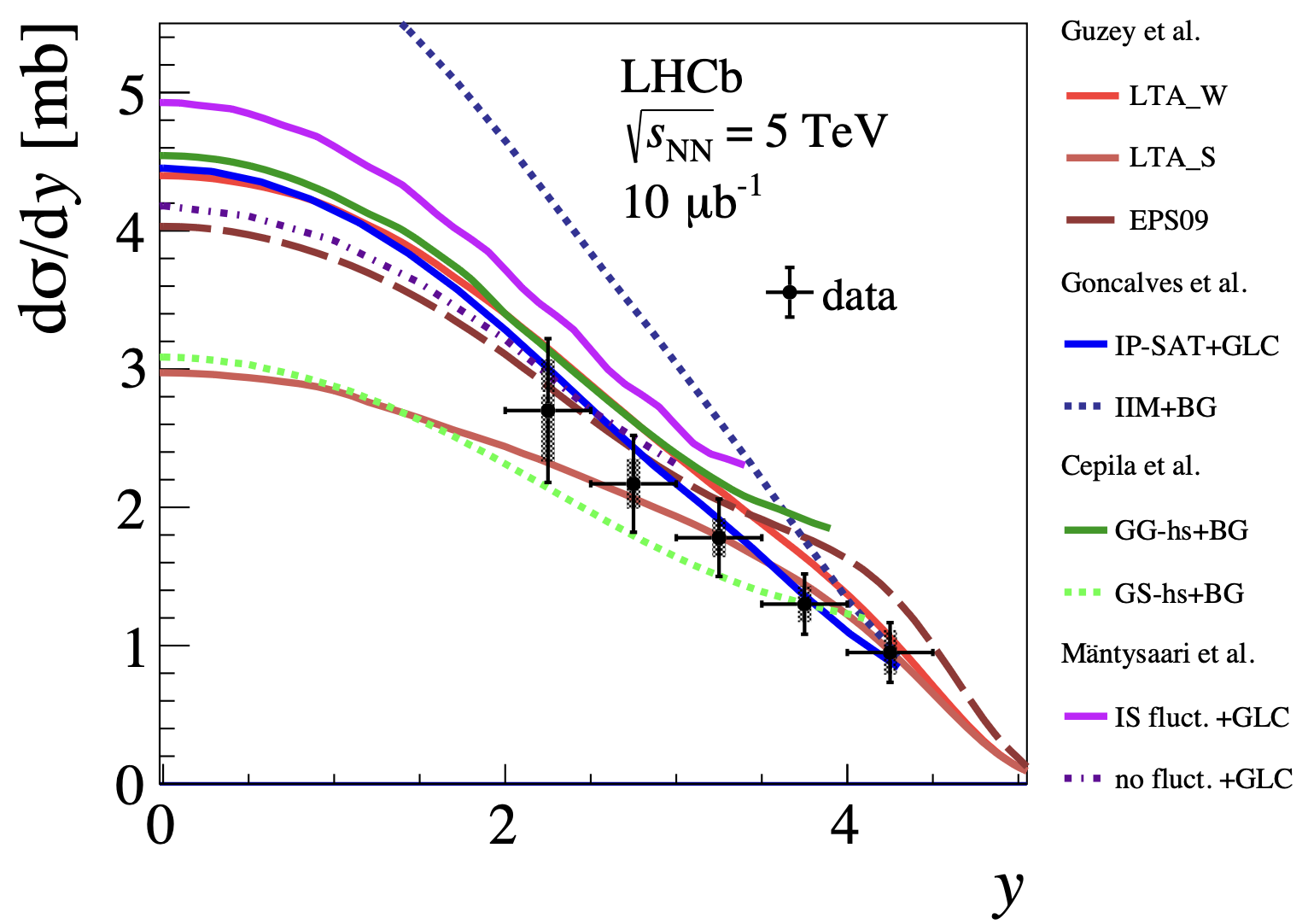}
  \caption{Differential 
 cross-section as a function of rapidity for coherent \(J/\psi\) production compared to different phenomenological predictions~\cite{PhysRevC.93.055206,PhysRevC.84.011902,PhysRevD.96.094027,PhysRevC.97.024901,MANTYSAARI2017832}. The measurements are shown as points, where the inner and outer error bars represent the statistical and the total uncertainties, respectively. This includes the uncertainty on the luminosity and is therefore highly correlated~\cite{LHCb:2022PbPbUPC5TeV}.}
  \label{fig:upc-lhcb5}
\end{figure}

For coherent \(J/\psi\) photoproduction, CMS studies the coherent \(J/\psi\) cross-section that disfavors impulse-approximation expectations in \(\mathrm{PbPb}\) collisions at \({2.76}~\text{TeV}\) using forward-rapidity dimuons (Figure~\ref{fig:upc-cms276})~\cite{CMS:2017PbPbUPC276}. 
LHCb performs the same class of measurement in \(\mathrm{PbPb}\) collisions at \({5}~\text{TeV}\) at forward rapidity with better acceptance at low transverse momentum, obtaining rapidity-differential coherent cross-sections that indicate sizable nuclear effects and extend the \(x\) reach (Figure~\ref{fig:upc-lhcb5})~\cite{LHCb:2022PbPbUPC5TeV}.

\vspace{-6pt}

\begin{figure}[H]
  \includegraphics[width=0.5\linewidth]{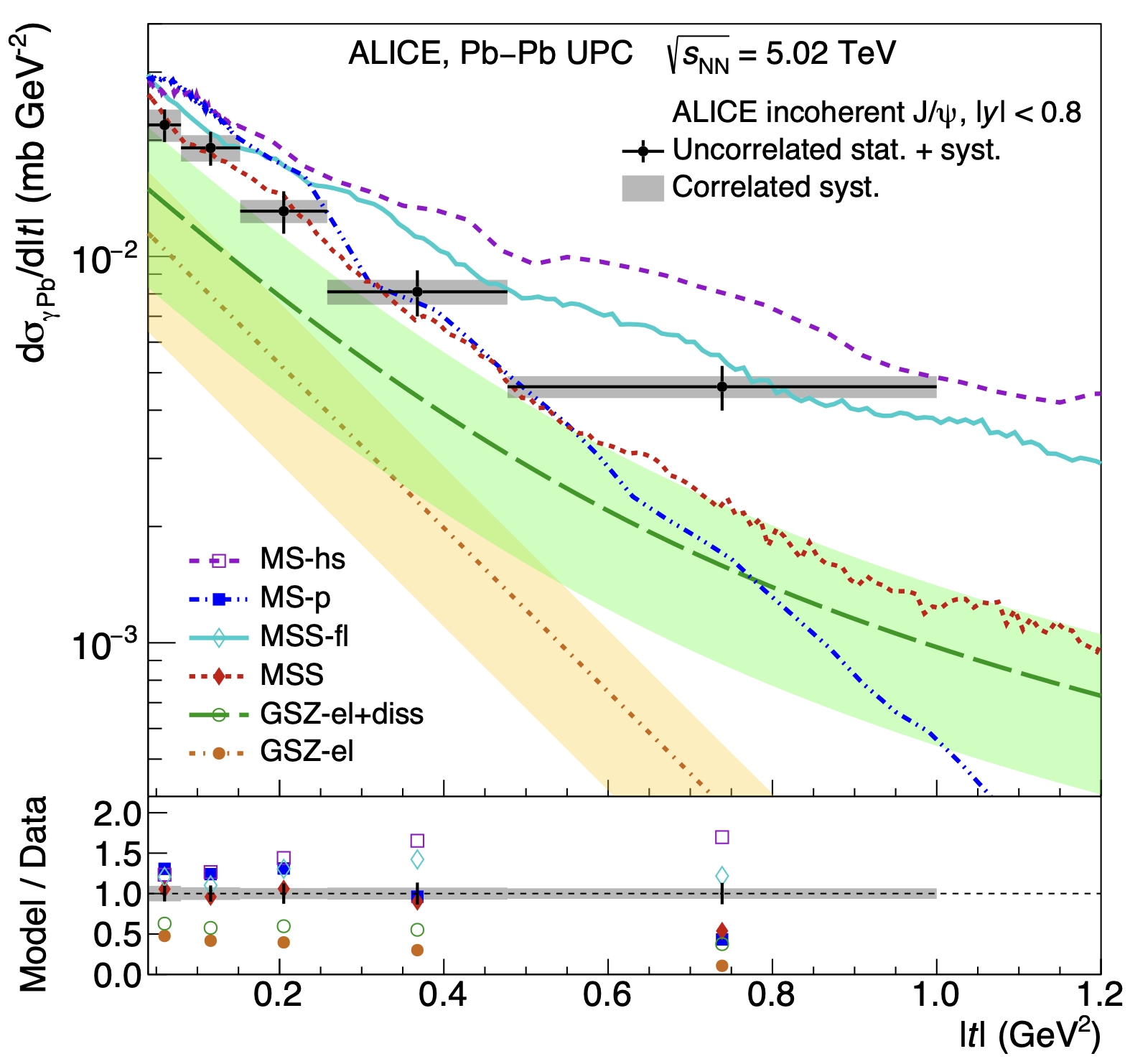}
  \caption{Cross-section for the incoherent photoproduction of \(J/\psi\) vector mesons in ultraperipheral \(\mathrm{PbPb}\) collisions at \({5.02}~{\text{TeV}}\) measured at midrapidity~\cite{ALICE:2024PbPbUPCIncohT}. The uncorrelated uncertainty (statistical and systematic added in quadrature) is indicated with the vertical bar, while the correlated uncertainty by the grey band. The width of each \(|t|\) range is given by the horizontal bars. The lines show the predictions of the different models described in the text. The bottom panel presents the ratio of the integral of the predicted to that of the measured cross-section in each \(|t|\) range. The relative uncertainties on the ratios calculated from GSZ are {45}{\%}.}
  \label{fig:upc-alice-incoh}
\end{figure}

Incoherent \(J/\psi\) photoproduction and the \(|t|=p_T^2\) spectrum are determined by ALICE at midrapidity using \(\mathrm{PbPb}\) collisions at \({5.02}~\text{TeV}\). The measured \(d\sigma/d|t|\) requires nucleon- and subnucleon-scale fluctuations to describe the observed shape (Figure~\ref{fig:upc-alice-incoh})~\cite{ALICE:2024PbPbUPCIncohT}.

\begin{figure}[H]
  \includegraphics[width=\linewidth]{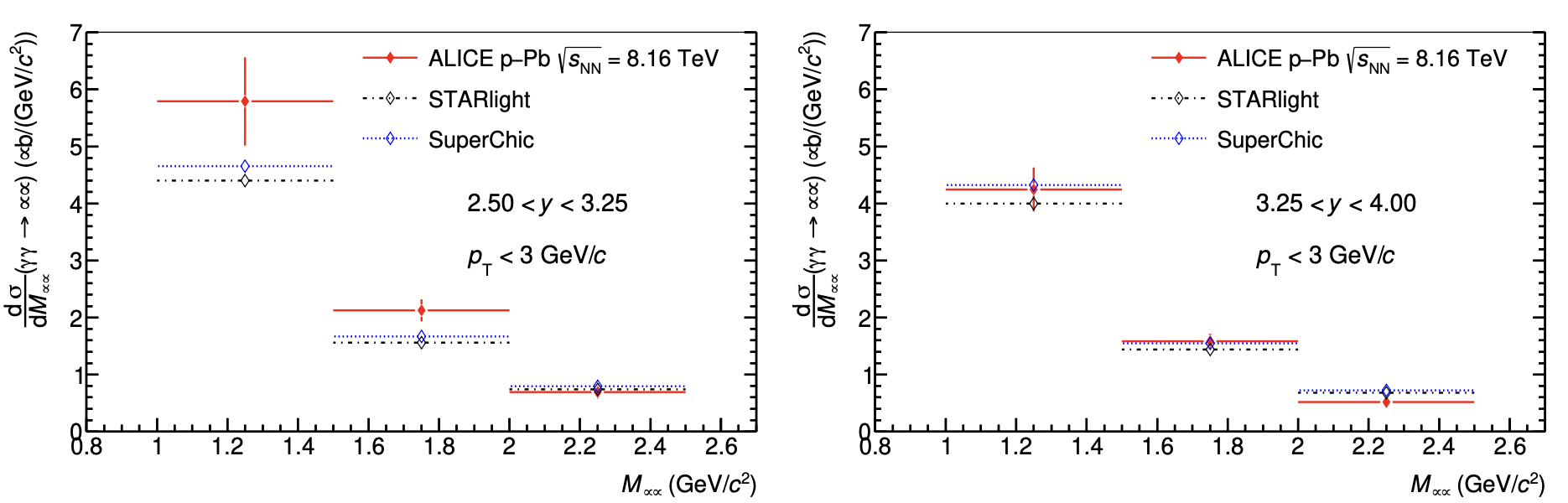}
  \caption{Differential cross-sections for exclusive \(\gamma\gamma\to\mu^{+}\mu^{-}\) production measured by ALICE in \(p\mathrm{Pb}\) UPCs at \({8.16}~{\text{TeV}}\) as a function of \(M_{\mu\mu}\), for \(2.5<y<3.25\) (\textbf{left}) and \(3.25<y<4\) (\textbf{right})~\cite{ALICE:2023pPbUPCExclDiss}. The vertical error bars represent the statistical and systematic uncertainties summed in quadrature. The results are compared with the predictions from STARlight~\cite{KLEIN2017258,PhysRevC.60.014903} and from SuperChic~\cite{Harland_Lang_2020}.}
  \label{fig:upc-alice-ppb}
\end{figure}

Exclusive and dissociative photoproduction are measured by ALICE in \(p\mathrm{Pb}\) UPCs at \({8.16}~\text{TeV}\) using forward detectors to enforce exclusivity and to separate \(\gamma p\to J/\psi p\), \(\gamma p\to J/\psi p^\ast\), and \(\gamma\gamma\to\mu^+\mu^-\) (Figure~\ref{fig:upc-alice-ppb})~\cite{ALICE:2023pPbUPCExclDiss}.

\section{Multi-Quarkonium Production}\label{sec4}

When single-quarkonium studies encounter challenges, investigation into multi-quarkonium production may shed new light on QCD and parton structure research. Compared to single-quarkonium research, the double quarkonium study provides a broader framework for theory--experiment comparison by providing additional measurable parameters like the separation in rapidity between the quarkonia $\Delta y$, the invariant mass of two quarkonia $m({q \bar q}_{1}{q \bar q}_{2})$, etc. 

Additionally, research on multi-quarkonium production may provide valuable insights into multi-body dynamics. The multi-quarkonium candidates produced in collisions can originate not only from single parton scattering (SPS), but also from double parton scattering (DPS) or even triple parton scattering (TPS) if three quarkonia are involved. A crucial parameter in multi-body dynamics within particle physics is the effective cross-section ($\sigma_{\mathrm{eff}}$), which relates the composite process to its constituent subprocesses:
\begin{equation} \label{Eq sigma eff}
    \sigma^{{q \bar q}_{1}{q \bar q}_{2}}_{\mathrm{DPS}}=\frac{m}{2}\frac{\sigma^{{q \bar q}{1}}_{\mathrm{SPS}}\sigma^{{q \bar q}_{2}}_{\mathrm{SPS}}}{\sigma_{\mathrm{eff}}}\ ,
\end{equation}
\noindent where $\sigma^{{q \bar q}_{1}{q \bar q}_{2}}_{\mathrm{DPS}}$ is the DPS cross-section of ${q \bar q}_{1}{q \bar q}_{2}$ production and $\sigma^{{q \bar q}_{1}}_{\mathrm{SPS}}$($\sigma^{{q \bar q}_{2}}_{\mathrm{SPS}}$)is the SPS cross-section of ${q \bar q}_{1}$(${q \bar q}_{2}$) production. Factor $m$ is set to $1$ for identical final-state particles \mbox{$({q \bar q}_1 = {q \bar q}_2)$} and $2$ for non-identical particles $({q \bar q}_1 \neq {q \bar q}_2)$. The effective cross-section is assumed to be universal across different processes and energy scales~\cite{dEnterria:2017yhd}. The effective cross-section plays a crucial role in multi-parton scattering (MPI) description~\cite{Belitsky:2005qn} and MC sample generation~\cite{Skands:2014pea,Alwall:2007st}.

New physics opportunities also emerge in multi-quarkonium study. Tetraquark states with fully heavy constituents which have not yet been observed experimentally~\cite{Yuan:2018inv}, such as $T_{c\bar{c}c\bar{c}}$, $T_{b\bar{b}b\bar{b}}$, are anticipated to be found in multi-quarkonium decay channels~\cite{Wu:2016vtq,Wang:2020mew,Celiberto:2024beg}. Such discoveries would provide novel insights into heavy flavor physics.

Several investigations into multi-quarkonium production during LHC Run-2 will be reviewed in this section, including cross-section measurement and the search for tetraquark states. In all studies discussed in this section, quarkonia are reconstructed via ${q \bar q}\to\mu^{+}\mu^{-}$ decay channel for its advantage of clearer signal and considerable branching ratio.

\subsection{Cross-Section Measurement}

\subsubsection{Di-Quarkonium Production Cross-Section Measurement in $pp$ Collisions} 

Several di-quarkonium production processes have been observed in $pp$ collisions at $13\rm\ TeV$ during the LHC Run-2 period, encompassing both well-studied channels and previously unexplored ones. The $J/\psi J/\psi$ production cross-section, which had been extensively measured before at lower energies~\cite{CMS:2014cmt, LHCb:2011kri, ATLAS:2016ydt}, has now been remeasured by the LHCb~\cite{LHCb:2023ybt} and ALICE collaborations~\cite{ALICE:2023lsn}. Particular attention has also been given to the cross-section measurement of previously scarce channels, like $\Upsilon(\mathrm{1S})\Upsilon(\mathrm{1S})$ measured by the CMS collaboration~\cite{CMS:2020qwa} and $J/\psi\Upsilon$ measured by the LHCb collaboration~\cite{LHCb:2023qgu}.

\begin{figure}[H]
	\includegraphics[width=0.90\textwidth]{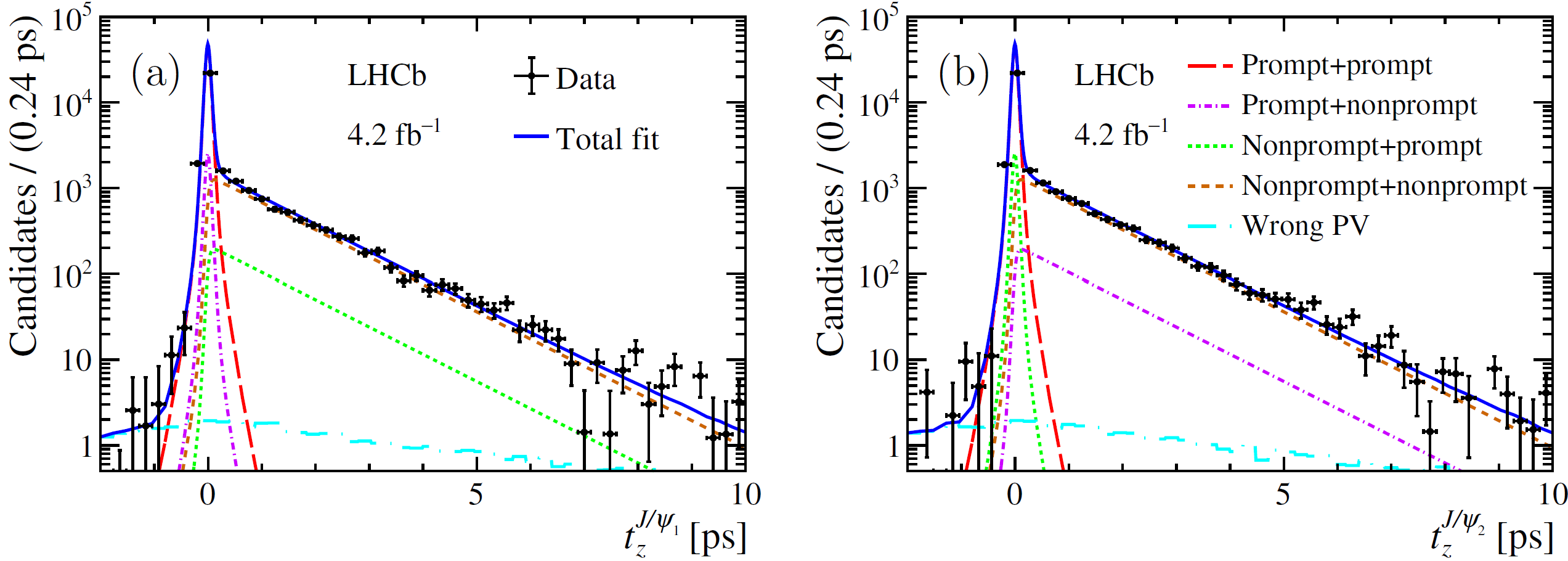}
	\caption{Projection of fit on the lifetime of OS muon pair dimensions with the LHCb $J/\psi J/\psi$ measurement. The black solid line represents the dataset, and the blue solid line represents the total fit~\cite{LHCb:2023ybt}.}
	\label{F LHCb cross-section lifetime}
\end{figure}

The strategies for variant studies followed similar approaches. Combinatorial background was considered the main background and was excluded through a 2D fit to the invariant mass distributions of opposite-sign (OS) muon pairs. The non-prompt background was also taken into account for $J/\psi$ production. A 2D fit to the lifetime ($\ell$) distributions of OS muon pairs has been applied to the combinatorial background subtracted data (using $sPlot$ method~\cite{Pivk:2004ty}) to determine the non-prompt fraction in the $J/\psi J/\psi$ study of the LHCb collaboration as illustrated by Figure~\ref{F LHCb cross-section lifetime}. However, the non-prompt fraction was only roughly estimated with the single-quarkonium measurement result in the $J/\psi J/\psi$ study of the ALICE collaboration and $J/\psi\Upsilon$ study of the LHCb collaboration.

The production cross-section was calculated after the signal yield was extracted from fits to the data:
\begin{equation} \label{Eq cross-section}
    \sigma(pp\to {q \bar q}_{1}{q \bar q}_{2})=\frac{N_{{q \bar q}_{1}{q \bar q}_{2}}}{\mathcal{L}\times \mathrm{Br}({q \bar q}_{1}\to\mu^{+}\mu^{-})\times \mathrm{Br}({q \bar q}_{2}\to\mu^{+}\mu^{-})}\ ,
\end{equation}
\noindent where $\sigma(pp\to {q \bar q}_{1}{q \bar q}_{2})$ is the production cross-section of di-quarkonium ${q \bar q}_{1}{q \bar q}_{2}$, $N_{{q \bar q}_{1}{q \bar q}_{2}}$ is the acceptance efficiency corrected signal yield, $\mathcal{L}$ is the integrated luminosity, and $\mathrm{Br}({q \bar q}_{1(2)}\to\mu^{+}\mu^{-})$ is the branching ratio of the ${q \bar q}_{1(2)}\to\mu^{+}\mu^{-}$ process. The total cross-sections of different channels and corresponding fiducial regions are summarized in Table~\ref{T cross-section}. The differential cross-section was also measured in various kinematic intervals in a similar way (not available in the $J/\psi J/\psi$ study of the ALICE collaboration).

\begin{figure}[H]
	\includegraphics[width=0.90\textwidth]{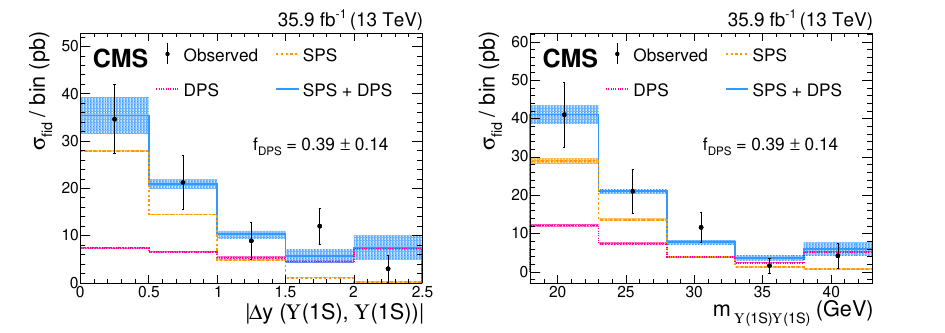}
	\caption{Differential cross-section distribution of $\Delta y$ (\textbf{left}) and $m\left(\Upsilon(\mathrm{1S})\Upsilon(\mathrm{1S})\right)$ (\textbf{right}) dimensions (black dots) with the CMS $\Upsilon(\mathrm{1S})\Upsilon(\mathrm{1S})$ measurement. SPS and DPS distributions are also represented by the orange and magenta dashed lines. Fits were applied on these two dimensions separately to extract the DPS contribution~\cite{CMS:2020qwa}.}
	\label{F CMS YY cross-section DPS}
\end{figure}

SPS and DPS contributions can be separated through their distinct kinematic distributions, although different approaches were employed across various studies. For the $J/\psi J/\psi$ study of the LHCb collaboration, the DPS contribution was extracted in the DPS dominant region ($1.8<\Delta y<2.5$) and then extrapolated to the full region using the DPS $\Delta y$ distribution acquired from a data-driven event-mixing DPS sample. The DPS contributions of other dimensions were obtained with the DPS weight acquired in the $\Delta y$ dimension. A similar strategy was adopted by the CMS collaboration in their $\Upsilon(\mathrm{1S})\Upsilon(\mathrm{1S})$ study but utilizing MC samples instead of the data driven method. Both SPS and DPS distributions can be acquired from MC samples, and fits were applied in the full phase space. The DPS fraction was estimated separately from two dimensions as $(39\pm14)\%$ when using $\Delta y$ and $(27\pm22)\%$ when using $m\left(\Upsilon(\mathrm{1S})\Upsilon(\mathrm{1S})\right)$ as illustrated by Figure~\ref{F CMS YY cross-section DPS}. The DPS contribution was estimated by subtracting the theoretically calculated SPS cross-section from the total cross-section in the $J/\psi\Upsilon$ study of the LHCb collaboration. Meanwhile, the ALICE collaboration assumed that the DPS process has 100\% contribution in their $J/\psi J/\psi$ study, given the limited statistics that prevented differential cross-section measurements in their study.

The effective cross-sections can be calculated from the DPS cross-section according to Equation~(\ref{Eq sigma eff}), and the results are summarized in Table~\ref{T cross-section}. 
Unfortunately, no suitable $\mathrm{SPS}\to\Upsilon(\mathrm{1S})$ production cross-section measurement is currently available for the $\Upsilon(\mathrm{1S})\Upsilon(\mathrm{1S})$ study of the CMS collaboration. A rough estimation can be obtained by extrapolating the $\Upsilon(\mathrm{1S})$ measurement results of either the CMS collaboration~\cite{Chatrchyan:2017xx} or LHCb collaboration~\cite{Lansberg:2019adr,LHCbCEPCharmonium13TeV2018} as $\sim4\rm\ mb$ or $\sim3\rm\ mb$.
\vspace{-6pt}

\begin{table}[H]
	\caption{Summary of total production cross-section ($\sigma(pp\to {q \bar q}_{1}{q \bar q}_{2})$) and effective cross-section ($\sigma_{\mathrm{eff}}$) measurement results of various di-quarkonium studies. $y$ and $p_{T}$ (in $\rm GeV/c$) represent rapidity and transverse momentum of quarkonium. The uncertainties are statistical followed by systematic. The third uncertainty of the total cross-section is from the branching ratio, and the third uncertainty of the effective cross-section is from the theory calculation. The di-quarkonium measurement results from the LHC at lower energy are also listed for reference. The estimation of $\sigma_{\mathrm{eff}}$ using the result of LHCb at $7\rm\ TeV$ can be difficult~\cite{Lansberg:2014swa}.}
	\label{T cross-section}
		\setlength{\tabcolsep}{2pt}
		\small
        \renewcommand{\arraystretch}{1.5}

		\begin{tabularx}{\fulllength}{C C C C C }
		  \toprule
            \textbf{Collaboration} & \textbf{Channel(}\boldmath{${q \bar q}_{1}{q \bar q}_{2}$}\textbf{)} & \textbf{Fiducial Region} & \boldmath{$\sigma(pp\to {q \bar q}_{1}{q \bar q}_{2})$} & \boldmath{$\sigma_{\mathrm{eff}}$} \textbf{[mb]} \\
            \midrule
            \multirow{3}{*}{\begin{tabular}{@{}c@{}}LHCb \\ (13 TeV)\end{tabular}} & $J/\psi J/\psi$~\cite{LHCb:2023ybt} & $p_{T}<14$, $2.0<y<4.5$ & $16.36\pm0.28\pm0.88\rm\ nb$ & $13.1\pm1.8\pm2.3$ \\
            \cline{3-3}
            & $J/\psi\Upsilon(\mathrm{1S})$~\cite{LHCb:2023qgu} & $p_{T}(J/\psi/\Upsilon)<10/30$,  & $133\pm22\pm7\pm3\rm\ pb$ & $26\pm5\pm2^{+22}_{-3}$\\
            & $J/\psi\Upsilon(\mathrm{2S})$~\cite{LHCb:2023qgu} & $2.0<y<4.5$ & $76\pm4\pm7\rm\ pb$ & $14\pm5\pm1^{+7}_{-1}$\\
            \cline{3-3}
            
            ALICE & \multirow{2}{*}{$J/\psi J/\psi$~\cite{ALICE:2023lsn}} & \multirow{2}{*}{$2.5<y<4.0$}  & \multirow{2}{*}{$7.3\pm1.7^{+1.9}_{-2.1}\rm\ nb$} & \multirow{2}{*}{$6.7\pm1.6\pm2.7$} \\
            (13 TeV) & & & & \\
            
            CMS  & \multirow{2}{*}{$\Upsilon(\mathrm{1S})\Upsilon(\mathrm{1S})$~\cite{CMS:2020qwa}} & \multirow{2}{*}{$y<2.0$} & \multirow{2}{*}{$79\pm11\pm6\pm3\rm\ pb$} & $\sim$3--4\\
            (13 TeV) & & & & ~\cite{Chatrchyan:2017xx,Lansberg:2019adr,LHCbCEPCharmonium13TeV2018} \\
            
            LHCb & \multirow{2}{*}{$J/\psi J/\psi$~\cite{LHCb:2011kri}} & \multirow{2}{*}{$p_{T}<10$, $2.0<y<4.5$} & \multirow{2}{*}{$5.1\pm1.0\pm1.1\rm\ nb$} & \multirow{2}{*}{-} \\
            (7 TeV) & & & &\\
            
            \cline{3-3}
            \multirow{3}{*}{\begin{tabular}{@{}c@{}}CMS \\ (7 TeV)\end{tabular}} & \multirow{3}{*}{$J/\psi J/\psi$~\cite{CMS:2014cmt}} & $p_{T}>6.5$, $|y|<1.2$  & \multirow{3}{*}{$1.49\pm0.07\pm0.13\rm\ nb$}& 
            \multirow{3}{*}{\begin{tabular}{@{}c@{}}$8.0\pm2.0\pm2.9$ \\ ~\cite{Lansberg:2019adr}\end{tabular}}\\
             & & $4.5<p_{T}<6.5$, $1.2<|y|<1.43$ & & \\
             & & $p_{T}>4.5$, $1.43<|y|<2.2$ & & \\
            \cline{3-3}
            
            ATLAS & \multirow{2}{*}{$J/\psi J/\psi$~\cite{ATLAS:2016ydt}} & $p_{T}>8.5$, $|y|<1.05$ & $15.6\pm1.3\pm1.2\rm\ pb$ & \multirow{2}{*}{$6.3\pm1.6\pm1.0$}\\
            (8 TeV) & & $p_{T}>8.5$, $1.05<|y|<2.1$ & $13.5\pm1.3\pm1.1\rm\ pb$& \\
		  \bottomrule
		\end{tabularx}
		
\end{table}

\begin{figure}[H]
	\includegraphics[width=0.45\textwidth]{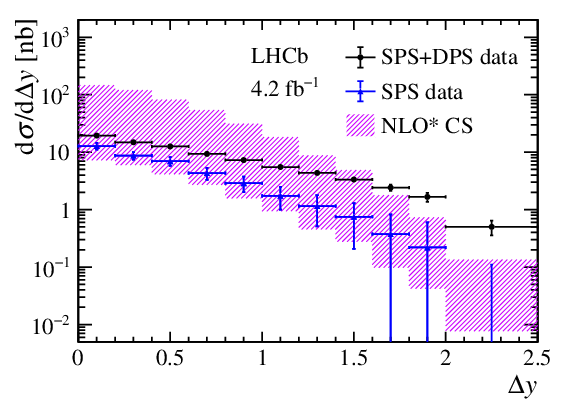}
     \includegraphics[width=0.45\textwidth]{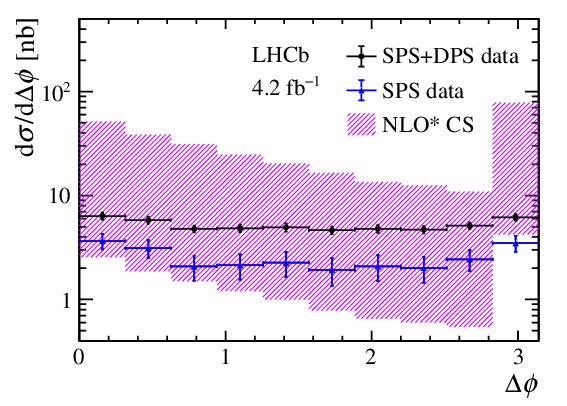}
	\caption{Differential cross-section (black dots), SPS cross-section (blue dots) with the LHCb $J/\psi J/\psi$ measurement, and NLO* CS prediction (purple bands) in $\Delta y$ (left) and $\Delta\phi$ (right) dimensions~\cite{LHCb:2023ybt}. }
	\label{F LHCb cross-section comparison}
\end{figure}

In the $J/\psi J/\psi$ study by the LHCb collaboration, the kinematic distributions of the SPS process were obtained after the DPS contribution was subtracted. These data were then compared to predictions from the next-to-leading-order color-singlet (NLO* CS) model. Two representative examples are illustrated here in the $\Delta y$ dimension and $\Delta\phi$ dimension (separation in the azimuthal angle between two $J/\psi$s) as shown in Figure~\ref{F LHCb cross-section comparison}. From the comparison, the estimated SPS cross-section distributions are generally consistent with the theory predictions, given the sizable  theoretical uncertainties. However, a discrepancy in the trend is observed for some kinematic variables, such as $\Delta\phi$.

A comprehensive methodology has been established by the LHCb's study of $J/\psi J/\psi$ production, incorporating background subtraction, SPS/DPS separation, and cross-section calculation. This approach demonstrated significant improvements over previous measurements~\cite{CMS:2014cmt, LHCb:2011kri, ATLAS:2016ydt}. Although the general framework was similar to the measurement made by the ATLAS collaboration at $8\rm\ TeV$~\cite{ATLAS:2016ydt}, the LHCb study benefited from substantially greater statistics, enabling more differential cross-section measurements and smaller uncertainties. A general consistency with theory prediction in differential cross-section distribution can be noticed. However, two key limitations should not be ignored. First, all the $J/\psi J/\psi$ production cross-section measurements at $13\rm\ TeV$ were restricted to the forward fiducial region. Complementary measurement in the central fiducial region is urgently needed since the differential cross-section distribution at the high $p_{T}$ region is expected to be particularly discriminating for testing QCD models~\cite{Hu:2017pat}. Second, neither cross-section measurement accounts for potential contributions from resonant states. This is a significant omission because several resonances that decay into the $J/\psi J/\psi$ final state and could therefore contribute considerably to the production cross-section have recently been discovered~\cite{LHCb:2020bwg,CMS:2023owd,CMS:2025xwt,ATLAS:2023bft}.

The substantial increase in statistics during the LHC Run-2 period has made precise measurements of production cross-sections for di-quarkonium states involving bottomonia feasible. Bottomonium production measurements are of particular importance for NRQCD study, as the $b$ quark is much heavier than the $c$ quark, and a system composed of $b$ quarks adheres closely to the non-relativistic approximation. Another advantage of bottomonium production study is that non-prompt background is absent in this case. Breakthroughs include the LHCb's observation of $J/\psi\Upsilon(\mathrm{1S})$ production in $pp$ collisions, achieving a significance of 7.9 standard deviations, alongside the first measurement of $\Upsilon(\mathrm{1S})\Upsilon(\mathrm{1S})$ production cross-section conducted by the CMS collaboration. Nevertheless, accumulating additional data remains imperative to enable more precise calculation, as well as the measurement of pair production of heavier $\Upsilon$ states.

Another critical area requiring advancement lies in improving the precision of theoretical calculations. Theory calculations are of great significance for experiment measurements, particularly for SPS/DPS separation~\cite{CMS:2020qwa}, theory--experiment comparison~\cite{LHCb:2023ybt}, effective cross-section calculation~\cite{LHCb:2023qgu}, etc. Currently, the limited accuracy of theoretical predictions represents a significant bottleneck across these applications, particularly notable in the studies discussed in this review. 


\subsubsection{$J/\psi J/\psi$ Production Cross-Section Measurement in $p\mathrm{Pb}$ Collisions} 

Another investigation into $J/\psi J/\psi$ production during the LHC Run-2 period was performed by the CMS collaboration, focusing on $p\mathrm{Pb}$ collisions~\cite{CMS:2024wgu} at a center-of-mass energy of $8.16\rm\ TeV$ with an integrated luminosity of 174.6$\ \mathrm{nb}^{-1}$. The measurement was conducted within the fiducial region defined by $p_{T}(J/\psi)>6.5\rm\ GeV/c$ and $\left|y(J/\psi)\right|<2.4$.

\begin{figure}[H]
	\includegraphics[width=0.90\textwidth]{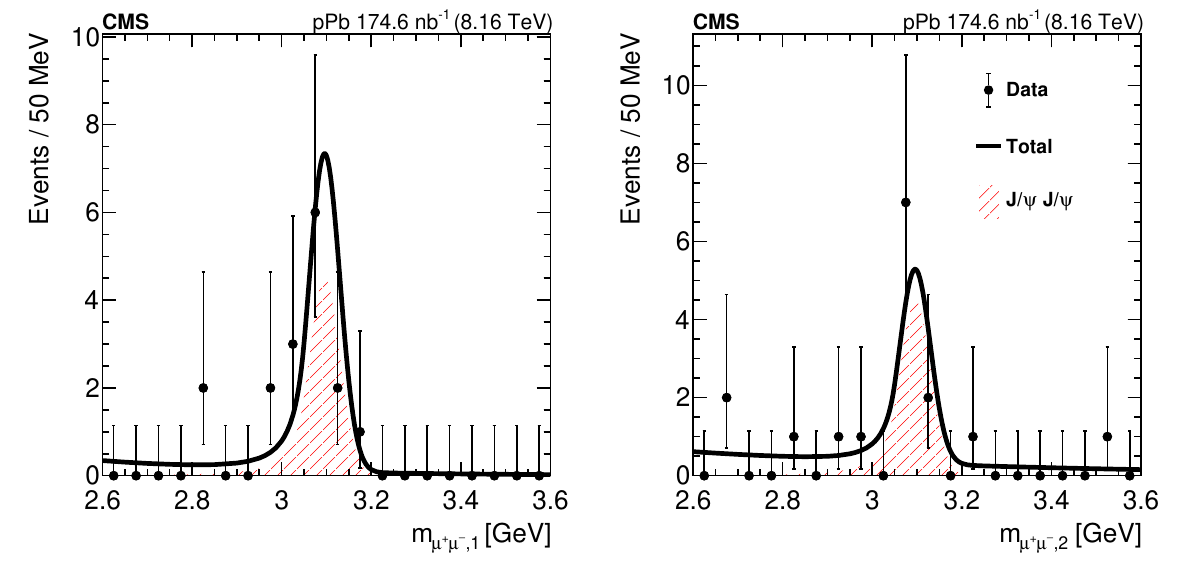}
	\caption{Projection 
 of fit on the invariant mass of OS muon pair dimensions with the CMS $p\mathrm{Pb}\to J/\psi J/\psi$ measurement. The black solid line represents the total fit, and the red area represents the signal yield~\cite{CMS:2024wgu}.}
	\label{F CMS JJ cross-section fit}
\end{figure}

Sixteen candidates were identified after event selection. All candidates exhibited vertices close to the primary vertex, implying their promptly produced origin. A 2D fit to the invariant mass of OS muon pairs was applied to exclude the combinatorial background as illustrated by Figure~\ref{F CMS JJ cross-section fit}. In total, $8.5\pm3.4$ signal events were extracted from the fit, from which the total production cross-section was calculated to be
\begin{equation}
    \sigma(p\mathrm{Pb}\to J/\psi J/\psi)=22.0\pm8.9(\mathrm{stat.})\pm1.5(\mathrm{syst.})\rm\ nb\ .
\end{equation}
\vspace{-18pt}

\begin{figure}[H]
	\includegraphics[width=0.40\textwidth]{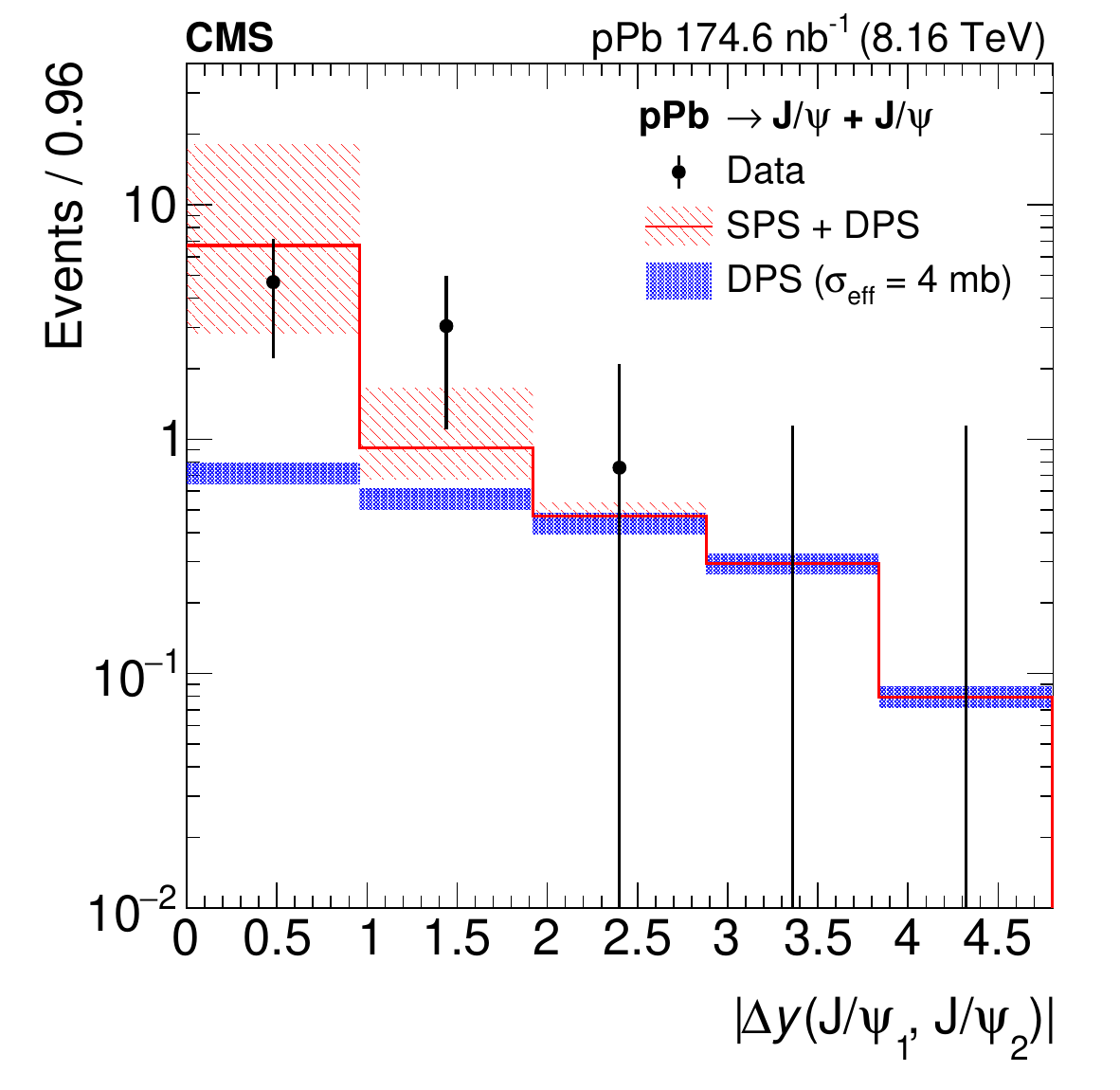}
	\caption{Event 
 distribution of $\Delta y$ (black dots), fitted DPS distribution (blue bands) and fitted total distribution (red bands) with the CMS $p\mathrm{Pb}\to J/\psi J/\psi$ measurement~\cite{CMS:2024wgu}.}
	\label{F CMS JJ cross-section DPS}
\end{figure}

Despite the quite limited statistics, DPS and SPS contributions were managed to be separated. Event distribution on the $\Delta y$ dimension was investigated, and the DPS contribution was extracted with a fit procedure, where the DPS fit template was acquired using a data-driven method (similar to the $pp\to J/\psi J/\psi$ measurement of the LHCb collaboration~\cite{LHCb:2023ybt}). The fit is presented in Figure~\ref{F CMS JJ cross-section DPS}. The effective cross-section for $p\mathrm{Pb}$ collisions was calculated according to Equation~(\ref{Eq sigma eff}) as
\begin{equation}
    \sigma^{p\mathrm{Pb}}_{\mathrm{eff}}=0.53^{+\infty}_{-0.2}\rm\ b\ ,
\end{equation}
\noindent where $\sigma^{p\mathrm{Pb}}_{\mathrm{SPS}\to J/\psi J/\psi}$ was estimated through a theory calculation. The effective cross-section for $pp$ collisions is available with a relationship:
\begin{equation}
    \sigma^{p\mathrm{Pb}}_{\mathrm{eff}}=\frac{A\sigma^{pp}_{\mathrm{eff}}}{1+\sigma^{pp}_{\mathrm{eff}}F_{p\mathrm{Pb}}/A}\ ,
\end{equation}
\noindent where $A=208$ is the $p\mathrm{Pb}$ mass number and factor $F_{p\mathrm{Pb}}\approx A^{1/3}/14\pi\rm\ mb^{-1}$~\cite{CMS:2024wgu}. $\sigma^{pp}_{\mathrm{eff}}$ was estimated to be
\begin{equation}
    \sigma^{pp}_{\mathrm{eff}}=0.53^{+\infty}_{-1.5}\rm\ mb\ ,
\end{equation}
\noindent and this estimation could be translated into a lower limit with 95\% confidence level as
\begin{equation}
    \sigma^{pp}_{\mathrm{eff}}>1.0\rm\ mb\ .
\end{equation}

The significance of the signal reaches 5.3 standard deviations, representing the first observation of the $J/\psi J/\psi$ candidate in $p\mathrm{Pb}$ collisions. Increased statistics are expected to significantly improve both the signal extraction and effective cross-section calculation. A comparison between this study and a $pp\to J/\psi J/\psi$ measurement at a similar collision energy (which is absent for now) may provide valuable information about nuclear structure and quarkonium production.

\subsection{Tetraquark States Search}

Investigation into potential tetraquark states decaying to multi-quarkonium final states during the LHC Run-2 and Run-3 periods focused on the $J/\psi J/\psi$ channel ($T_{c\bar{c}c\bar{c}}$). The LHCb~\cite{LHCb:2020bwg}, CMS~\cite{CMS:2023owd,CMS:2025xwt,CMS:2025fpt}, and ATLAS~\cite{ATLAS:2023bft} collaborations successively pursued this investigation using the similar strategy. The ATLAS and CMS collaborations also extended their searches to include the $J/\psi \psi(\mathrm{2S})$ channel~\cite{ATLAS:2023bft,CMS:2025vnq} for possible resonances.

\begin{figure}[H]
	\includegraphics[width=0.50\textwidth]{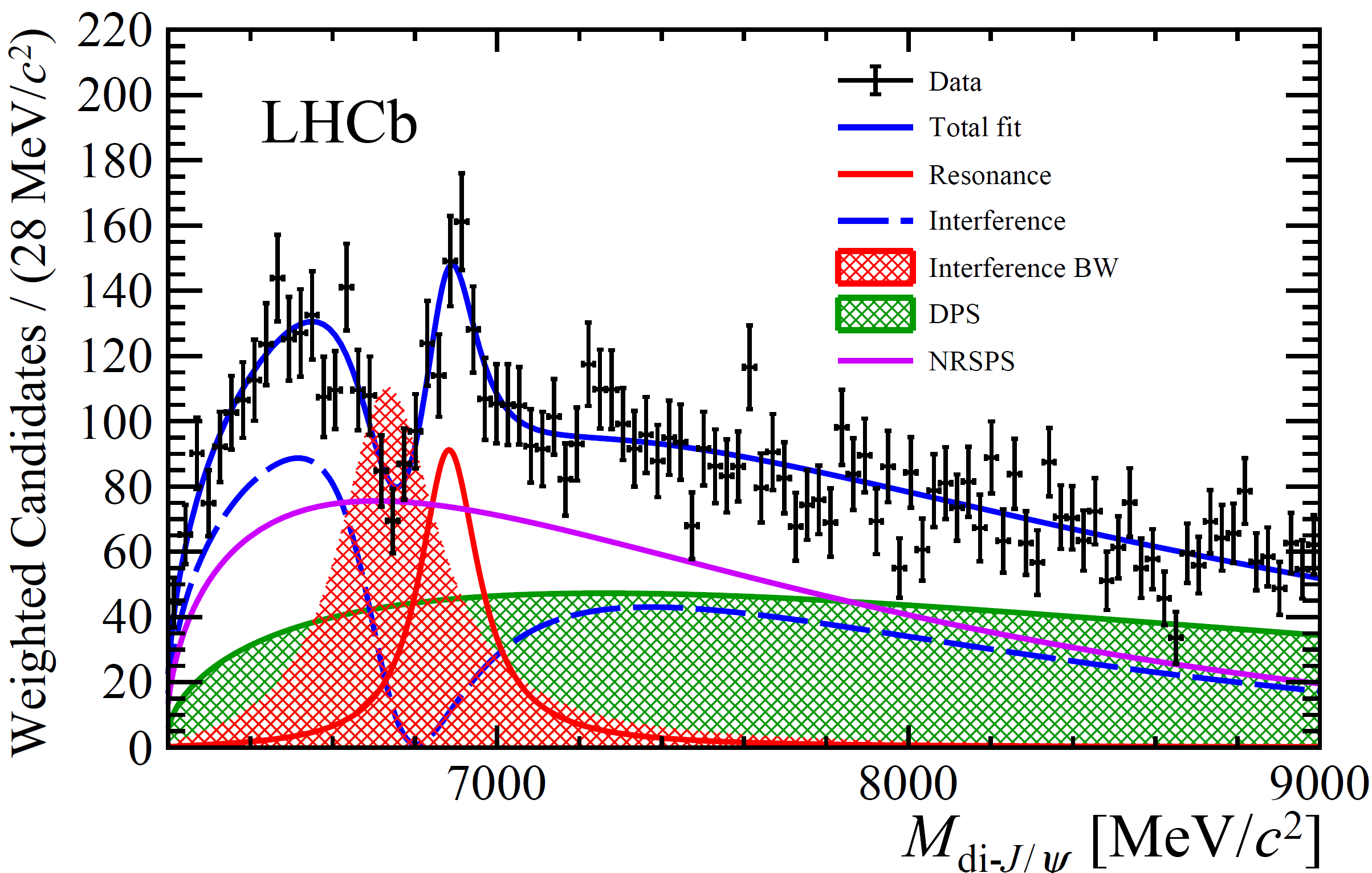}
	\caption{Event distribution of $m\left(J/\psi J/\psi\right)$ of the LHCb collaboration and fit with interference~\cite{LHCb:2020bwg}.}
	\label{F LHCb resonance}
\end{figure}

\begin{figure}[H]
	\includegraphics[width=0.50\textwidth]{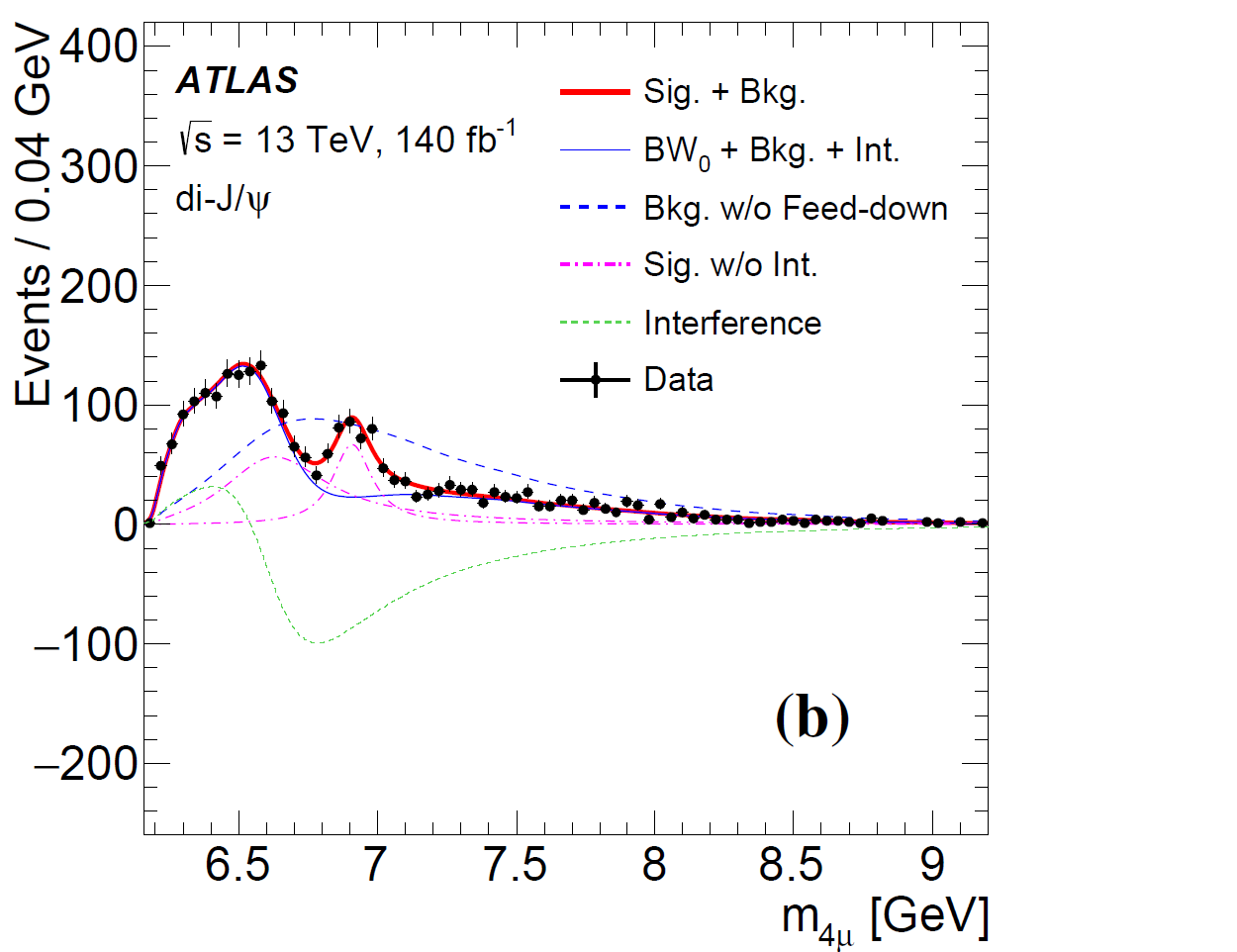}
     \includegraphics[width=0.40\textwidth]{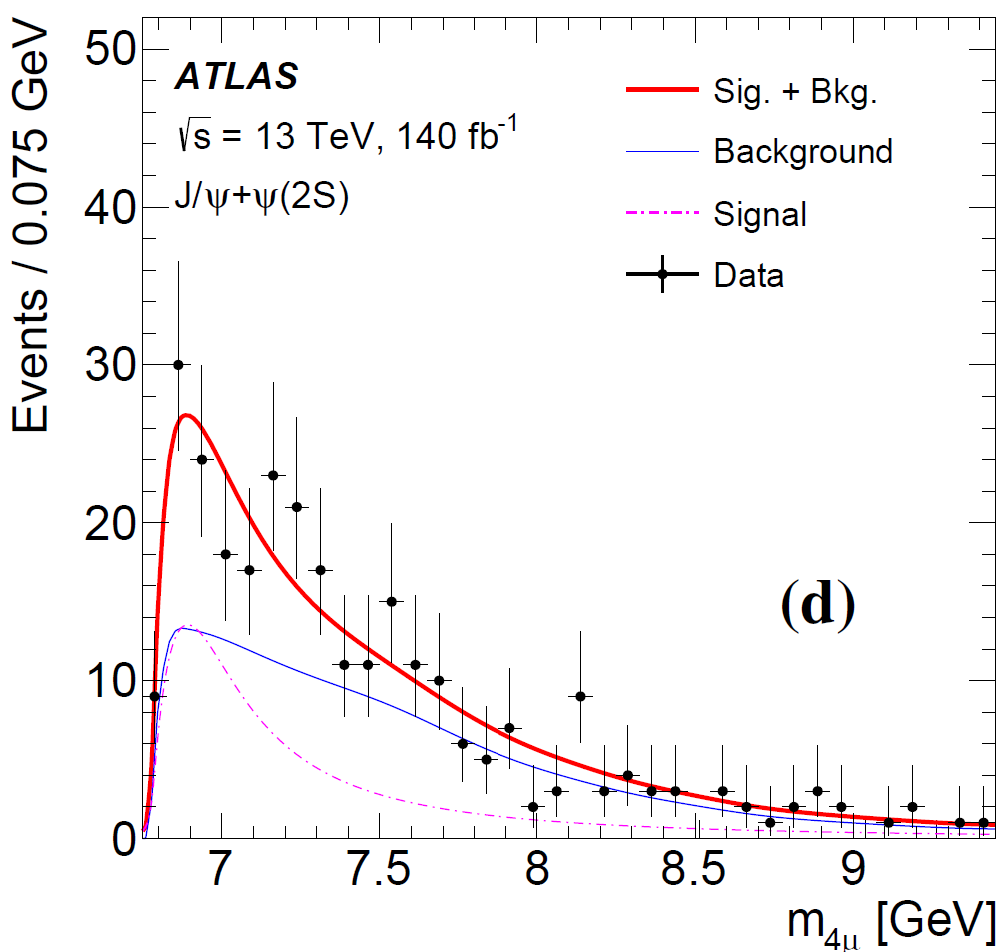}
	\caption{Event 
 distribution of $m\left(J/\psi J/\psi\right)$ (\textbf{left}) and fit with interference (between the first resonance and SPS) and $m\left(J/\psi \psi(\mathrm{2S})\right)$ (\textbf{right}) and fit with one resonance assumption of the ATLAS collaboration~\cite{ATLAS:2023bft}.}
	\label{F ATLAS resonance}
\end{figure}

\begin{figure}[H]
	\includegraphics[width=0.46\textwidth]{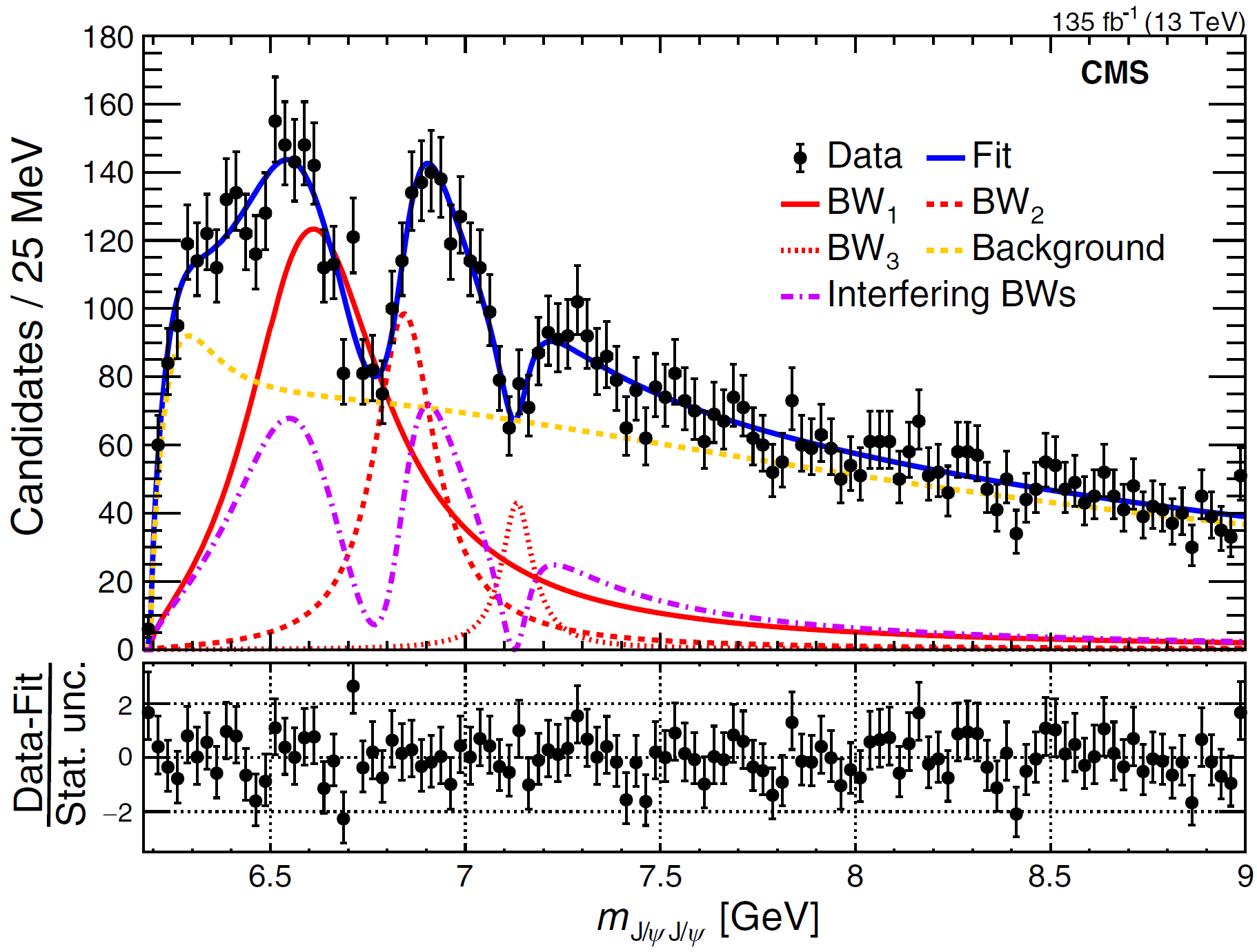} 
        \includegraphics[width=0.50\textwidth]{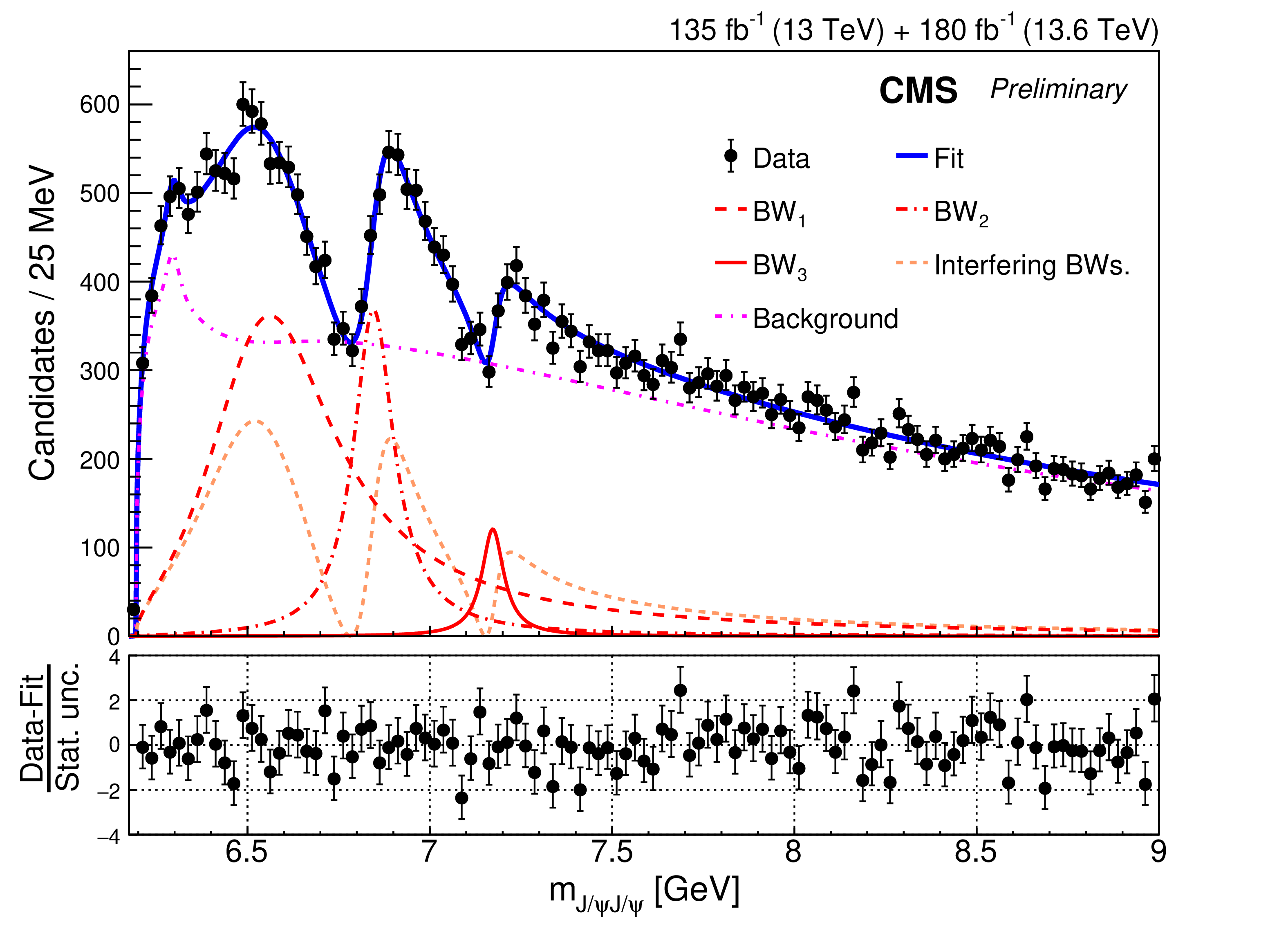}
	\caption{Event 
 distribution of $m\left(J/\psi J/\psi\right)$ of the CMS collaboration and fit with interference.\linebreak \textbf{Left}: CMS Run-2 data~\cite{CMS:2023owd}; \textbf{Right}: CMS Run-2 and Run-3 data~\cite{CMS:2025xwt}.}
	\label{F CMS resonance}
\end{figure}

\begin{figure}[H]
	\includegraphics[width=0.50\textwidth]{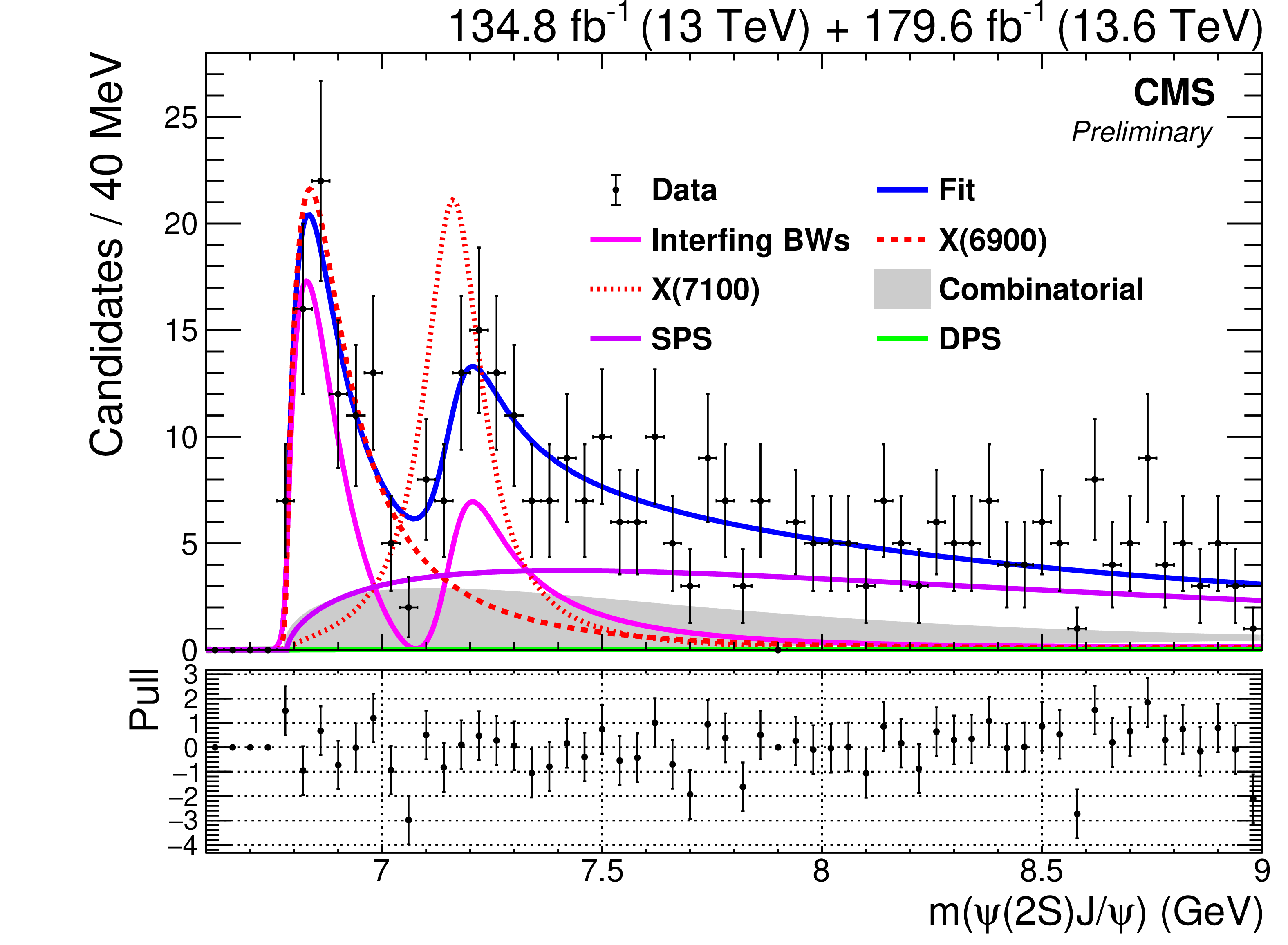}
	\caption{Event 
 distribution of $J/\psi \psi(\mathrm{2S})$ of the CMS collaboration and fit with interference~\cite{CMS:2025vnq}.}
	\label{BPH-22-004-Figure002}
\end{figure}

All collaborations employed a similar strategy: identifying $J/\psi J/\psi$ and $J/\psi \psi(2\mathrm{S})$ candidates and examining their invariant mass distributions. Clear excesses were observed by all three experiments, though features varied due to differences in detector resolution, trigger, and integrated luminosity. In the $J/\psi J/\psi$ channel, both LHCb and ATLAS observed the $X(6900)$ resonance alongside a threshold enhancement (Figure~\ref{F LHCb resonance} for LHCb; Figure
~\ref{F ATLAS resonance}, left, for ATLAS). CMS, however, reported a more complex structure featuring a threshold enhancement and three distinct resonances, $X(6600)$, $X(6900)$, and $X(7100)$, as shown in Figure~\ref{F CMS resonance}). For the $J/\psi \psi(2\mathrm{S})$ channel, ATLAS and CMS both detected two excesses (Figure~\ref{F ATLAS resonance}, right, for ATLAS; Figure~\ref{BPH-22-004-Figure002} for CMS), with CMS establishing $X(6900)$ at the $5\sigma$ discovery level.

\vspace{-6pt}

\vspace{-6pt}

\vspace{-6pt}

The mass and width of the resonances were determined through fits to the
$m\left(J/\psi J/\psi\left(\psi(\mathrm{2S})\right)\right)$ dimension. SPS, DPS, combinatorial backgrounds, and feed-down from heavier charmonium states were considered the principal backgrounds, and interference was considered by all three collaborations. Discrepancies in interference strategy can be noticed among different collaborations. The LHCb collaboration incorporated interference between the threshold enhancement and SPS, while the CMS collaboration assumed the interference happened among three candidate resonances. The ATLAS collaboration has explored multiple interference scenarios. For the $J/\psi J/\psi$ channel, interference among the threshold and two resonances as well as between the first resonance and SPS was attempted, while no interference was considered for the $J/\psi \psi(\mathrm{2S})$ channel.

Three collaborations reported results using different fitting strategies. Table~\ref{T resonance} summarizes the masses, widths, and significances of the $X(6600)$, $X(6900)$, and $X(7100)$ resonances from these studies. The consistency in the reported properties across all collaborations provides robust evidence for these potential resonances. 

\begin{table}[H]
	\caption{Summary of the masses, widths, and significances for $T_{c\bar{c}c\bar{c}}$s reported by the LHCb~\cite{LHCb:2020bwg}, \mbox{CMS~\cite{CMS:2025xwt,CMS:2025vnq}} and ATLAS collaborations~\cite{ATLAS:2023bft}. The uncertainties are statistical followed by systematic. }
	\label{T resonance}
		\setlength{\tabcolsep}{2pt}
		\small
        \renewcommand{\arraystretch}{1.5}

		\begin{tabularx}{\fulllength}{C C C C C C C}
		  \toprule
            \multicolumn{2}{c}{\multirow{2.25}{*}{\boldmath{$T_{c\bar{c}c\bar{c}}$}}} & \textbf{LHCb} & \multicolumn{2}{c}{\textbf{CMS}} & \multicolumn{2}{c}{\textbf{ATLAS}}\\
            \cmidrule{3-7}
            \multicolumn{2}{c}{} & \boldmath{$J/\psi J/\psi$} & \boldmath{$J/\psi J/\psi$} & \boldmath{$J/\psi \psi(\mathrm{2S})$} & \boldmath{$J/\psi J/\psi$} & \boldmath{$J/\psi \psi(\mathrm{2S})$}\\
		  \midrule 
           \multirow{3}{*}{$X(6600)$} & $m(\rm MeV)$ & \multirow{3}{*}{-} & $6593^{+15}_{-14}\pm25$ & \multirow{3}{*}{-} & \multirow{3}{*}{-} & \multirow{3}{*}{-} \\
            & $\Sigma(\rm MeV)$ &  & $446^{+66}_{-54}\pm87$ & & & \\
            & Significance & & $15.2 \sigma$ & & & \\
            \midrule 
            \multirow{3}{*}{$X(6900)$} & $m(\rm MeV)$ & $6886\pm11\pm11$ & $6847^{+10}_{-10}\pm15$ & $6876^{+46+110}_{-29-110}$ & $6910\pm10\pm10$ & $6960\pm50\pm30$ \\
            & $\Sigma(\rm MeV)$ & $168\pm33\pm69$ & $135^{+16}_{-14}\pm14$ & $253^{+290+120}_{-100-120}$ & $150\pm30\pm10$ & $510\pm170^{+110}_{-100}$\\
            & Significance & $5.1 \sigma$ & $16.7 \sigma$ & $7.9 \sigma$ & $>$5.0$\sigma$ & $4.3 \sigma$ \\
            \midrule 
            \multirow{3}{*}{$X(7100)$} & $m(\rm MeV)$ &\multirow{3}{*}{-} & $7173^{+9}_{-10}\pm13$ & $7169^{+26+74}_{-52-70}$ & \multirow{3}{*}{-} & $7220\pm30^{+10}_{-40}$\\
            & $\Sigma(\rm MeV)$ & & $73^{+18}_{-15}\pm10$ & $154^{+110+140}_{-82-160}$ & & $90\pm60^{+60}_{-50}$\\
            & Significance & & $7.7 \sigma$ & $4.0 \sigma$ & & $3.0 \sigma$ \\
		  \bottomrule
		\end{tabularx}
		
\end{table}

These observed resonances in the $J/\psi J/\psi\left(\psi(\mathrm{2S})\right)$ decay channels represent the first experimental discovery of a family of all-charm tetraquarks $T_{c\bar{c}c\bar{c}}$~\cite{PhysRevD.109.054034,sym12111869,ZHU2021115393,PhysRevLett.126.132001,RevModPhys.90.015004}. Increased statistics will be essential to improve the description of the data and the characterization of these states~\cite{LIU20242802,CPC.41.111201}. 

\begin{figure}[H]
	\includegraphics[width=0.65\textwidth]{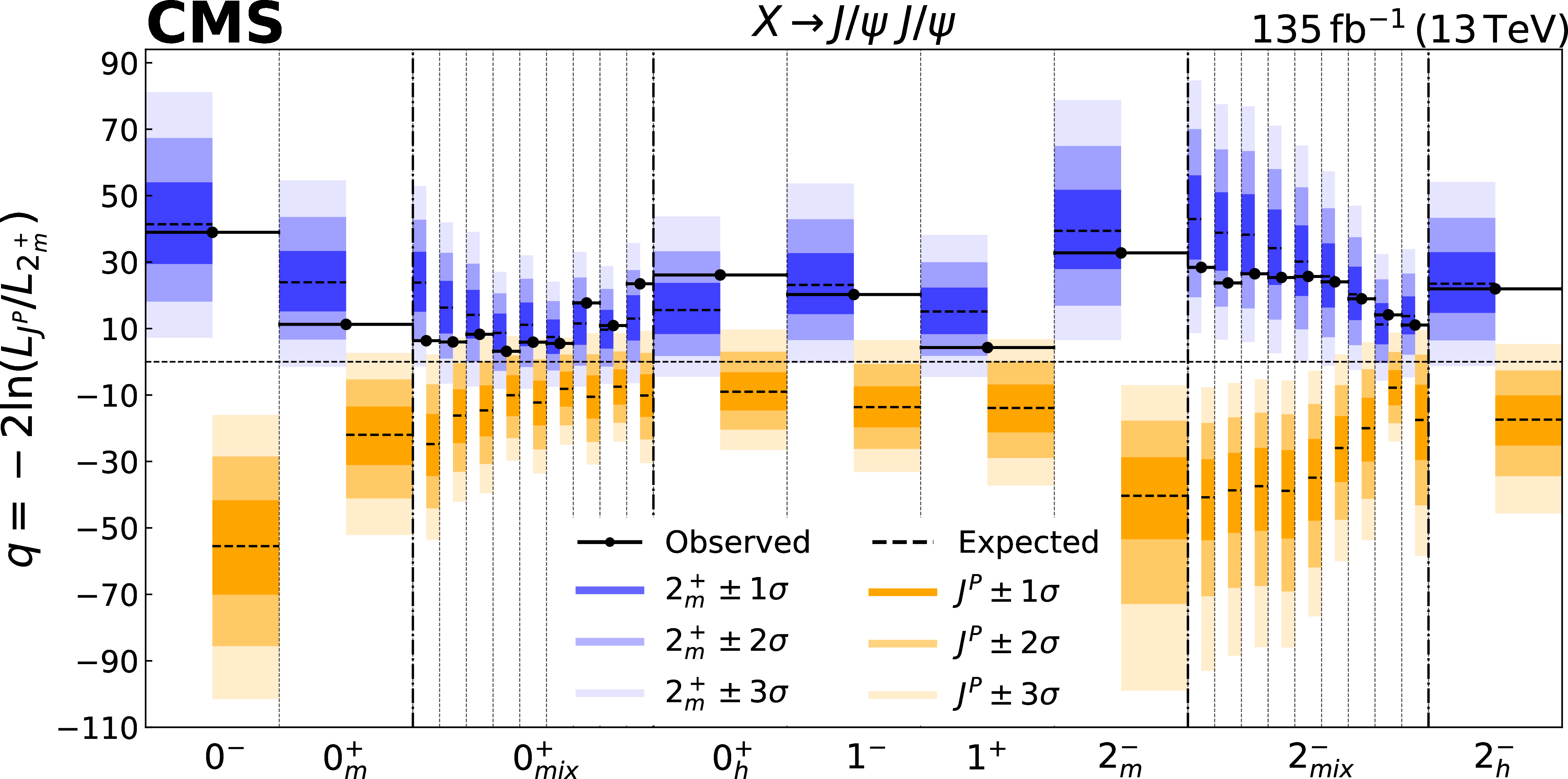}
	\caption{Statistical tests of various $J^{P}$ hypotheses against the $2^{+}$ model from the CMS experiment~\cite{CMS:2025fpt}.}
	\label{BPH-24-002-Figure004}
\end{figure}

Additionally, the measurement of the quantum numbers of these newly discovered all-charm tetraquarks is particularly crucial to testify the current interference assumptions and help constrain theoretical models of the tetraquark structure. The quantum numbers measured by the CMS Collaboration confirm positive parity ($P = +1$) and positive charge conjugation ($C = +1$) for these states. The spin $J$ is determined to be $2\hbar$, while values of $0\hbar$ and $1\hbar$ are excluded at the 95\% and 99\% confidence levels, respectively, as shown in Figure~\ref{BPH-24-002-Figure004}. This $J^{PC} = 2^{++}$ assignment implies the possible internal configurations of these tetraquarks and favors the tightly bound scenario~\cite{CMS:2025fpt}. 

\subsection{Triple $J/\psi$ Candidate Search}

Another experimental effort focused on searching for triple produced $J/\psi$ candidates in $pp$ collisions at $13\rm\ TeV$. This final state offers a unique opportunity to observe the TPS process. This research is also of particular value for the measurement of the effective cross-section since the SPS contribution to this final state is expected to be highly \mbox{suppressed~\cite{Shao:2012iz,Shao:2015vga,PhysRevLett.122.192002,dEnterria:2016ids}}, allowing the DPS contribution to be more clearly isolated.

\begin{figure}[H]
	\includegraphics[width=0.30\textwidth]{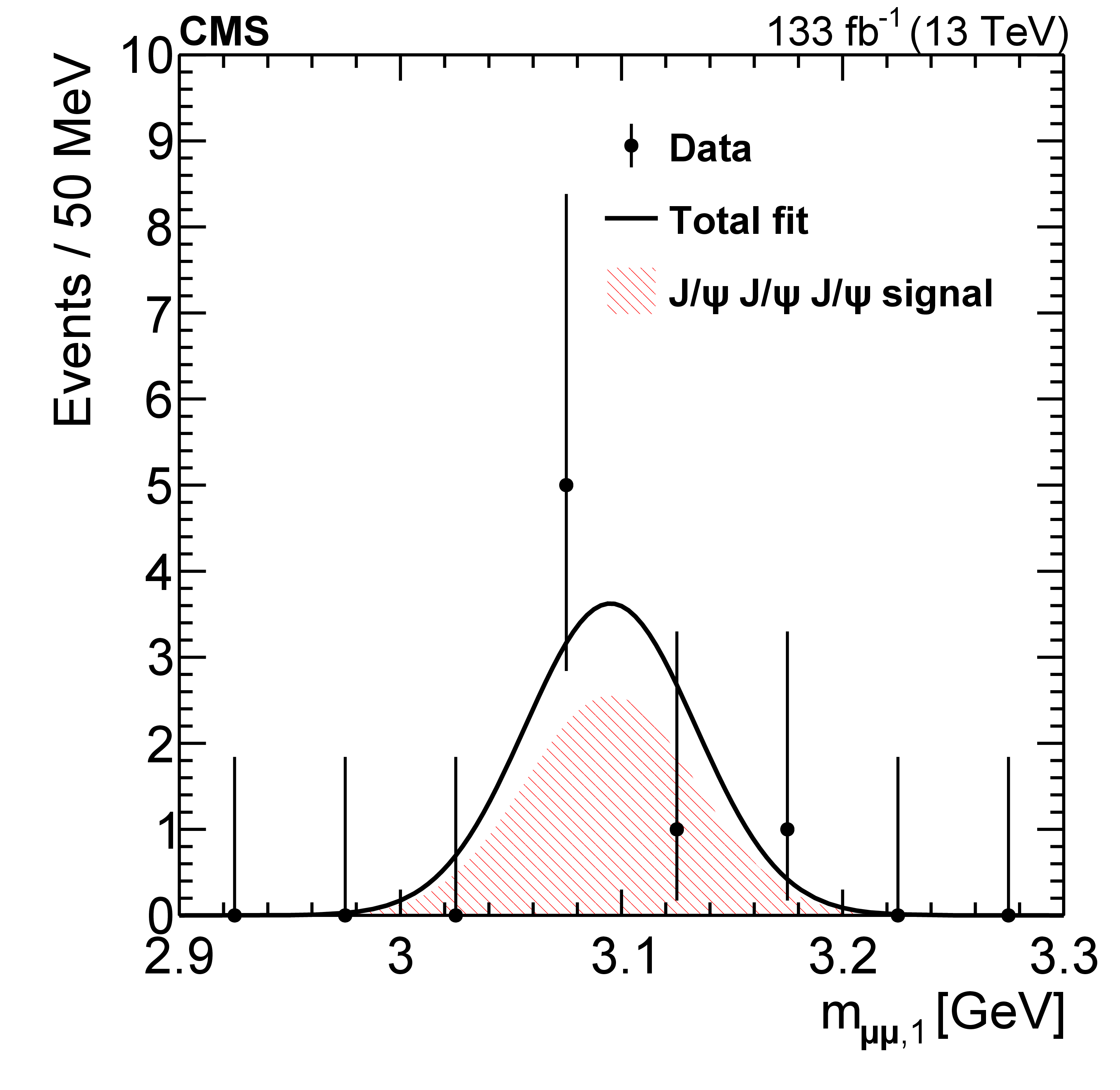}
     \includegraphics[width=0.30\textwidth]{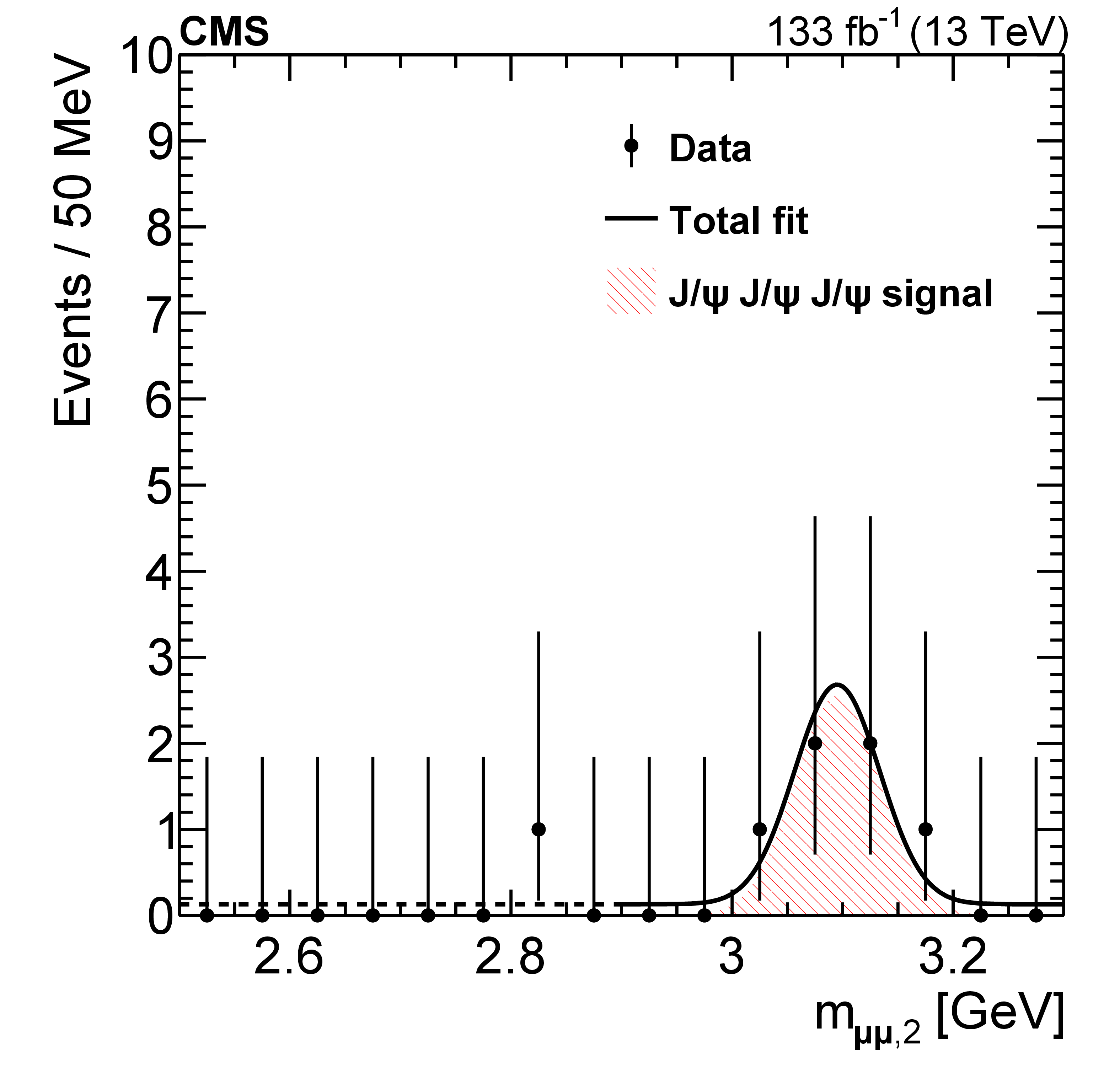}
     \includegraphics[width=0.30\textwidth]{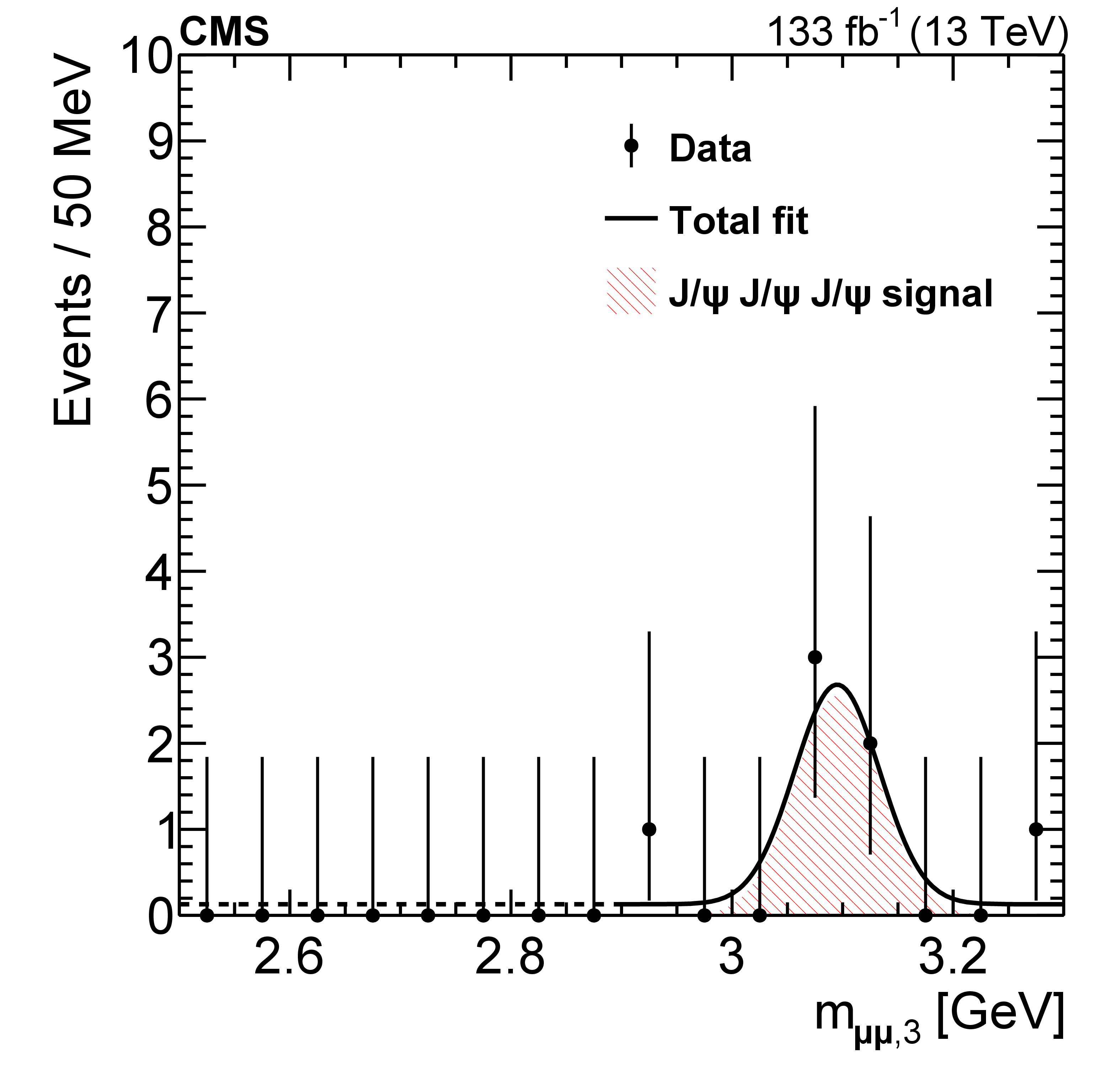}
	\caption{Projection 
 of fit on the mass of OS muon pair dimensions with the CMS triple $J/\psi$ measurement. The black solid line represents the total fit, and the red area represents the signal yield~\cite{CMS:2021qsn}.}
	\label{F CMS 3J fit}
\end{figure}

The first search for the triple $J/\psi$ production was conducted by the CMS collaboration utilizing a dataset of 133$\ \mathrm{fb}^{-1}$ at a center-of-mass energy of 13 TeV~\cite{CMS:2021qsn}. Six triple $J/\psi$ candidates were found after event selection. A 3D fit was performed to the invariant mass of OS muon pairs to extract the signal yield from the combinatorial background as illustrated by Figure~\ref{F CMS 3J fit}.

The total production cross-section was calculated to be
\begin{equation}
    \sigma(pp\to3J/\psi)=272^{+141}_{-104}(\mathrm{stat.})\pm17(\mathrm{syst.})\rm\ fb\ .
\end{equation}

\begin{figure}[H]
	\includegraphics[width=0.70\textwidth]{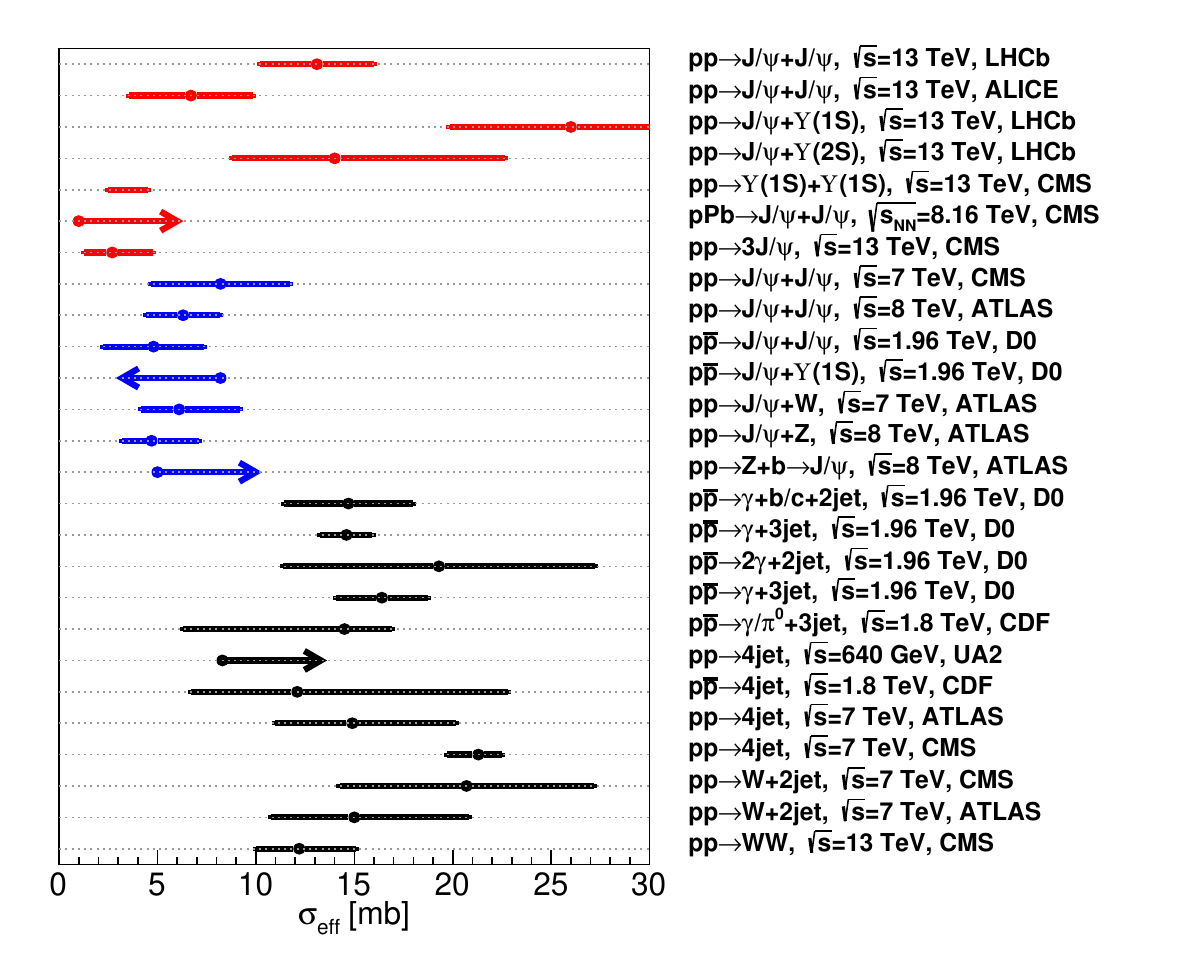}
	\caption{Effective cross-section measurement results of this review (red) and previous \mbox{studies~\cite{CMS:2024wgu, Lansberg:2019adr}} (blue and black).}
	\label{F SigmaEff}
\end{figure}

The cross-section included the non-prompt production since the non-prompt component was not excluded in this study. The sparation of SPS, DPS and TPS processes was conducted through a theory study. The production cross-section of SPS ($\sigma_{\mathrm{SPS}}^{3J/\psi}$) was theoretically estimated, and the contributions from the DPS and TPS were computed as
    
\begin{equation}
    \begin{gathered}
    \sigma_{\mathrm{DPS}}^{3J/\psi}=\frac{\sigma_{\mathrm{SPS}}^{\mathrm{2p}}\sigma_{\mathrm{SPS}}^{\mathrm{1p}}+\sigma_{\mathrm{SPS}}^{\mathrm{2p}}\sigma_{\mathrm{SPS}}^{\mathrm{1np}}+\sigma_{\mathrm{SPS}}^{\mathrm{1p}}\sigma_{\mathrm{SPS}}^{\mathrm{1p1np}}+\sigma_{\mathrm{SPS}}^{\mathrm{1p}}\sigma_{\mathrm{SPS}}^{\mathrm{2np}}+\sigma_{\mathrm{SPS}}^{\mathrm{1p1np}}\sigma_{\mathrm{SPS}}^{\mathrm{1np}}+\sigma_{\mathrm{SPS}}^{\mathrm{2np}}\sigma_{\mathrm{SPS}}^{\mathrm{1np}}}{\sigma_{\mathrm{eff}}^{\mathrm{DPS}}}\ ,\\
	\sigma_{\mathrm{TPS}}^{3J/\psi}=\frac{\frac{1}{6}\left[ (\sigma_{\mathrm{SPS}}^{\mathrm{1p}})^{3}+(\sigma_{\mathrm{SPS}}^{\mathrm{1np}})^{3}\right]+\frac{1}{2}\left[(\sigma_{\mathrm{SPS}}^{\mathrm{1p}})^{2}\sigma_{\mathrm{SPS}}^{\mathrm{1np}}+(\sigma_{\mathrm{SPS}}^{\mathrm{1np}})^{2}\sigma_{\mathrm{SPS}}^{\mathrm{1p}}\right] }{\left(\sigma_{\mathrm{eff}}^{\mathrm{TPS}}\right)^{2}}\ ,
    \end{gathered}
\end{equation}

\noindent where $\mathrm{p}$ ($\mathrm{np}$) represents promptly (non-promptly) SPS produced $J/\psi$, $\sigma_{\mathrm{eff}}^{\mathrm{DPS}}$ is the DPS effective cross-section, and $\sigma_{\mathrm{eff}}^{\mathrm{TPS}}=1.22\sigma_{\mathrm{eff}}^{\mathrm{DPS}}$~\cite{dEnterria:2016ids} is the TPS effective cross-section. By summing $\sigma_{\mathrm{SPS}}^{3J/\psi}$, $\sigma_{\mathrm{DPS}}^{3J/\psi}$ and $\sigma_{\mathrm{TPS}}^{3J/\psi}$, the total production cross-section can be expressed as a function of $\sigma_{\mathrm{eff}}^{\mathrm{DPS}}$, and the effective cross-section is determined to be
\begin{equation}
    \sigma_{\mathrm{eff}}^{\mathrm{DPS}}=2.7^{+1.4}_{-1.0}(\mathrm{exp.})^{+1.5}_{-1.0}(\mathrm{theo.})\rm\ mb\ ,
\end{equation}
\noindent where the first uncertainty originated from the uncertainty in the production cross-section measurement, and the second one represented the uncertainty of theoretical calculation. The effective cross-section measurement results summarized in this review alongside previous studies~\cite{CMS:2024wgu} can be found in Figure~\ref{F SigmaEff}. Despite the assumption that the process and energy scale universality for the effective cross-section, experimental evidence has demonstrated a strong dependence on the nature of the final state in previous studies. Quarkonium final state measurements systematically yielded values around $5\rm\ mb$ (blue dots in Figure~\ref{F SigmaEff}), contrasting sharply with the approximately $15\rm\ mb$ values characteristic of jets and electroweak bosons final state measurements (black dots in Figure~\ref{F SigmaEff}). 

Nevertheless, quarkonium studies during the LHC Run-2 period (red dots in Figure~\ref{F SigmaEff}) apparently presented a trend to resolve this longstanding discrepancy. The studies of $J/\psi J/\psi$ of the ALICE collaboration, $\Upsilon(\mathrm{1S})\Upsilon(\mathrm{1S})$, $p\mathrm{Pb}\to J/\psi J/\psi$, and 3$J/\psi$ of the CMS collaboration had a customarily small value. In contrast, studies about $J/\psi J/\psi$ and $J/\psi\Upsilon$ of the LHCb collaboration reported relatively large values, approaching those results extracted from jets and electroweak bosons. However, these higher LHCb measurements require careful evaluation given their significant tension with established results. For example, the effective cross-section extracted from process $pp\to J/\psi J/\psi$ by previous studies concentrated around $5\rm\ mb$ (CMS at $7\rm\ TeV$~\cite{Lansberg:2014swa} and ATLAS at $8\rm\ TeV$~\cite{ATLAS:2016ydt}). It cannot be explained by the energy difference since ALICE's study at $13\rm\ TeV$ also maintains a relatively small value. And obvious difference can also be noticed by comparing the $J/\psi\Upsilon(\mathrm{1S})$ channel between the LHCb's study at $13\rm\ TeV$ and the D0's study at $1.96\rm\ TeV$~\cite{Shao:2016wor}. These systematic variations underscore the critical need for both the enhancement of experimental precision and deeper theoretical understanding of effective cross-section dynamics.

The CMS collaboration's pioneering search for triple $J/\psi$ production represented the first observation of this process in $pp$ collisions, with a significance of 5.8 standard deviations. A new prospect for TPS dynamics study has been presented, and an excellent precision of the effective cross-section measurement has been provided due to the strongly suppressed SPS process characteristic of this final state. Nevertheless, the measurement precision is expected to improve substantially with future data accumulation.

\section{Summary and Prospects}

Extensive measurements of charmonium and bottomonium production in proton–proton collisions at center‐of‐mass energies from {7} {TeV} to {13.6} {TeV} have provided precise differential cross-sections and polarization parameters. These results reveal limited polarization effects, challenging leading‑order NRQCD predictions and motivating higher‑order calculations and refined long‑distance matrix elements. The LHC Run-3 and the future HL‑LHC upgrades will allow not only tighter constraints on conventional quarkonium but also a detailed study of top‐quark pair quasi-bound states. Building on the initial observations of $\eta_{t\bar{t}}$ near threshold by CMS and ATLAS~\cite{CMS:2025kzt,ATLAS:2025mvr}, Run-3 will employ higher‐precision differential angular analyses to extract the spin–parity and production dynamics of toponium resonance. Theoretical efforts in threshold resummation, potential non-relativistic QCD, and lattice inputs will also be critical to predict level splittings, decay widths and Yukawa‐coupling modifications. Achieving percent‐level precision on the toponium mass, width, and production cross-section will test QCD in the extreme non-perturbative regime, constrain the top-quark Yukawa interaction, and probe for deviations signaling new physics at the highest mass scales.

In $\mathrm{Pb}\mathrm{Pb}$ and $p\mathrm{Pb}$ collisions, the sequential suppression of $\Upsilon(\mathrm{nS})$ states and constraints on cold nuclear‑matter effects have been established, offering insights into coor screening and medium properties in heavy‑ion environments. In LHC Run‑3, oxygen--oxygen ($\mathrm{O}\mathrm{O}$) and neon--neon ($\mathrm{Ne}\mathrm{Ne}$) collisions data will enable detailed studies of quarkonium suppression across intermediate system sizes.

With unprecedented luminosity and significant advances in analysis techniques, investigations into multi-quarkonium final states have been massively conducted during the LHC Run-2 period. Some production cross-sections that have not been measured before are measured, like $J/\psi\Upsilon(\mathrm{2S})$ study of the LHCb collaboration~\cite{LHCb:2023qgu} and $\Upsilon(\mathrm{1S)}\Upsilon(\mathrm{1S})$ study of the CMS collaboration~\cite{CMS:2020qwa}; some resonances that have not been found before are found, like $T_{c\bar{c}c\bar{c}}$ study of the LHCb~\cite{LHCb:2020bwg}, CMS~\cite{CMS:2023owd,CMS:2025vnq,CMS:2025xwt,CMS:2025fpt} and ATLAS~\cite{ATLAS:2023bft} collaborations; some processes that have not been observed before are observed, like $p\mathrm{Pb}\to J/\psi J/\psi$ study~\cite{CMS:2024wgu} and triple $J/\psi$ study~\cite{CMS:2021qsn} of the CMS collaboration; even some production cross-sections that have been measured before are remeasured with more advanced methodology, like $J/\psi J/\psi$ study of the LHCb collaboration~\cite{LHCb:2023ybt}. 

While these studies have significantly advanced our understanding, they have also revealed numerous outstanding questions, such as the properties of $T_{c\bar{c}c\bar{c}}$ states, the discrepancy in the effective cross-section measurement results and the associated production of heavier states of $\Upsilon$. Considerable opportunities remain for further progress for both experiment and theory. For example, increased statistics are necessary to enable more accurate studies of rare processes, such as triple $J/\psi$, $J/\psi J/\psi \phi$, $J/\psi \Upsilon \phi$, and $\Upsilon\Upsilon$ productions; a more careful treatment is also crucial for improving measurement precision, for instance, contributions from the resonances should be considered in future $J/\psi J/\psi$ production cross-section measurement. Theoretical improvements, especially in calculation precision, are equally crucial, particularly for MC production, SPS/DPS process separation, cross-section estimation, and theory--experiment comparison. The anticipated luminosity upgrade during the HL-LHC period promises to deliver the statistical precision needed for transformative advances in multi-quarkonium studies.

The study of quarkonium production in $pp$, $p\mathrm{Pb}$, and PbPb collisions at the LHC has entered an era of NRQCD physics, where both conventional spectroscopic measurements and searches for exotic states provide stringent tests of NRQCD in various conditions. The interplay between experimental advancements and theoretical developments continues to deepen our understanding of strong interactions. As the community enter the Run-3 and future HL-LHC era, the field stands poised to address fundamental questions about hadron formation, quark--gluon plasma properties, and the potential emergence of new physics through precision studies of quarkonium systems. These developments will not only illuminate NRQCD study but may also reveal unexpected connections to other frontiers \mbox{of physics}.

\vspace{6pt} 

\authorcontributions{Conceptualization, Z.H.; methodology, Z.H.; investigation, J.L., Y.Z. and Z.H.; writing—original draft preparation, J.L. and Y.Z.; writing—review and editing, J.L., Y.Z., X.C., C.W. and Z.H. All authors have read and agreed to the published version of the manuscript.}

\funding{This work is partially supported by Tsinghua University Initiative Scientific Research Program, Natural Science Foundation of China under Grants No. 11975011, No. 12061141002 and No. 125B1008, Ministry of Science and Technology of China under Grant No. 2024YFA1610501. Y. Z. is supported by the Beijing Natural Science Foundation under Grant No. QY24227. X. C. and C. W. are supported by the Beijing Natural Science Foundation under Grant No. QY25010.}

\dataavailability{The data will be available on request.} 

\conflictsofinterest{The authors declare no conflicts of interest.} 


\reftitle{References}

\end{document}